\documentclass[final,pre,twocolumn,superscriptaddress,floatfix,longbibliography]{revtex4-1}
\usepackage[english]{babel}
\usepackage{latexsym}
\usepackage{amsmath}
\usepackage{color,soul}
\usepackage{fancyhdr}
\usepackage{graphicx}
\usepackage{wrapfig}
\usepackage{hhline}
\usepackage{dcolumn}
\usepackage[normalem]{ulem} 

\newcommand{\rev}[1]{\textcolor{black}{#1}}

\newcommand{\revtwo}[1]{\textcolor{black}{#1}}

\newcommand{\revthree}[1]{\textcolor{black}{#1}}

\newcommand{\gio}[1]{\textcolor{black}{#1}}

\usepackage{xr}
\usepackage{graphicx}
\graphicspath{{./}{./Figures/}{./SupplementalFigs/}}

\begin{document}



\title{Individuality and universality in the growth-division laws of
  single \emph{E.~coli} cells.}

\author{Andrew S. Kennard}
\affiliation{Cavendish Laboratory, University of
  Cambridge, Cambridge CB3 0HE, U.~K.}
\affiliation{Biophysics Program, Stanford University, Stanford, CA, 94305, USA}

\author{Matteo Osella}
\affiliation{Dipartimento di Fisica and INFN, University of
  Torino, V. Pietro Giuria 1, Torino, I-10125, Italy}

\author{Avelino Javer}
\affiliation{Cavendish Laboratory, University of
  Cambridge, Cambridge CB3 0HE, U.~K.}

\author{\rev{Jacopo Grilli}}
%
\affiliation{Department of Ecology and Evolution,
  University of Chicago, 1101 E 57th st., Chicago, IL, 60637, USA}
\affiliation{Dipartimento di Fisica e Astronomia `G. Galilei',
Universit\`a di Padova, via Marzolo 8, Padova, 35131, Italy}

\author{\rev{Philippe Nghe}} 
\affiliation{FOM Institute AMOLF, Science Park 104 1098 XG Amsterdam,
  The Netherlands}
\affiliation{ Laboratoire de Biochimie, UMR 8231 CNRS/ESPCI, \'Ecole
  Sup\'erieure de Physique et de Chimie Industrielles, 10 rue Vauquelin,
  75005 Paris, France}

\author{Sander J. Tans} \affiliation{FOM Institute AMOLF, Science Park
  104 1098 XG Amsterdam, The Netherlands}

\author{Pietro Cicuta}
\affiliation{Cavendish Laboratory, University of
  Cambridge, Cambridge CB3 0HE, U.~K.}

\author{Marco {Cosentino Lagomarsino}}
\affiliation{Sorbonne Universit\'es, UPMC Univ Paris 06, UMR 7238,
  Computational and Quantitative Biology, 15 rue de l'\'{E}cole de
  M\'{e}decine Paris, France}
\affiliation{CNRS, UMR 7238, Paris, France}

\date{\today}

\begin{abstract}
  \revtwo{ The mean size of exponentially dividing \textit{E.~coli}
    cells
    in different nutrient conditions is known to depend on the mean
    growth rate only.  However, the joint \emph{fluctuations} relating
    cell size, doubling time and individual growth rate are only
    starting to be characterized.
    Recent studies in bacteria 
    reported a universal trend where the spread in both size and
    doubling times is a linear function of the population means of
    these variables.
    Here, we combine experiments and theory and use scaling concepts
    to elucidate the constraints posed by the second observation on
    the division control mechanism and on the joint fluctuations of
    sizes and doubling times.
    We found that scaling relations based on the means both collapse
    size and doubling-time distributions across different conditions,
    and explain how the shape of their joint fluctuations deviates
    from the means.
    Our data on these joint fluctuations highlight the importance of
    cell individuality: single cells do not follow the dependence
    observed for the means between size and either growth rate or
    inverse doubling time.
    Our calculations show that these results emerge from a broad class
    of division control
    mechanisms 
    requiring a certain scaling form of the so-called "division hazard
    rate function", which defines the probability rate of dividing as
    a function of measurable parameters. This ``model free'' approach
    gives a rationale for the universal body-size distributions
    observed in microbial ecosystems across many microbial species,
    presumably dividing with multiple mechanisms.
    Additionally, our experiments show a crossover between fast and
    slow growth in the relation between individual-cell growth rate
    and division time, which can be understood in terms of different
    regimes of genome replication control.
}
\end{abstract}



\maketitle

\section{Introduction}


How is the size of a cell at division determined in different
environments and conditions? This simple question lies at the
foundations of our understanding of cellular growth and
proliferation~\cite{Tzur2009,Leslie2011}.
For some fast-growing bacteria, part of the question was answered
between 1958 and 1968, through a series of key studies starting from
the seminal work of Schaechter, Maal\o{}e and
Kjeldgaard~\cite{SCHAECHTER1958}.  Quoting these authors, size (mass),
as well as DNA and RNA content, ``could be described as exponential
functions of the growth rates afforded by the various media at a given
temperature.''
Remarkably, these laws for the dependency of mass and intracellular
content on population growth rate are fully quantitative, and suggest
the possibility of a theory of bacterial physiology, in the way this
term is intended by physicists~\cite{Bremer1996,Scott2010a}.
\rev{Mean} growth rate results as the sole \rev{``state variable''},
not unlike thermodynamic intensive properties such as pressure or
concentration. Specifically, the exponent of the Schaechter \emph{et
al.} curve for size has been related to the control of replication
initiation~\cite{Cooper1968,Donachie1968}, which is a key regulation
step in the cell cycle.

The understanding summarized above, however, solely relates to the
\emph{average} behavior of, e.g., \emph{E.~coli} cells within large
colonies.  A population can be made of between a handful to billions
of cells, each of which will exhibit individual growth and division
dynamics, where diversity depends both on fluctuations of the
perceived environment and on inherent stochasticity in the decision
process underlying cell division.  One has then to address how such a
heterogeneous collective of growing cells behaves in order to give
rise to the Schaechter-Maal\o{}e-Kjeldgaard ``growth law''. \rev{
  Thinking of mean growth rate as a ``control parameter'', i.e. a
  scalar variable that the cells may individually measure in their
  decisions about cell division}, one key aspect is whether each cell
is individually ``aware'' of the mean growth conditions to regulate
its individual cell division dynamics, or if it simply responds to
individual-cell parameters. These two scenarios imply different
relationships between the three main observed quantities: cell size,
individual growth rate and interdivision time (the two latter
quantities cease to be equivalent for single cells), \rev{e.g.,
  whether cells dividing at the same rate in different conditions will
  divide at similar sizes or tend to have similar growth rates.}
%
%
Early experimental efforts to capture this behavior were limited in
precision and statistics~\cite{Koch1962,Schaechter1962}. Furthermore,
such ``non-molecular'' approaches rapidly came to be considered
old-fashioned in favor of the rising paradigm of molecular
biology~\cite{Cooper1993}.
Today, the characterization of the fluctuations of cell growth and
division across growth conditions remains a largely open question,
with potential impact for our general understanding of cell
proliferation and its molecular determinants.  Additionally, advances
in hardware and computational power have made it possible to
efficiently collect high-resolution and high-quality data resolved at
the single cell level.
\revthree{Recent studies on
  \emph{E.~coli}~\cite{Campos2014,Taheri-Araghi2014} and
  \emph{B.~subtilis}~\cite{Taheri-Araghi2014} have found a near
  constancy of the size extension in a single cell cycle (a so-called
  ``adder'' mechanism of division control, see also
  refs.~\cite{Deforet2015,Tanouchi2015}). There is disagreement on
  whether the same feature is present in
  \emph{C.~crescentus}~\cite{Campos2014,Taheri-Araghi2014,Iyer-Biswas2014a,Jun2015}.
  One study~\cite{Taheri-Araghi2014} reports a universal trend in
  \emph{E.~coli} where the spread in both size and doubling times is a
  linear function of the population means of these variables, but also
  that the joint fluctuations of size and growth rate in a given
  condition deviates from the mean law followed by the means of these
  quantities. While the authors show that both behaviors are captured
  by a suitable ``adder'' model, they do not address a possible link
  between them, nor an origin that is more general than the specific
  division mechanism they consider. }

\revthree{ Here, we use a generic scaling theoretical analysis, and a
  set of high-throughput experiments to fully characterize the joint
  fluctuations of individual~\emph{E.~coli} cell size, growth rate and
  doubling times in a considerable range of growth conditions, and
  show their links with the scaling properties.
  Our experiments confirm the findings of
  ref.~\cite{Taheri-Araghi2014} on the universal scaling and
  ``individual'' joint fluctuations of size and growth rate, as well
  as uncovering novel behavior. Most importantly, the joint
  fluctuations of growth rate and doubling times also show the same
  ``individualism'', whereby two individuals with the same
  interdivision time, but coming from two populations with different
  average growth rate typically do not follow the same behavior in
  growth rate (and deviate more strongly in faster growth conditions).
  Our general theoretical approach shows that the diversity in
  individual cell-size and the scaling are linked by the control of
  cell division across conditions.
  Specifically, we calculate the condition under which the control of
  cell size varies with mean growth rate in such a way that the
  observed scaling behavior for the distributions of cell size and
  doubling time is respected, and show that this generally leads to
  the observed joint fluctuation patterns of doubling times and cell
  size.
  Importantly, while our results are compatible with a near-adder
  division control, we show theoretically that the link between
  scaling and fluctuations holds beyond this specific mechanism, and
  cannot be regarded as evidence in favor of a specific mechanism.}

\revtwo{In the following we will first introduce the experiment
  (Fig.~\ref{fig:Schematic}) and approach the problem from the point
  of view of the resulting data (Fig.~\ref{fig:Rescaling} to
  \ref{fig:SchaechterFigure}). Subsequently, we will introduce the
  theoretical approach and show how it unifies the interpretation of
  the experimental results shown in Fig.~\ref{fig:Rescaling} (collapse
  of size and doubling-time distributions) and
  \ref{fig:SchaechterFigure} (joint fluctuations of growth and
  size). The link between all these results is shown in
  Fig.~\ref{fig:SizeControl2}. Furthermore,
  Fig.~\ref{fig:AlphaDistribution} and \ref{fig:AlphaInvTau} report
  measurements on individual growth rate and interdivision times that
  are not described by the current theories.  }

\section{Results}

\subsection*{Reliable high-throughput collection of cell division
  cycles }

\begin{figure*}[t]
  \includegraphics[width=0.6\textwidth]{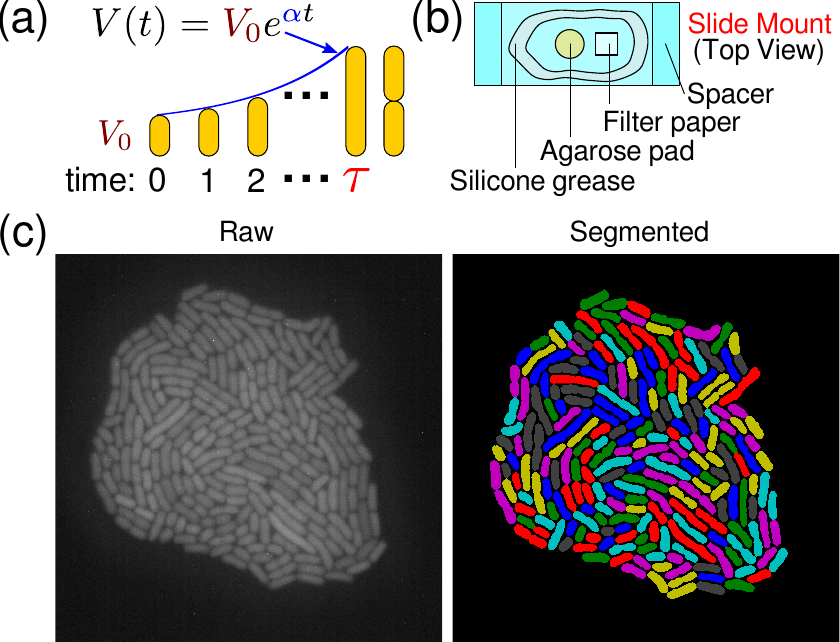}
  \caption{Description of experimental procedure. \textbf{(a)}
    Schematic of the data collected about each cell. Initial and final
    volumes $V_0$ and $V_f$ were estimated from the initial and final
    lengths of the cell and the width of the cell averaged across its
    life. The interdivision time $\tau$ was defined as the number of
    frames between two divisions, multiplied by the time between
    frames. Since cell growth was well-described by exponential
    growth~\cite{Osella2014,Wang2010a}, the growth rate $\alpha$ was
    defined by fitting the length of the cell to an
    exponential. \textbf{(b)} Schematic of the agarose pad growth
    environment. An agarose pad infused with a given growth media was
    placed on a cover slip, along with a piece of wet filter paper. A
    dilute bacterial suspension was placed on the agarose pad, sealed
    with silicone grease, and covered with a second cover slip.  The
    cover slip ``sandwich'' was placed on the microscope for viewing
    (see Methods).  \textbf{(c)} Example of the raw and processed
    data. The left panel is a representative ``raw'' image of a 
    microcolony after several generations of continuously observed
    growth. The right panel is the result of the segmentation
    algorithm applied to the raw image (see Methods).}
\label{fig:Schematic}
\end{figure*}

By using agarose pad microscopy we grew and imaged a large set of
colonies in media of varying nutrient quality
(Fig.~\ref{fig:Schematic}).  Specifically, we report five
physiological conditions from a total of four different nutrient
conditions split across two (similar) strains, in the following
referred to as P5-ori and MRR (see Methods).
A custom-made protocol involving automated imaging and efficient
segmentation algorithms (see Methods) gave us wide samples of full
cell cycles, typically order ten thousand for each condition,
including multiple biological replicates.
Since, as we mentioned in the introduction, doubling time and growth
rate are not equivalent variables for single cells, it is important to
define a consistent terminology. Fig.~\ref{fig:Schematic}a illustrates
the variables measured in our experiment.
Since growth in time of single cells is well-described by an
exponential~\cite{Osella2014,Wang2010a}, the growth rate $\alpha$ is
defined by an exponential fit.  The interdivision time $\tau$ is
defined as the time interval between two divisions. The inverse
interdivision time defines a ``rate'' or ``frequency'' of cell
division for a given cell, which can be naturally compared to
$\alpha$. Since we also consider a division hazard rate function $h$,
which defines how the probability per unit time of dividing changes
with internal cell-cycle variables such as instantaneous and initial
size, we reserve the wording division rate for $h$, \revtwo{and refer
  explicitly to inverse interdivision time otherwise}.
Finally, $V_0$ and $V_f$ are defined as the estimated spherocylinder
volume from the initial and final lengths of the cell and the average
width of the cell.
Since we monitored cell volume fluctuations across a range of
conditions, the changes in cell width made it necessary to estimate
cell volume by measuring both width and length of cells (see Methods).

\revtwo{During the analysis, we controlled for a possible dependency
  of growth parameters on position within the colony, finding that
  doubling times and growth rates of single cells are not dependent on
  colony position.  However, we found that measured cell sizes on the
  outer edge of a colony appeared larger, due to an image segmentation
  bias (Supplementary Fig.~\ref{fig:S-size-bias-colny-edge}, see
  Methods for a discussion). Removing these outer-most
  cells from the analysis did not affect the results.}

Colonies grown on agarose in microscope slides are known to show
dependency of growth rates on both time and cell position in the
colony.  To avoid problems of non-steady growth we designed and
optimized our experiment in order to prepare and keep the cells in
conditions that were as close as possible to steady growth.
Importantly, both the total cell volume and the total number of cells
grew exponentially (Supplementary
Fig.~\ref{fig:S-balanced_microcol_growth}) ---consistent with previous
reports~\cite{Wang2010a,Osella2014,Iyer-Biswas2014a,Wakamoto2013,Grant2014}
---and the growth rates of total colony volume and cell number are in
good agreement.
\revtwo{Further, colony growth rates in agar compared well with bulk
  growth rates (Supplementary Fig.~\ref{fig:S-BulkGrowthCompare}) with
  the exception of one condition, in which growth on agar was faster
  than bulk growth.
  As we shall show, however, these deviations from classic behavior in
  a single condition do not affect the statistics of cell size
  fluctuations.}

\begin{table}
  \centering
  \begin{tabular}{|c|c|c|c|}
    \hline
    Condition & $\langle V_0\rangle$ ($\mu$m$^3$) & $\langle\tau\rangle$ (min) &$\langle\alpha\rangle$ (doublings / hr) \\
    \hline
    P5ori Glc & 1.4 & 68.2 & 0.9 \\
    \hline
    P5ori CAA & 1.6 & 38.2 & 1.5 \\
    \hline
    P5ori RDM & 3.4 & 24.6 & 2.4 \\
    \hline
    MRR Glc   & 2.3 & 31.3 & 1.8 \\
    \hline
    MRR LB    & 6.5 & 20.9 & 2.9 \\
    \hline
  \end{tabular}
  \caption{Average values of the main parameters: initial size
    ($V_0$), interdivision time ($\tau$), and growth rate ($\alpha$)} 
  \label{tab:avgvals}
\end{table}

\revtwo{We analyzed 2,000-20,000 cells in each condition (Table
  \ref{tab:avgvals}) that satisfied various technical requirements
  (completely tracked over their whole cell cycle, did not cross the
  image border, had a positive growth rate, etc).
  We also verified that the area growth of microcolonies corresponded
  very well with the average growth rate of segmented cells, and that
  the distributions of all measured variables agreed with manually
  curated data, hence the divisions that the segmentation algorithm
  failed to capture did not create any relevant bias in the data
  (Supplementary Fig.~\ref{fig:S-BulkGrowthCompare}).}
The initial size distributions changed gradually with generation, at
least in part due to the segmentation problems for cells close to colony
edges mentioned above (Supplementary
Fig.~\ref{fig:S-bias_and_steadiness}, and Methods section C.4).  This
change was noticeable, but small relative to the variability present
within any one generation.
  To control for the effect of this time-dependency on results, we
  analyzed the cells from the range of generations in which the main
  growth variables are most steady as well as the full data set
  (Supplementary Fig.~\ref{fig:S-steadyness_by_gen}). \revtwo{From the
    2,000-20,000 cell divisions for each condition, about 1000-6000
    were in the steadier interval of generations (Supplementary
    Table~\ref{tab:DataSummary})~\cite{Wang2010a}. Subsequent analysis
    reported here refers to these data.}  Importantly, Supplementary
  Fig.~\ref{fig:S-unfiltered-results}, \revtwo{which reports our main
    plots on the joint fluctuations of cell size and doubling times
    for the unrestricted set, shows that the the subsample has the
    same statistical properties as the whole, and hence shows} that
  our conclusions do not depend on this restriction.

\subsection*{Single-cell size and interdivision times
rescale with growth rate.}

\begin{figure*}[t]
  \includegraphics[width=0.7\textwidth]{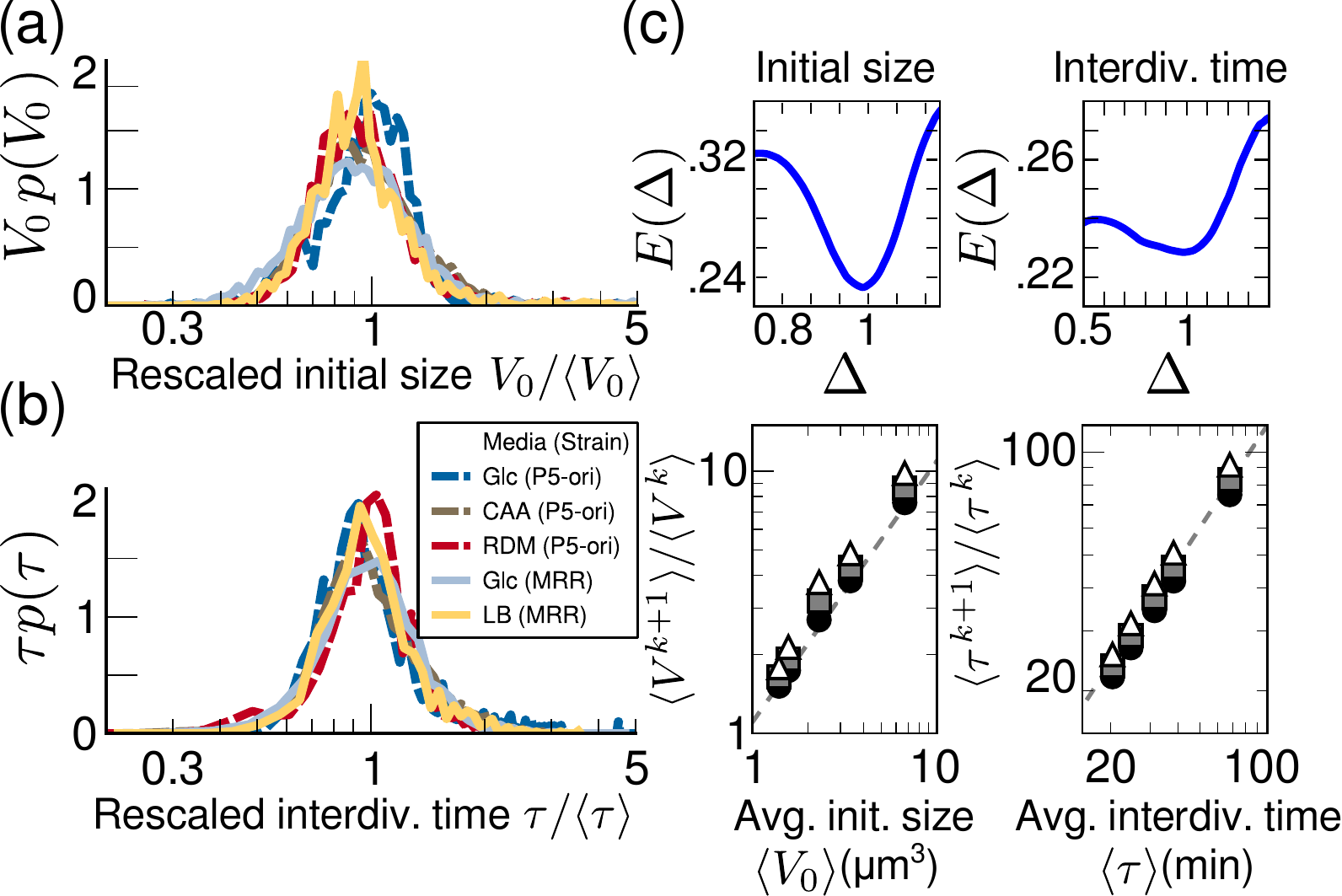}
  \caption{\emph{Escherichia coli} cell size and interdivision time
    distributions have a common scaling form across growth
    conditions. \textbf{(a)} Histograms of initial cell size (from
    $n>950$ cell cycles each) in different nutrient conditions
    (represented by different curves) with different mean growth
    rates, rescaled according to mean initial cell size $V_0$ as
    $p(V_0) = \frac{1}{V_0}F(V_0/\langle
    V_0\rangle)$~\cite{Giometto2013}. In this and later figures,
    nutrient conditions are: M9 + Glucose 0.4\% (Glc), M9 + casamino
    acids 0.5\% + Glucose 0.4\% (CAA), Neidhardt's rich defined media
    (RDM) \cite{Neidhardt1974}, and LB. See Methods for exact
    formulations. P5-ori is the shorthand for a BW25113 derivative
    strain described in the Methods, and MRR is the strain described
    in \cite{Elowitz2002}. \textbf{(b)} same plot as in (a), but for
    the interdivision time distribution. \textbf{(c)} \textit{Top
      panels}: the minimum of the functional $E(\Delta)$
    \cite{Giometto2013,Bhattacharjee2001} \rev{is a measure of the
      most parsimonious scaling exponent $\Delta$; when} evaluated for
    the distributions of initial size (left) and interdivision time
    (right), it suggests that the best estimate for the scaling
    exponent is near to one. \rev{(For the definition of $E(\Delta)$,
      see Methods.)}  \textit{Bottom panels}: linear scaling of
    successive moment ratios for the distributions of initial size
    (left) and interdivision time (right) confirms the linear scaling
    behavior. For a quantity $X$ (either initial size or interdivision
    time), filled circles represent the value of $\langle
    X^2\rangle/\langle X\rangle$ for each condition; grey squares
    represent $\langle X^3\rangle/\langle X^2\rangle$; open triangles
    represent $\langle X^4\rangle/\langle X^3\rangle$. A dashed line
    with slope one is shown as a guide to the eye.  }
  \label{fig:Rescaling}
\end{figure*}

We first considered the distributions of three main observables:
interdivision time $\tau$, growth rate $\alpha$ (obtained from fitting
an exponential to the curve of length \emph{vs.} time, see Methods),
and initial size $V_0$. \revthree{Since the distribution of initial
  sizes at one generation is defined by the distribution of final
  sizes in the previous one, in steady growth with binary cell
  division and equal daughter cell sizes, the distribution of final
  sizes divided by two has to match the distribution of initial
  sizes.}. We verified that this was the case in our data
(Supplementary Fig.~\ref{fig:S-steadyV0distr}, \revthree{since
  daughters are nearly equal\cite{Osella2014}, this condition is
  applicable to our data}).


The distribution of newly-divided cell size is right skewed, and
symmetric when plotted on a log scale, resembling a log-normal or a
Gamma distribution (Fig.~\ref{fig:Rescaling}a).  This is one of the
most consistently reported features of \emph{E.~coli}
size~\cite{Henrici1928,Schaechter1962,
  Trueba1982,Kubitschek1983,Akerlund1995,Wakamoto2005,Stewart2005,Wang2010a,Mannik2012,Taheri-Araghi2014,Campos2014}.
We found that the distribution of interdivision time $\tau$ was also
positively skewed (Fig.~\ref{fig:Rescaling}b), and resembles a
Gaussian on a logarithmic scale. This point has been discussed in the
recent
literature~\cite{Iyer-Biswas2014a,Soifer2014,Taheri-Araghi2014,Campos2014}.
Both initial size and doubling time distributions across all five
growth conditions collapse when rescaled by their means
(Fig.~\ref{fig:Rescaling}a,b).  This feature was reported early on for
\emph{E.~coli} cell sizes~\cite{Trueba1982}, and very recently also
for doubling times~\cite{Iyer-Biswas2014a} in \emph{Caulobacter
  crescentus} cells growing at different temperatures but constant
nutrient conditions.

We tested a finite-size scaling form of these
distributions~\cite{Giometto2013}
\begin{equation}
  p(x) = \frac{1}{x^{\Delta}}F\left( \frac{x}{\langle
      x\rangle^{1/(2-\Delta)}} \right),
  \label{eq:FSS}
\end{equation}
where $p(x)$ is the distribution of a quantity of interest $x$ ($\tau$
or $V_0$), $\Delta$ is a scaling exponent, and $F(\xi)$ is the
functional shape seen in the distribution of $x$, assumed to be
constant for all conditions~\cite{Giometto2013,Fisher1972}.  \gio{Note
  that the normalization condition for Eq.~\eqref{eq:FSS} requires
  some constraint on the cutoff of the upper or lower tail of the
  distribution in order to be compatible with $\Delta \neq 1$
  (discussed in the supplementary information of
  ref.~\cite{Giometto2013}). }
\rev{Eq.~\eqref{eq:FSS} is a postulate of self-similarity (stating
  that under a suitable rescaling a set of different curves are in
  fact the same), classically introduced by Fisher in the context of
  critical phenomena in statistical physics, justified by behavior of
  a thermodynamic system near a critical
  point~\cite{cardy1988}. However, in the past decades, it found
  application very broadly, for example in ecology, including
  microbial size
  spectra~\cite{Giometto2013,Rinaldo2002,Banavar2007,Banavar1999}.}
Using a quantitative method to assess the most parsimonious value for
$\Delta$~\cite{Bhattacharjee2001} \revtwo{ based on a cost function
  $E(\Delta)$ measuring the goodness of the collapse (see Methods)},
we obtained values very close to unity for this parameter
(Fig.~\ref{fig:Rescaling}c).  This suggests---as proposed in
ref.~\cite{Giometto2013}---that these size distributions can be
described by a single parameter: their mean.  Finally, we found that
the scaling prediction that the ratios of successive moments of the
size distributions should scale with the mean is
verified~(Fig.~\ref{fig:Rescaling}c).


\begin{figure*}[t]
  \includegraphics[width=0.45\textwidth]{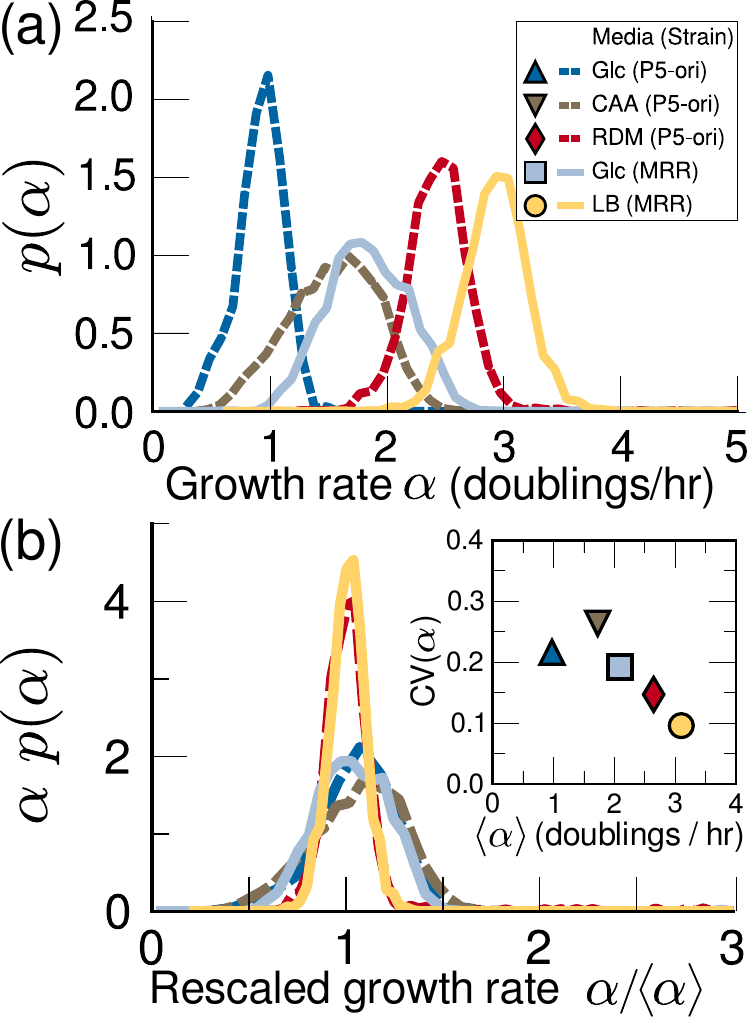}
  \caption{The distribution of single-cell growth rates is symmetric
    and does not show linear scaling with respect to the
    mean. \textbf{(a)} Distributions of growth rates $\alpha$ in
    different conditions.  \textbf{(b)} Growth rate distributions
    rescaled by the means, as in Fig.~\ref{fig:Rescaling}ab, do not
    collapse. \textit{Inset}: coefficient of variation of $\alpha$
    distributions for each experimental condition, as a function of
    average growth rate.}
  \label{fig:AlphaDistribution}
\end{figure*}

In contrast with initial size and doubling time, the distribution of
the single-cell growth rates $\alpha$ was more symmetric, and roughly
compatible with a Gaussian in all conditions
(Fig.~\ref{fig:AlphaDistribution}a), with the two faster growth
conditions visibly distinct from the rest when the distributions were
rescaled linearly as a test of the finite-size scaling hypothesis
(Fig.~\ref{fig:AlphaDistribution}b).  Notably, the coefficient of
variation (CV) of the growth rate decreases in faster growth
conditions, consistent with recent results~\cite{Kiviet2014}, and
hence the distribution does not show a simple linear scaling with the
mean across all conditions
(Fig.~\ref{fig:AlphaDistribution}b). We also tested scaling with other
exponents \rev{with the same goodness-of-collapse measurement used for
  the initial size and interdivision time \gio{but the results gave
    poorer collapse}.  Indeed, the miminum value of $E(\Delta)$ was
  noticeably higher than for the other variables (Supplementary
  Fig.~\ref{fig:Alpha_GoodnessofScaling})}. The most parsimonious
scaling exponent for the \revtwo{growth} rate distribution was
$\Delta=0.82$.

\subsection*{Increased deviations from mean-cell behavior at 
faster growth conditions.}


\begin{figure*}[t]
  \includegraphics[width=0.7\textwidth]{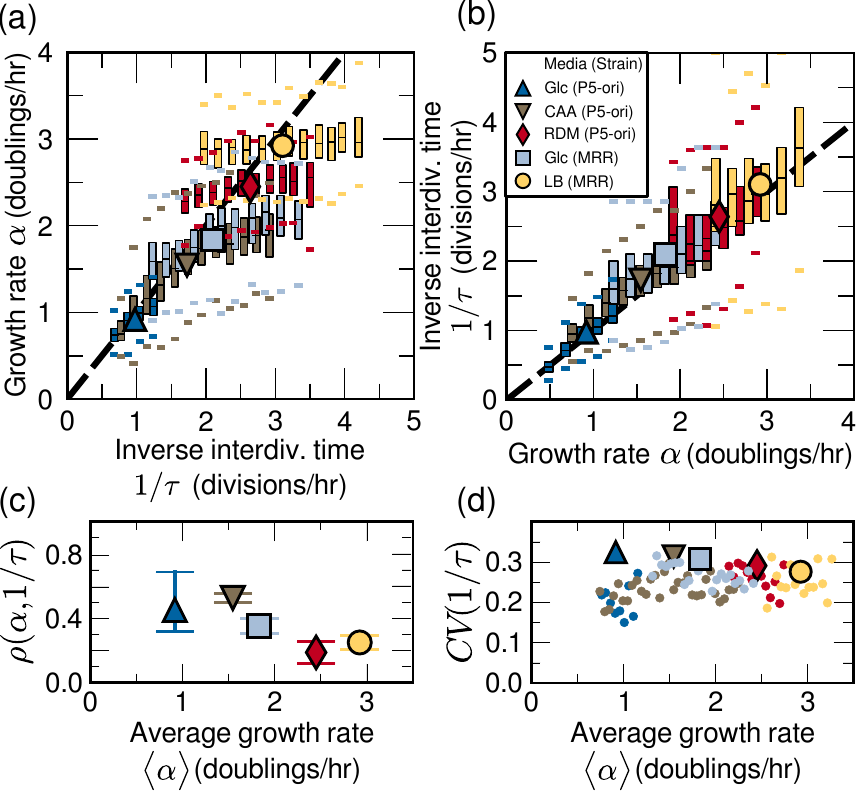}
  \caption{ Heterogeneous behaviour of growth rates and interdivision
    time of single cells. \textbf{(a)} Box plots of growth rate
    $\alpha$ vs. the inverse interdivision time $1/\tau$, showing that
    growth rates of cells with similar division frequencies (inverse
    interdivision times) are similar across slow growth conditions,
    while this correlation is lost in faster growth conditions.
    \textbf{(b)} Plot as in (a), but binned instead by growth rate
    $\alpha$, showing that the mean expected equality of division and
    doubling rates is restored at the single cell level.  Bin width is
    0.2 units of $1/\tau$ (in divisions / hr) or $\alpha$ (doublings /
    hr). Boxes are the inner quartile range and whiskers represent
    data within 1.5 times the inner quartile range; bins represent at
    least 50 cells. Large symbols represent population averages. Black
    dashed lines have a slope of one as a guide to the
    eye. \textbf{(c)} Pearson correlation coefficient between $\alpha$
    and $1/\tau$ across growth conditions. Error bars represent
    bootstrapped 95\% confidence intervals. \textbf{(d)} Coefficient
    of variation (CV) of $1/\tau$ distribution as a function of growth
    rate. Large symbols represent the whole population CV; dots
    represent CV binned by $\alpha$ (bin width 0.05 doublings / hr,
    each dot represents at least 50 cells). Discrepancy between the
    large and small dots reveals heterogeneity. }
  \label{fig:AlphaInvTau}
\end{figure*}

Next, we asked how the growth process of cells influenced cell
division.
To explore this question, we first analyzed the relation between
inverse doubling times $1/\tau$ (i.e., ``division frequencies'') and
growth rates $\alpha$ of single cells.  Fig.~\ref{fig:AlphaInvTau}a
shows boxplots of growth rates for cells with different inverse
doubling times.
As expected ---on average--- growth rate and inverse doubling time
still follow the expected trend $y=x$.
This is also confirmed by binning the same data by $\alpha$
(Fig.~\ref{fig:AlphaInvTau}b).
Conversely, the behavior of the fluctuations around this mean
evidenced by Fig.~\ref{fig:AlphaInvTau}a is different between slow and
fast growth conditions. Indeed, in faster growth conditions, cells
that divide at a given rate either because of stochasticity or carbon
source, can have very different growth rates. More specifically,
Fig.~\ref{fig:AlphaInvTau}a shows a transition in behavior at
intermediate growth rates between roughly 1.5 and 2 divisions per
hour, or equivalently at a crossover time scale \rev{of roughly 30}
minutes. This crossover is demonstrated by the slope of the plot
gradually switching from the straight line $y=x$ (expected for the
population means) to a completely flat slope, and by a drop in the
Pearson correlation (Fig.~\ref{fig:AlphaInvTau}c) between the two
variables, possibly because cells have less time to adapt their
division to transient environmental fluctuations. A similar
  crossover is visible in Fig.~\ref{fig:AlphaDistribution}, although
  the measured quantity is not the same.

Several additional observations suggest a crossover. The correlation
between inverse doubling time $1/\tau$ and initial size $V_0$ is
stronger when $\langle 1/\tau \rangle $ is less than 2 divisions per
hour (Supplementary Fig.~\ref{fig:S-Pearsonate}), 
and the correlation between $\alpha$ and $V_0$ is low except when
$\langle \alpha \rangle$ is about 1.5-2 doublings per hour
(Supplementary Fig.~\ref{fig:S-Pearsonate}).
Finally, for slower growth conditions, the CV of inverse doubling
times of a population deviates from the CV of data binned by $\alpha$,
indicating that cells with similar individual growth rates have a more
homogeneous division frequency in slow-growth conditions, while in
faster conditions the variability in their inverse interdivision times
increasingly matches the population behavior
(Fig.~\ref{fig:AlphaInvTau}d).
Taken together, these data clearly indicate that to characterize
individual cell behavior one needs to specify both mean population
growth rate and a deviation from the mean.


\begin{figure*}[t]
  \includegraphics[width=0.7\textwidth]{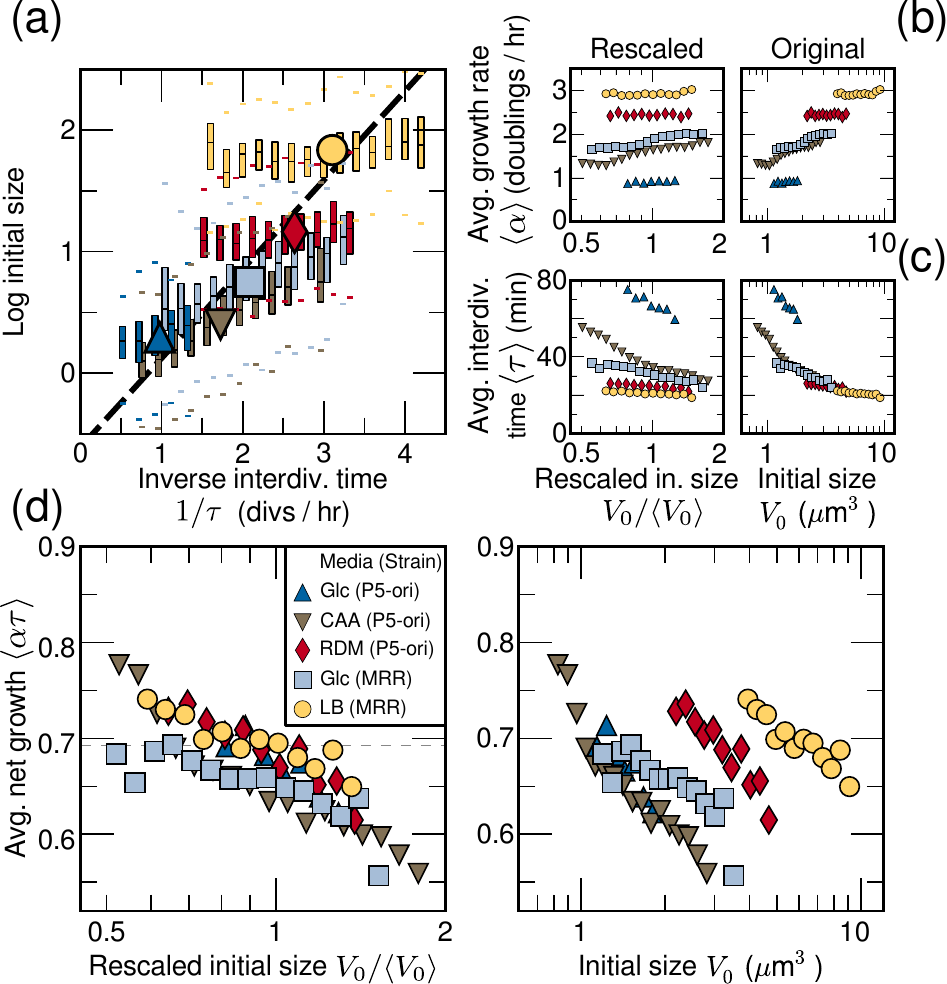}
  \caption{\revtwo{Joint fluctuations of interdivision times and size,
      and cell division control.}
    \textbf{(a)} Box plots of the logarithm of initial cell size
    binned by inverse interdivision time. Bins are as in
    Fig.~\ref{fig:AlphaInvTau}a. Large symbols represent the
    population averages. The black dashed line represents exponential
    fit of population averages and is compatible with the
    Schaechter-Maal\o{}e-Kjeldgaard result. \rev{The fluctuations
      around this mean deviate from the law, more strongly for faster
      growth conditions.}  Single-cell growth rate \textbf{(b)}, and
    interdivision time \textbf{(c)} do not show any collapse when
    plotted as a function of rescaled initial size $V_0/\langle
    V_0\rangle$, but average net growth $\alpha\tau$ (commonly used to
    proxy for size control in cell division~\cite{DiTalia2007}) does
    \textbf{(d)}. Each point represents the average value of the
    corresponding quantity binned by size (bins are constant on
    log-scale, and each bin is 0.02 units of $\log(V_0/\langle
    V_0\rangle)$). Left panels in b,c,d are rescaled versions of the
    right panels.}
\label{fig:SchaechterFigure}
\end{figure*}


Diversity of cell behavior is also evident on the single-cell analogue
of the plot from Schaechter, Maal\o{}e, and Kjeldgaard of cell size
\emph{vs} growth rate $\alpha$ or inverse of doubling time $1/\tau$
(Fig.~\ref{fig:SchaechterFigure}a and Supplementary
Fig.~\ref{fig:S-Schaechter_alpha}).  As previously discussed, inverse
doubling time (division frequency) is equivalent to growth rate only
when averaged over a population in steady-state growth conditions
(i.e., $\langle \alpha\rangle =\log(2)/\langle \tau\rangle$), but the
two quantities represent (in principle) independent variables at the
single-cell level.  Fig.~\ref{fig:SchaechterFigure}a and Supplementary
Fig.~\ref{fig:S-Schaechter_alpha} show that fixing \rev{either
  variable, the} deviations from the population behavior become
stronger in faster growth conditions;
\rev{furthermore, the Schaechter-Maal\o{}e-Kjeldgaard ``growth law''
  (stating that for balanced growth, mean cell size increases
  exponentially when plotted against the mean of the growth rate or
  the reciprocal of the mean doubling time) does not appear to hold at
  the single cell level in even the slowest conditions}.
These findings indicate that also the laws coupling individual cell
growth to division (hence to cell size) cannot be extrapolated from
the population averages, seemingly in contrast with the universal
features of size and doubling time fluctuations.
On the other hand, the average sizes of cells growing in different
conditions in our data are fully compatible with the \rev{expected
  trend} \rev{(Fig.~\ref{fig:SchaechterFigure} and Supplementary
  Fig.~\ref{fig:SchaechterLinLin})}.
%
%


\subsection*{\revtwo{Fluctuations in cell size and interdivision times
 are linked to cell division control.}}

The roles of individual growth rate and doubling time in setting cell
division size may be profoundly different.
\revtwo{The slope of the plots in Fig.~\ref{fig:SchaechterFigure}a
  (and Supplementary Fig.~\ref{fig:S-Schaechter_alpha}) may be
  interpreted as a test for how much a cell that is born larger or
  smaller than average compensates for this error by modulating its
  growth or interdivision time.} Equivalently, the changes in size
control at different growth rates are shown directly by scatter plots
of doubling time $\tau$ and single-cell growth rate $\alpha$ versus
logarithmic initial size $\log V_0$
(Fig.~\ref{fig:SchaechterFigure}b,c).
Consistently with previous results~\cite{Osella2014}, these plots show
little correlation between initial size and growth rates
(Fig.~\ref{fig:SchaechterFigure}b) and significant anticorrelation
between initial size and interdivision time
(Fig.~\ref{fig:SchaechterFigure}c),
suggesting that the control of cell size should be mostly effected by
modulating doubling times rather than growth rate. Additionally,
the slopes of these plots show variability across conditions even when
rescaled by mean initial size, reinforcing the idea that the extent of
this doubling time modulation varies in the different conditions along
the Schaechter-Maal\o{}e-Kjeldgaard curve.
To test how this is compatible with the observed universal scaling of
initial size distributions, we considered another way to quantify size
control in cell division, comparing the amount of relative growth
within a time interval versus the cell size at the entrance of the
interval (Fig.~\ref{fig:SchaechterFigure}d, often referred to as a
``size-growth plot'')~\cite{Osella2014,Skotheim2013a,DiTalia2007}.
The slope of this plot is normally considered a proxy of how much cell
division depends on cell size.
%
Fig.~\ref{fig:SchaechterFigure}d shows the average net growth $\langle
\alpha \tau \rangle$ \emph{vs} initial size.  These curves show a
common slope and, analogously to the size distributions, they collapse
when rescaled by the mean initial size in each condition. 
Note that this is possible only because the correlation of $\alpha$
with $1/\tau$ is nonzero and varies across conditions; one extreme
case is LB, where the trend of both $\alpha$ and $1/\tau$ with initial
size is very weak, but the trend in Fig.~\ref{fig:SchaechterFigure}d
is the same as in other conditions.
These results are consistent with a mechanism of cell size control
that modulates the division time, such that the scaling is maintained,
or, equivalently, operated by a mechanism that contains a single
intrinsic length scale~\cite{Iyer-Biswas2014a}.
\revtwo{Our measurements are also consistent with the nearly constant
  added volume in each cell cycle reported recently for
  \emph{E. coli}~\cite{Jun2015} (Supplementary
  Fig.~\ref{fig:Sadder}).}

%
%
%

\begin{figure*}[t]
  \includegraphics[width=0.5\textwidth]{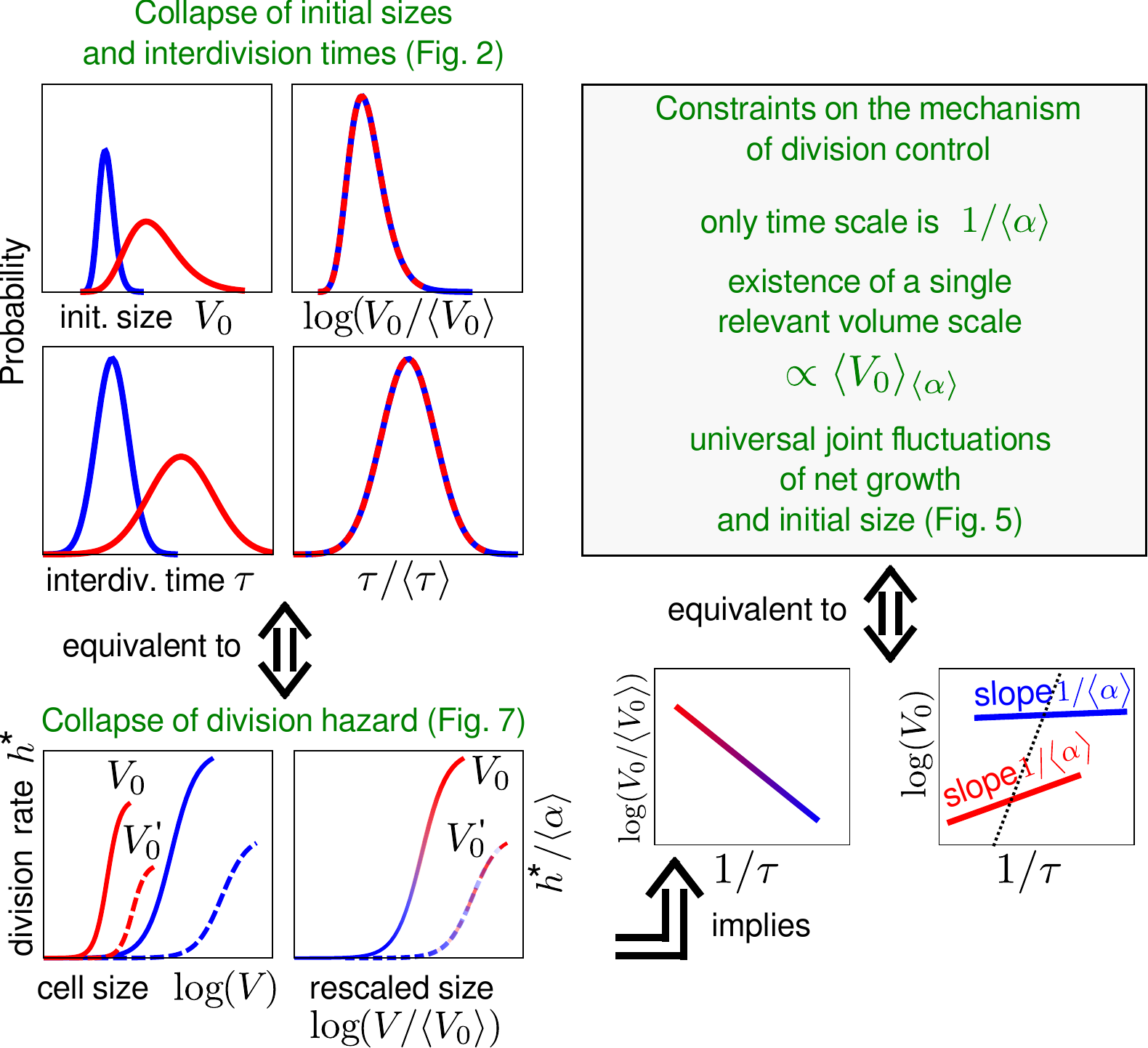}
  \caption{ \revthree{Theoretical analysis showing that division
      control across different growth conditions is intimately linked
      to the universal size and doubling-time distributions.  Scheme
      of the theoretical result, which unifies the findings of
      Fig.~\ref{fig:Rescaling} and \ref{fig:SchaechterFigure}. In the
      cartoons, different colors refer to different conditions. The
      collapse of initial sizes and doubling times are equivalent to
      the collapse property of the division rate
      (Eq.~\eqref{eq:collapseh}), when plotted as a function of size
      rescaled by the average initial size, and with $h^*$ rescaled by
      growth rate. The collapse of the size-growth plot in
      Fig.~\ref{fig:SchaechterFigure}d is a consequence of these
      properties. The theory predicts that the slope of the
      fluctuations around the Shaechter-Maal\o{}e-Kjeldgaard law should
      be the inverse of the mean growth rate. } }
\label{fig:SizeControl2}
\end{figure*}

\subsection*{\revtwo{Theoretical constraints posed by finite-size
    scaling on division control.}}

\revtwo{To address the relationship between scaling, cell division
  control, and individuality in fluctuations observed in our data, we
  used a theoretical approach (Fig.~\ref{fig:SizeControl2}).  The
  framework we employed generally describes cell division through the
  growth-division process in terms of a division hazard rate function
  $h^*$\cite{Osella2014,Taheri-Araghi2014}. The hazard division rate
  is defined as the probability per unit time that a cell divides,
  given the values of the available state variables (e.g., current
  size, cell-cycle time, etc).  This general description allows us to
  show that the collapse of initial size and doubling time
  distributions and the fluctuations around the
  Schaechter-Maal\o{}e-Kjeldgaard law can be explained as a common result
  of the division control mechanism.}

\begin{figure*}[t]
  \includegraphics[width=0.7\textwidth]{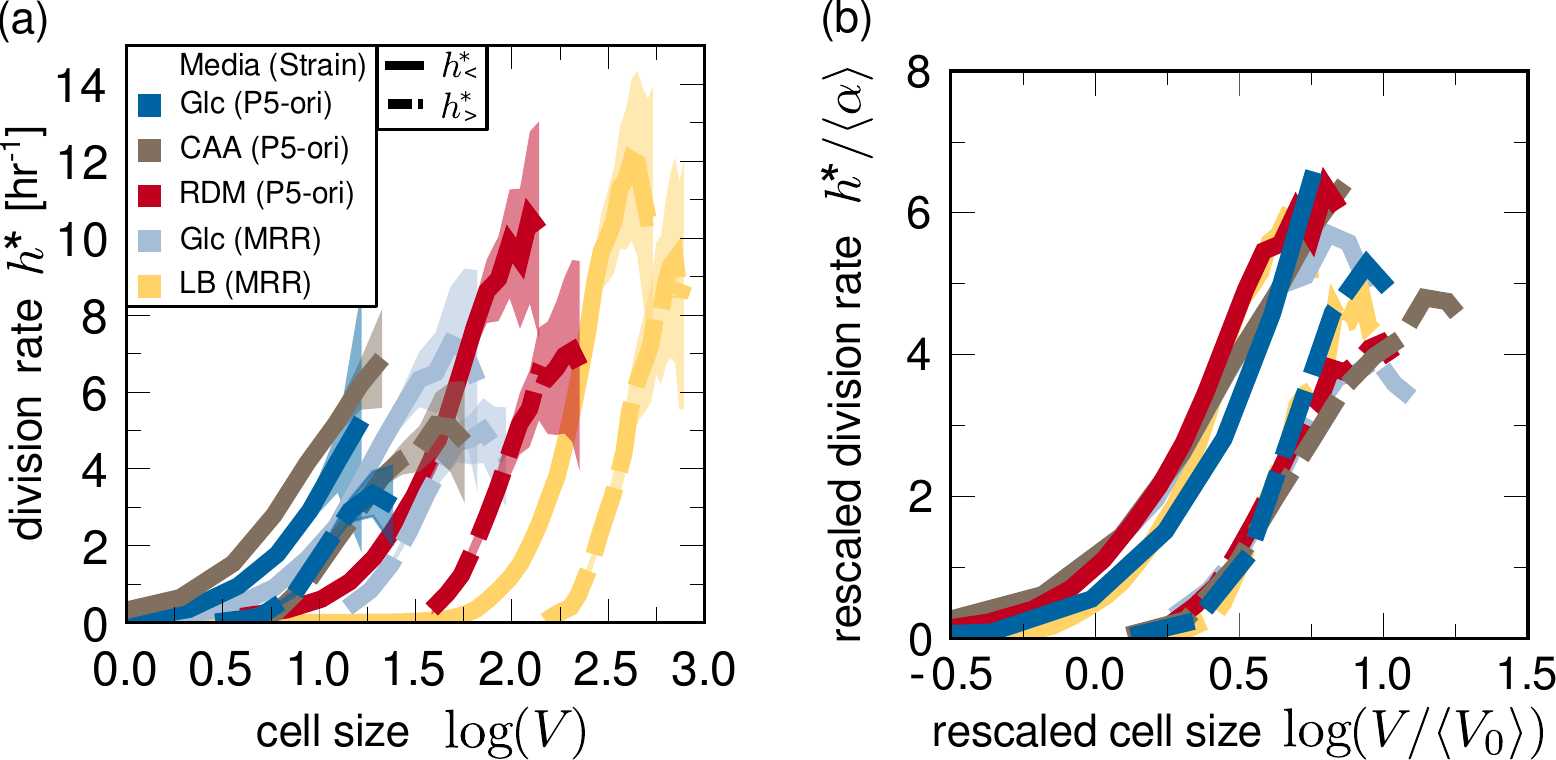}
  \caption{ \revthree{ Division hazard rates inferred from data follow
      the predicted collapse properties. \textbf{(a)} Hazard rate
      plotted as a function of size, conditional on initial size.
      $h^*_<(V)$ (solid lines) is the rate of cell division for cells
      whose initial size was smaller than the average initial size;
      $h^*_>(V)$ (dashed lines) is the rate of cell division for cells
      whose initial size was larger than the average initial size (the
      curves would be the same if division depended only on current
      size~\cite{Osella2014,Taheri-Araghi2014}). \textbf{(b)} the
      \revtwo{empirical} functions $h^*_{\langle \alpha
        \rangle}(V,V_0)$ follow the collapse property of
      Eq.~\eqref{eq:collapseh} (see also Supplementary
      Fig.~\ref{fig:SizeControl}).  } }
\label{fig:SizeControl3}
\end{figure*}

Specifically, we assumed a division hazard rate of the form
$h^*_{\langle \alpha \rangle}(V,V_0)$ (for a population with given
mean growth rate $\langle \alpha \rangle$), and asked under which
conditions this hazard function can generate the observed scaling
behavior of the doubling-time and initial-size distributions.  This
assumption includes \revtwo{\emph{as a particular case} ``adder''}
models where the control variable is a size difference
$V-V_0$~\cite{Campos2014,Taheri-Araghi2014,Soifer2014} as well as
models where elapsed time from cell division is a control variable
instead of $V_0$, provided the distribution of growth rates is
sufficiently peaked~\cite{Osella2014}.
\revtwo{To understand this, note that $h^*$ can be a function of all
  the state variables $(t,V_0,V,\alpha)$, but under the constraint of
  exponential growth $V_f=V_0 e^{\alpha\tau}$, different choices of
  parameters become equivalent.}
\revtwo{(The full calculation, as well as further details about the
  formulation of the model, are reported in the Appendix). The essence
  of the calculation is that the initial size distribution
  $\rho_{\langle \alpha \rangle}(V_0)$ can be obtained as a functional
  of $h^*$ by solving the model. One can then impose the finite-size
  scaling condition on $\rho_{\langle \alpha \rangle}$ and derive the
  consequences for $h^*$.  
  This gives the condition}
\rev{
\begin{equation}
  h^*_{\langle \alpha \rangle}(V,V_0) = 
\langle \alpha
        \rangle  f \left( \frac{V}{\langle V_0 \rangle_{\langle \alpha
        \rangle}} \, , \, 
    \frac{V_0}{\langle V_0 \rangle_{ \langle \alpha
        \rangle}}     \right) \ .
\label{eq:collapseh}
\end{equation} }
In other words, our theoretical calculations show that under the
condition stated by Eq.~\eqref{eq:collapseh} (i.e., the scaling form
of the division hazard rates from different conditions), the
observed scaling behavior for doubling times and initial sizes
(Fig.~\ref{fig:Rescaling}) hold, and are equivalent.
\revtwo{To test
  Eq.~\eqref{eq:collapseh}, i.e. the collapse of the division hazard
  rate $h^*_{\langle \alpha \rangle}(V,V_0)$, with data, we used} direct
inference from the histograms of dividing cells. \revtwo{The procedure
  is described in detail in ref.~\cite{Osella2014} and in the
  Appendix, and is based on the fact that, as in a Poisson process,
  the division hazard rate $h^*$ is mathematically related to
  conditional histograms of undivided cells.}
Fig.~\ref{fig:SizeControl3}ab  shows that
the condition given in Eq.~\eqref{eq:collapseh} is verified in our
data.

\subsection*{\revtwo{The theory justifies the increased deviations of
    fluctuations from means in faster growth conditions.}}

Furthermore, the dependencies of the division hazard rate
\revtwo{determine} the slope and collapse of the size-growth plot
(Appendix and Fig.~\ref{fig:SchaechterFigure}d). Since the size-growth
plot is also related to the heterogeneus behavior in the growth of
single cells (Fig.~\ref{fig:SchaechterFigure}a and Supplementary
Fig.~\ref{fig:S-Schaechter_alpha}), this \revtwo{shows} that, while
apparently in contrast, the universal behavior of the fluctuations and
the deviations of single cells from the Schaechter-Maal\o{}e-Kjeldgaard
behavior are in fact two sides of the same coin.
\revtwo{ This link can be derived directly from
  Eq.~\eqref{eq:collapseh}, as we report in the Appendix (see also
  Fig.~\ref{fig:SizeControl2}).  Here, we support it with the
  following simple quantitative argument, valid for small
  fluctuations.  Fig.~\ref{fig:SchaechterFigure}d implies that $\alpha
  \tau \approx \log2 - 1/b \log(V_0 / \langle V_0
  \rangle_{\langle\alpha\rangle}) $, where $b \simeq 2$ is a
  dimensionless constant (the horizontal dashed line in the plot is
  $\log 2$). However, Fig.~\ref{fig:SchaechterFigure}b shows that most
  of the correlation with size is contained in $\tau$. One can then
  suppose that $\alpha \tau \approx \langle\alpha \rangle \tau$. From
  these two conditions, one gets that
  \begin{equation}
    (1/\tau) (\log2- \frac{1}{b} \log (V_0/ \langle V_0
    \rangle_\alpha)) \approx \langle\alpha \rangle \ . 
\end{equation}
Assuming small fluctuations (see also the Appendix,
Eq.~~\eqref{eq:fig5aformulafinal}), the first term in the left-hand
side of this equation can be written as a mean, plus a fluctuation
term, $\langle \alpha \rangle + \delta_{1/\tau}$, while the second
term is interpreted as a fluctuation of logarithmic size $\delta_{\log
  V_0}$. Assuming small fluctuations, one immediately has that
\begin{equation}
 \delta_{\log V_0}  \approx \frac{ \delta_{1/\tau}}   { \langle\alpha \rangle}
 \  .
\end{equation}
In other words, the fluctuations in logaritmic cell size around the
Schaechter-Maal\o{}e-Kjeldgaard law should become shallower in faster
growth conditions, coherently with the observed trend in
Fig.~\ref{fig:SchaechterFigure}a.}

In conclusion, the joint universality in doubling time and size
distributions can be explained by a generic division control mechanism
based on a single length scale.  \revtwo{Importantly,
  Eq.~\eqref{eq:collapseh} shows that other mechanisms, and not only
  the adder principle, may exhibit both scaling and individuality in
  the fluctuations.  Thus, scaling and individuality are more general
  and not evidence or simple consequence of near-adder
  behavior\cite{Taheri-Araghi2014,Jun2015}. The Appendix also shows
  how this result holds using specific examples of division control
  models~\cite{Osella2014}.
  For an adder, Eq.~\eqref{eq:collapseh} translates into the
  additional constraint that the division rate $h^*_{\langle \alpha
    \rangle}(V-V_0)$ should be a function of $\frac{V-V_0}{\langle V_0
    \rangle_{\langle \alpha \rangle}}$, which immediately implies the
  prediction that, in each condition, each moment of order $k$ of the
  distribution of added size should be proportional to the $k$-th power
  of $\langle V_0 \rangle_{\langle \alpha \rangle}$ (since this is the
  only relevant length scale), while stationarity of the process
  requires that only their means are equal (as stated in
  ref.~\cite{Taheri-Araghi2014}). }

\section{Discussion and Conclusions}


\revtwo{Our study shows that single cells from a given condition with
  a defined average division rate deviate from the
  Schaechter-Maal\o{}e-Kjeldgaard ``growth law'' (which states that mean
  cell size grows exponentially with mean growth rate), with stronger
  trend for faster growth conditions.}
A similar \revtwo{``individuality'' in cell behavior} relates growth
rate to cell division: at slow growth, individual cells appear to
adapt their doubling time to match their individual growth rate (thus
behaving like a small colony). Conversely, at fast growth the
correlation between inverse doubling times and individual growth rates
decreases visibly.  A crossover time scale \rev{around 30 minutes} is
seen across the data, marking the transition between these two
regimes.
In analogy with the standard interpretation linking the
Schaechter~\emph{ et al.} law with the control of replication
initiation~\cite{Cooper1968,Donachie1968}, one can speculate that this
characteristic time may be connected to replication time: for example,
at fast growth, variability in interdivision times might be more
dependent on DNA replication, which becomes increasingly challenging
in presence of overlapping rounds, while other determinants of cell
division might be more relevant in slow
growth. 
A connection between fluctuations in growth variables and multifork
replication is also consistent with the qualitatively different
correlations between $\alpha$ and $1/\tau$ observed in our work
compared to that recently shown in ref.~\cite{Iyer-Biswas2014a}, since
\emph{C. crescentus} does not use multifork replication.  Iyer-Biswas
\emph{et al.} found that $1/\alpha$ and $\tau$ were well correlated in
all growth rates they observed, similarly to our data from slowly
growing \emph{E.~coli}, which likely are not undergoing multifork
replication.

\revtwo{ Our fast-growth results are consistent with findings on cells
  growing steadily in a microchemostat in rich growth
  conditions~\cite{Wang2010a,Osella2014} and in line with more recent
  microchemostat results~\cite{Taheri-Araghi2014} (Supplementary
  Fig.~\ref{fig:Sadder} and Fig~\ref{fig:SchaechterFigure}a) }
\revtwo{Finally, we compared our results to previously obtained data
  in three additional growth media, including poor carbon sources, in
  order to enhance the range of explored growth rates (Supplementary
  Fig.~\ref{fig:TansData}). These extra experiments were also in line
  with our main results, showing collapse of size and doubling time
  distributions, as well as increased deviations from mean behavior at
  faster growth rates. }


We now address the measurements of the distributions of the main
variables. The fact that the distribution of cell size is right skewed
is one of the most consistently reported features in the
\emph{E. coli} literature~\cite{Henrici1928,Schaechter1962,
  Trueba1982,Kubitschek1983,Akerlund1995,Wakamoto2005,Stewart2005,Wang2010a,Mannik2012},
and it has been derived theoretically using different assumptions
about the dynamics (or fluctuations) in the growth
process~\cite{Koch1962,Hosoda2011a,Amir2014,Osella2014,Robert2014,
  Giometto2013}.  The evidence on the shape of the doubling time
distribution has been less consistent, with some studies observing
that the distribution is weakly skewed and close to
Gaussian~\cite{Schaechter1962}, and other studies finding positive
skew in the
distribution~\cite{Powell1955,Voorn1998,Ullman2013,Iyer-Biswas2014a}.
%
The unskewed distribution of the growth rate $\alpha$ has previously
been reported for one growth
condition~\cite{Wang2010a,Osella2014}. The $\alpha$ distributions for
different mean values $\langle\alpha\rangle$ have been considered in
ref.~\cite{Taheri-Araghi2014}, which appeared while the present work
was under review. \revthree{Interestingly, while this work only tests
  the collapse with $\Delta = 1$, this seems to be good, in some
  contrast with our agar-derived data. An analytical form of this
  distribution has been obtained by a recent modeling
  study~\cite{Pugatch2015}. The functional form is Log-Fr\'echet, which
  appears to fit the published single-cell data better than other
  forms, and is not collapsible.  }


The linear finite-size-scaling form of the initial size and doubling
times distributions is consistent with recent results in
\emph{Caulobacter crescentus}~\cite{Iyer-Biswas2014a} for cells grown
at different temperatures. Earlier work had shown such a scaling for
size, but had not investigated doubling time~\cite{Trueba1982}.  Our
experiments extend the findings in \emph{C.~crescentus} to a
phylogenetically distant bacteria with a radically different cell
cycle, as well as a complementary perturbation (change of nutrient
conditions instead of temperature), showing that the scaling
properties of these distributions are unvaried for cells grown at the
same temperature in different media.
Interestingly, while the linear scaling suggests that the mean
behavior (the relative time/length scale) fully sets the shape of the
size distribution, the naive expectation would be that the
fluctuations around the mean size would also behave equally in
different conditions.  It is then interesting to ask how these
differing properties relate to the shape of the size and doubling-time
distributions.

An important standing question is what sets this markedly universal
scaling for both size and doubling times.
Iyer-Biswas and coworkers~\cite{Iyer-Biswas2014a,Iyer-Biswas2014}
employ an autocatalytic model for growth fluctuations to predict that,
within a cell cycle, cell sizes should not follow a multiplicative
random walk, but a multiplicative process where the noise scales as
the square root of size. Under these conditions, the growth dynamics
preserve the scaling of the size distribution, and provided that
binary division does not affect this property, scaling should be
observed.
This reasoning is robust and consistent with
data~\cite{Iyer-Biswas2014a}.  However, \revtwo{being focused mostly
  on growth} it does not fully address the possible role of cell
division in setting the shape of the distribution.

In our case, we are able to show theoretically that in such models,
finite-size scaling of the size and interdivision time distributions
is directly related to the collapse \rev{of the division hazard rate
  functions of different conditions,
  Eq.~\eqref{eq:collapseh}}. 
Since this would not necessarily be the case if the scaling were
purely determined by the cell growth process, we are led to surmise
that both growth and cell division contribute to the observed size and
doubling-time fluctuations.  
Considering the data, two different measurements of cell division
control---the size-growth plot between net growth and initial size
(Fig.~\ref{fig:SchaechterFigure}d) and our direct estimate of the
division hazard rate as a function of cell size--- show rescaling
collapse, suggesting that cell division control across conditions
contains the same universal scale observed in the size distributions.
Hence, since the size-growth plot is also directly related to the
fluctuations around the Schaechter-Maal\o{}e-Kjeldgaard curve
(Fig.~\ref{fig:SchaechterFigure}a), the outcome of this analysis
suggests that both the observed finite-size scaling and the
heterogeneity in single-cell behavior across conditions may have a
common explanation through cell division control.

\revtwo{It is important to frame this result in the current debate
  regarding the specific mechanism for the division control.}
Importantly, our theoretical result is achieved through a ``model
free'' approach, i.e., a generic argument that can apply to a wide
range of division mechanisms. Recent
works~\cite{Campos2014,Taheri-Araghi2014,Soifer2014} have
\revtwo{shown evidence} in favor of ``adder'' mechanisms of cell
division, where the division hazard rate depends on the volume added
by a cell $h^*(V,V_0)=h^*(V-V_0)$ \revtwo{and a nearly constant mean
  volume is added at each cell division}. Our analysis \revtwo{and our
  data} are compatible with this mechanism \revtwo{(Supplementary
  Fig.~\ref{fig:Sadder})}. However, our calculations (see Appendix)
also indicate that the scaling of size and doubling-time distributions
and the fluctuation behavior around the Schaechter-Maal\o{}e-Kjeldgaard
curve should not be regarded as a \revtwo{smoking gun for} an adder
mechanism. \revtwo{Indeed, different hazard rate functions than that
  of an adder can obey the scaling given by Eq.~\eqref{eq:collapseh}.
  Supplementary Fig.~\ref{figShd} and~\ref{figScollapse}, show
  specific examples of non-adder models with universal size and
  interdivision-time distributions}.

\revtwo{We conclude that the apparently contrasting universal behavior
  of the fluctuations, and the deviations of single cells from the
  Schaechter-Maal\o{}e-Kjeldgaard behavior, are in fact two sides of the
  same coin. They come from control of cell division, but they do not
  suffice to pinpoint a single specific mechanism of cell division
  control.}  The idea that division control plays a relevant role in
setting size and doubling time distributions is also supported by the
finding of Giometto and coworkers~\cite{Giometto2013}.  These authors
observe size scaling for a wide range of microorganisms \revtwo{in the
  context of a microbial ecosystem}, not all of which presumably grow
\revtwo{and divide in the same way, suggesting that the reason for the
  scaling behavior of sizes and doubling times should go beyond the
  specificity of a single
  mechanism}~\cite{Giometto2013,Jun2015,Iyer-Biswas2014a}.
\revtwo{Finally, we note} that our explanation \rev{of the link
  between size fluctuations and scaling behavior} does not include the
additional heterogeneous behavior \rev{that we found experimentally
  between} growth rates and doubling time
(Fig.~\ref{fig:AlphaInvTau}), and its crossover time scale.
A model fully accounting for fluctuations in both the growth and
division processes is still lacking, but the data reported here should
provide important clues to construct it.

\section{Materials and Methods}


\subsection*{Strains and Growth Conditions}
Two strains were used in this research: a GFP reporter strain of
BW25113 (gift of Dr. Bianca Sclavi) with \emph{gfp} and a kanamycin
resistance cassette fused to the $\lambda$ phage P5 promoter and
inserted near the \emph{aidB} gene and the origin of
replication---this strain is referred to as P5-ori. The second strain
was the MRR strain previously described in~\cite{Elowitz2002}.

Four different media were used: LB (Lennox formulation, Sigma L3022);
Neidhardt's rich defined media \cite{Neidhardt1974}, referred to here
as RDM (Teknova); and M9 (Difco, 238\,mM Na$_2$HPO$_4$, 110\,mM
KH$_2$PO4, 43\,mM NaCl, 93\,mM NH$_4$Cl, pH~$6.8\pm0.2$, supplemented
with 2\,mM MgSO$_4$ and 100\,$\mu$M CaCl$_2$ (Sigma)) with either
0.4\%\,w/v of Glucose (Sigma) or 0.4\%\,w/v Glucose and 0.5\%\,w/v
casamino acids (Difco) added. M9 media were prepared by autoclaving
separately M9 salts, MgSO$_4$, CaCl$_2$, and casamino acids, and
combining after autoclaving. Glucose was filter sterilized.
\rev{Additional data (Supplementary Fig.~\ref{fig:TansData}) were
  obtained as described in ref.~\cite{Kiviet2014}, for 3 different
  nutrient conditions: M9 + Acetate, M9 + Lactose, and Neidhardt's
  Rich Defined Media (RDM) + Glycerol, spanning growth rates from
  between 0.25 to 1.8 doubling per hour.}

Strains were temporarily stored on LB-agar plates with appropriate
selective antibiotic at 4$^{\circ}$C for up to one week. Prior to an
experiment, cultures were inoculated into LB with appropriate
selective antibiotics and incubated at 37$^{\circ}$C with shaking at
200\,rpm overnight (10-16\,hours). Cultures were then diluted
1000$\times$ into 10\,mL of growth medium without antibiotics in a 50\,mL
Ehrlenmeyer flask with a loosened cap for oxygen exchange, and
grown until early exponential phase (OD$_{600}\sim 0.05$)---3-10 hours
depending on the growth rate. The culture was diluted again into fresh
pre-warmed media and grown to $OD_{600}\sim0.05$, 2-6 hours depending
on growth rate.

\subsection*{Microscopy}
Agarose pads were cast using a custom-made mould, maintained at
35$^{\circ}$C. Sterile molten agarose (3\%\,w/v, Sigma) was
mixed 1:1 with pre-heated 2$\times$ growth media, poured onto a
coverslip placed in the mould, covered with a glass slide, and allowed
to cool. Agarose pad height was measured with a digital caliper to be
$0.48\pm0.04$ mm (standard deviation, $n=4$).

Immediately before starting the microscopy experiment, a disc was cut
out of the agarose pad using an 8\,mm biopsy punch and placed on a
coverslip heated to 37$^{\circ}$C.  0.18\,mm spacers were placed on
each end of the coverslip, and a piece of damp filter paper (approx
6\,mm square) was placed next to the agarose pad to decrease
evaporation. The pad was inoculated with 3\,$\mu$L of bacterial culture
diluted to $\sim.0006$\,OD units (approximately 1,000 cells total).
The pad and filter paper were sealed with air-permeable silicone
grease and a second coverslip was pressed on
top. 

The agarose pad-coverslip ``sandwich'' was transported to the
microscope on a metal block heated to $37^{\circ}$C to minimize
temperature shock. During the experiment the sample was heated by
direct thermal contact with the objective via the immersion oil. The
objective was maintained at 37$^{\circ}$C using a custom-built PID
controlling  an objective jacket from ALA Scientfic Instruments.

Cells were imaged using a Nikon Eclipse Ti-E inverted microscope
equipped with ``perfect focus'' autofocusing hardware and a $60\times$
oil objective (NA\,1.45). Images from the MRR strain and the Glucose
experiments of the P5-ori strain were taken with an Andor iXon
DU897\_BV EMCCD camera using EM gain, \revthree{and an additional
  Nikon 2.5x magnifying element (VM lens
  C-2.5x). The pixel size was measured to be
  .106 um/pixel, within the optimal Nyquist sampling regime. }  For the CAA and RDM experiments in the P5-ori strain a
Ximea MQ042MG-CM camera was used, \revthree{with spatial sampling of
  0.09167um/pixel, again within the Nyquist regime}. Fluorescence
images were taken with light from a blue LED passed through a GFP
filter (Semrock: excitation FF01-472/30, dichroic FF495-Di03, emission
FF01-520/35).  \revthree{Frequency and intensity of illumination were
  kept as low as possible, since fluorescence illumination has
  phototoxic effects, which may slow down cell growth and cause
  smaller cell sizes. In order to set these parameters, we performed
  preliminary control experiments measuring the area growth of
  microcolonies at different illumination frequency and LED
  brightness. Based on these controls, we chose conditions such that
  our maximum loss in growth was less than 10\%.}  When acquiring
images, light from the LED was always shone on cells for 0.3\,s.

In a given experiment multiple fields of view were observed:
custom-written microscope control software kept track of the locations
of the different fields of view and moved between them, acquiring an
image of each field of view at specified intervals. The time between
fields of view was chosen based on the growth rate so that on average
a cell would be imaged about 20-30\,times during a cell cycle. A
typical field of view contained 1-3 cells initially.

\subsection*{Data Analysis}
\subsubsection*{Segmentation and Tracking}
Segmentation was accomplished using custom-written Matlab scripts.  A
pre-processing step of dark-field subtraction was required for images
taken with Ximea, due to the lower camera sensitivity. Individual
micro-colonies were identified by calculating the image gradient using
the Sobel operator, and the threshold over the background using the
Otsu method. Individual cells were identified by filtering with a
logarithm of a Laplacian and using morphological operations.  Most of
the cells were segmented in the previous steps, except for
overlapping or recently divided cells.  To further segment overlapping
cells, we used a seeded version of the watershed method. The
segmentation mask of the preceding image was eroded to obtain the
seeds. To separate cells that were recently divided, we calculated the
mean intensity along the major axis of the candidate cells. If there
was a decrease in intensity in the center, the candidate cell was
divided in two.

\rev{To test how reliably our segmentation algorithm detects cell
  divisions, we investigated the asymmetry in daughter cell
  sizes. Because \emph{E.~coli} are known to divide symmetrically, if
  the segmentation algorithm is working the size of both daughter
  cells after division should be close to identical. We defined the
  ``division asymmetry'' as $L_{0}^{D1}/(L_{0}^{D1}+L_0^{D2})$, where
  $L_0^{D1}$ and $L_0^{D2}$ are the initial lengths of daughters 1 and
  2 after a division; if division is symmetric the division asymmetry
  score should be 0.5.  In all conditions the discrepancy between
  daughter cell sizes was very small (Supplementary
  Figure~\ref{fig:DivAsymmetry}), comparing favorably to that reported
  in other studies with other segmentation
  algorithms~\cite{Campos2014,Taheri-Araghi2014,Soifer2014},
  suggesting that our algorithm can reliably detect divisions.}

To track the lineages, we measured the overlap between labeled regions
in two consecutive frames. Since in these experiments the growth rate is
slow compared with the frame rate, most of the pixels identified for a
given cell in one frame will correspond to the same cell or its
daughters in the next frame. Therefore if we considered the labeled
pixels for a single cell identified in a given frame, in the next
frame they could contain either: 1)~only one label therefore being the
same cell; 2)~two labels, implying the cell divided, or 3)~zero or
more than two labels, meaning that there was a problem in the
segmentation and the lineages must be restarted.

\subsubsection*{Measurements}
The volume of a given cell was calculated (to leading order) assuming a
cylindrical shape with hemispherical caps according to $V(t) =
\frac{\pi}{4}\ell(t)\langle w\rangle^2$, where $\ell(t)$ is the length
of that cell at a particular time and $\langle w\rangle$ is the width
of the cell averaged over that cell's life. Length and width were
calculated as the major and minor axes of the ellipse with the same
normalized second central moments as the cell, as calculated by
MATLAB's \texttt{regionprops} command.

Interdivision time $\tau$ was calculated as the number of images
containing the cell, multiplied by the time elapsed between consecutive
images. To calculate the growth rate $\alpha$, linear regression
was performed on $\log_2(\ell(t))$, with $\alpha$ the inferred slope.

\subsubsection*{Image analysis filters}
\label{sec:Filters}
\revtwo{An essential part of the automated analysis pipeline is the
  quality control of data, avoiding false positives cells and tracks,
  while introducing no bias from filters.
Data on segmented objects were processed by technical filters
  removing segmented objects that are not cells, and excluded wrong or
  incomplete tracks.}  \revtwo{A cell was excluded from analysis on
  \emph{technical} grounds if
\begin{enumerate}
\item It was smaller than the cutoff size (cross section less than
  $\approx 0.46$ $\mu$m$^2$).
\item It was touching the border of the image.
\item It had no mother cell. This filter excludes the first cell,
  since its initial size is unknown, as well as other cells which
  emerge from just outside the field of view or due to errors in
  tracking.
  \item Its growth rate $\alpha$ was negative.
  \item An error in tracking occurred such that the cell was lost for
    at least a frame. This could be caused by mis-segmentation of a
    cell, or by overlap between two adjacent cells. A significant
    number of the total segmented cells were excluded by this
    criteria. We determined that this filter did not bias the
    distributions of the critical observables $\alpha$, $\tau$, and
    $V_0$.
  \end{enumerate}} 
\revtwo{Additionally, we scanned for false-positive detections of cell
  division giving unreasonable interdivision times. Inspection of
  several movies of such events revealed that these were often cells
  with tracking errors that had not been captured by the earlier
  technical filter.}  \revtwo{We also excluded cells for which the
  goodness-of-fit ($r^2$) value of cell growth to an exponential was
  less than 0.8 (these where outliers, since 85-90\% of cells had
  $r^2$ values larger than $0.9$). We verified that this affected
  $<10$\% of cells, mostly with erratic tracks due to wrong
  segmentation. This filter also reduced the spuriously low
  interdivision time population without biasing the remainder of the
  distributions. Finally, we eliminated objects with track length less
  than 8.6 minutes. Relatively few cells failed to pass this filter:
  between 0.1-6\% of cells in each condition passing all other filters
  were excluded due to their interdivision time---less than 2\%
  overall. This procedure completely eliminates the peak of tracks
  with implausibly short duration.}
\revtwo{Supplementary Table~\ref{tab:DataSummary} highlights how many
  cells were excluded by each filter.}

\subsubsection*{Selection of steady state cells}
As mentioned in the main text, to control for varying conditions on
the agarose pad, analysis was restricted to generations in which cell
size, interdivision time, and growth rate were relatively steady (see
Supplementary Fig.~\ref{fig:S-steadyness_by_gen}).  In most
experiments, the growth rate and interdivision time varied little over
the course of the experiment, while the initial size showed more
visible change. We have tried to diagnose the source of the change in
initial size (which occurs without concomitant changes in $\tau$ or
$\alpha$), but it remains elusive. Part of the effect is attributable
to the fact that cells on the outside edge of a colony appear larger
than cells on the inside (Supplementary
Fig.~\ref{fig:S-size-bias-colny-edge}). \revtwo{This effect only
  affects cells on the outermost ring in the colonies, and does not
  vary with time or correlate with concomitant variations of
  interdivision time or growth rate that could explain the increase in
  size}. Hence, a plausible explanation is the image segmentation bias
\revtwo{due to overall change in fluorescence signal in this area}.
Importantly, regardless of the source of this variability in initial
size, our main conclusions are not qualitatively changed when the
analysis is performed on cells from all generations (Supplementary
Fig.~\ref{fig:S-unfiltered-results}).
Alternative microfluidics devices~\cite{Wang2010a,Long2013} are more
laborious and fragile, and at the time of writing are giving us too
low experimental throughput.

\subsubsection*{Statistics and evaluation of goodness-of-collapse}
The goodness of scaling for the finite-size scaling ansatz of cell
size and interdivision time was calculated similarly to
\cite{Bhattacharjee2001,Giometto2013}. The distributions $p(x)$ were
smoothed using a Gaussian kernel, and then rescaled according to
\[p(x) = \frac{1}{x^{\Delta}}F\left( \frac{x}{\langle
    x\rangle^{1/2-\Delta}} \right)\] for varying $\Delta$. The
collapse of the distributions onto a single curve $F(x)$ was assessed
by calculating the function $E(\Delta)$, which is defined as the
average area enclosed by each pair of curves over
their common support. This functional was minimized for $\Delta$.
Bootstrapped confidence intervals were calculated using the
Bias-Corrected and Accelerated (BCa) bootstrap method \cite{Efron1987}
implemented in the Python \texttt{scikits.bootstrap} module. Data
points were repeatedly resampled with replacement to obtain the
bootstrapped sampling distribution.

\begin{acknowledgments}
  We thank A.~Giometto for useful discussions and feedback, J.~Kotar
  for support on the imaging, and M.~Panlilo, Q.~Zhang and N.~Walker
  for technical help.  This work was supported by the International
  Human Frontier Science Program Organization, grants RGY0069/2009-C
  and RGY0070/2014, and a Herchel Smith Harvard Postgraduate
  Fellowship (A.S.K.)
\end{acknowledgments}

\bibliography{CellSize3}

\begin{thebibliography}{51}%
\makeatletter
\providecommand \@ifxundefined [1]{%
 \@ifx{#1\undefined}
}%
\providecommand \@ifnum [1]{%
 \ifnum #1\expandafter \@firstoftwo
 \else \expandafter \@secondoftwo
 \fi
}%
\providecommand \@ifx [1]{%
 \ifx #1\expandafter \@firstoftwo
 \else \expandafter \@secondoftwo
 \fi
}%
\providecommand \natexlab [1]{#1}%
\providecommand \enquote  [1]{``#1''}%
\providecommand \bibnamefont  [1]{#1}%
\providecommand \bibfnamefont [1]{#1}%
\providecommand \citenamefont [1]{#1}%
\providecommand \href@noop [0]{\@secondoftwo}%
\providecommand \href [0]{\begingroup \@sanitize@url \@href}%
\providecommand \@href[1]{\@@startlink{#1}\@@href}%
\providecommand \@@href[1]{\endgroup#1\@@endlink}%
\providecommand \@sanitize@url [0]{\catcode `\\12\catcode `\$12\catcode
  `\&12\catcode `\#12\catcode `\^12\catcode `\_12\catcode `\%12\relax}%
\providecommand \@@startlink[1]{}%
\providecommand \@@endlink[0]{}%
\providecommand \url  [0]{\begingroup\@sanitize@url \@url }%
\providecommand \@url [1]{\endgroup\@href {#1}{\urlprefix }}%
\providecommand \urlprefix  [0]{URL }%
\providecommand \Eprint [0]{\href }%
\providecommand \doibase [0]{http://dx.doi.org/}%
\providecommand \selectlanguage [0]{\@gobble}%
\providecommand \bibinfo  [0]{\@secondoftwo}%
\providecommand \bibfield  [0]{\@secondoftwo}%
\providecommand \translation [1]{[#1]}%
\providecommand \BibitemOpen [0]{}%
\providecommand \bibitemStop [0]{}%
\providecommand \bibitemNoStop [0]{.\EOS\space}%
\providecommand \EOS [0]{\spacefactor3000\relax}%
\providecommand \BibitemShut  [1]{\csname bibitem#1\endcsname}%
\let\auto@bib@innerbib\@empty
\bibitem [{\citenamefont {Tzur}\ \emph {et~al.}(2009)\citenamefont {Tzur},
  \citenamefont {Kafri}, \citenamefont {LeBleu}, \citenamefont {Lahav},\ and\
  \citenamefont {Kirschner}}]{Tzur2009}%
  \BibitemOpen
  \bibfield  {author} {\bibinfo {author} {\bibfnamefont {Amit}\ \bibnamefont
  {Tzur}}, \bibinfo {author} {\bibfnamefont {Ran}\ \bibnamefont {Kafri}},
  \bibinfo {author} {\bibfnamefont {Valerie~S}\ \bibnamefont {LeBleu}},
  \bibinfo {author} {\bibfnamefont {Galit}\ \bibnamefont {Lahav}}, \ and\
  \bibinfo {author} {\bibfnamefont {Marc~W}\ \bibnamefont {Kirschner}},\
  }\bibfield  {title} {\enquote {\bibinfo {title} {Cell growth and size
  homeostasis in proliferating animal cells},}\ }\href@noop {} {\bibfield
  {journal} {\bibinfo  {journal} {Science}\ }\textbf {\bibinfo {volume}
  {325}},\ \bibinfo {pages} {167--71} (\bibinfo {year} {2009})}\BibitemShut
  {NoStop}%
\bibitem [{\citenamefont {Leslie}(2011)}]{Leslie2011}%
  \BibitemOpen
  \bibfield  {author} {\bibinfo {author} {\bibfnamefont {Mitch}\ \bibnamefont
  {Leslie}},\ }\bibfield  {title} {\enquote {\bibinfo {title} {Mysteries of the
  cell. how does a cell know its size?}}\ }\href {\doibase
  10.1126/science.334.6059.1047} {\bibfield  {journal} {\bibinfo  {journal}
  {Science}\ }\textbf {\bibinfo {volume} {334}},\ \bibinfo {pages} {1047--8}
  (\bibinfo {year} {2011})}\BibitemShut {NoStop}%
\bibitem [{\citenamefont {Schaechter}\ \emph {et~al.}(1958)\citenamefont
  {Schaechter}, \citenamefont {Maal\o{}e},\ and\ \citenamefont
  {Kjeldgaard}}]{SCHAECHTER1958}%
  \BibitemOpen
  \bibfield  {author} {\bibinfo {author} {\bibfnamefont {M}~\bibnamefont
  {Schaechter}}, \bibinfo {author} {\bibfnamefont {O}~\bibnamefont
  {Maal\o{}e}}, \ and\ \bibinfo {author} {\bibfnamefont {N~O}\ \bibnamefont
  {Kjeldgaard}},\ }\bibfield  {title} {{\selectlanguage {english}\enquote
  {\bibinfo {title} {Dependency on medium and temperature of cell size and
  chemical composition during balanced grown of salmonella typhimurium.}}\
  }}\href@noop {} {\bibfield  {journal} {\bibinfo  {journal} {J. Gen.
  Microbiol.}\ }\textbf {\bibinfo {volume} {19}},\ \bibinfo {pages} {592--606}
  (\bibinfo {year} {1958})}\BibitemShut {NoStop}%
\bibitem [{\citenamefont {Bremer}\ and\ \citenamefont
  {Dennis}(1996)}]{Bremer1996}%
  \BibitemOpen
  \bibfield  {author} {\bibinfo {author} {\bibfnamefont {Hans}\ \bibnamefont
  {Bremer}}\ and\ \bibinfo {author} {\bibfnamefont {Patrick~P}\ \bibnamefont
  {Dennis}},\ }\bibfield  {title} {\enquote {\bibinfo {title} {Modulation of
  chemical composition and other parameters of the cell by growth rate},}\ }in\
  \href@noop {} {\emph {\bibinfo {booktitle} {Escherichia coli and
  Salmonella}}},\ \bibinfo {editor} {edited by\ \bibinfo {editor}
  {\bibfnamefont {F.~C.}\ \bibnamefont {Neidhardt}}}\ (\bibinfo  {publisher}
  {ASM press Washington, DC},\ \bibinfo {year} {1996})\ pp.\ \bibinfo {pages}
  {1553--69}\BibitemShut {NoStop}%
\bibitem [{\citenamefont {Scott}\ \emph {et~al.}(2010)\citenamefont {Scott},
  \citenamefont {Gunderson}, \citenamefont {Mateescu}, \citenamefont {Zhang},\
  and\ \citenamefont {Hwa}}]{Scott2010a}%
  \BibitemOpen
  \bibfield  {author} {\bibinfo {author} {\bibfnamefont {Matthew}\ \bibnamefont
  {Scott}}, \bibinfo {author} {\bibfnamefont {Carl~W}\ \bibnamefont
  {Gunderson}}, \bibinfo {author} {\bibfnamefont {Eduard~M}\ \bibnamefont
  {Mateescu}}, \bibinfo {author} {\bibfnamefont {Zhongge}\ \bibnamefont
  {Zhang}}, \ and\ \bibinfo {author} {\bibfnamefont {Terence}\ \bibnamefont
  {Hwa}},\ }\bibfield  {title} {\enquote {\bibinfo {title} {Interdependence of
  cell growth and gene expression: origins and consequences},}\ }\href@noop {}
  {\bibfield  {journal} {\bibinfo  {journal} {Science}\ }\textbf {\bibinfo
  {volume} {330}},\ \bibinfo {pages} {1099--102} (\bibinfo {year}
  {2010})}\BibitemShut {NoStop}%
\bibitem [{\citenamefont {Cooper}\ and\ \citenamefont
  {Helmstetter}(1968)}]{Cooper1968}%
  \BibitemOpen
  \bibfield  {author} {\bibinfo {author} {\bibfnamefont {S}~\bibnamefont
  {Cooper}}\ and\ \bibinfo {author} {\bibfnamefont {C~E}\ \bibnamefont
  {Helmstetter}},\ }\bibfield  {title} {\enquote {\bibinfo {title} {{Chromosome
  replication and the division cycle of \emph{Escherichia coli} B/r}},}\
  }\href@noop {} {\bibfield  {journal} {\bibinfo  {journal} {J. Mol. Biol.}\
  }\textbf {\bibinfo {volume} {31}},\ \bibinfo {pages} {519--40} (\bibinfo
  {year} {1968})}\BibitemShut {NoStop}%
\bibitem [{\citenamefont {Donachie}(1968)}]{Donachie1968}%
  \BibitemOpen
  \bibfield  {author} {\bibinfo {author} {\bibfnamefont {William~D}\
  \bibnamefont {Donachie}},\ }\bibfield  {title} {\enquote {\bibinfo {title}
  {{Relationship between Cell Size and Time of Initiation of DNA
  Replication}},}\ }\href@noop {} {\bibfield  {journal} {\bibinfo  {journal}
  {Nature}\ }\textbf {\bibinfo {volume} {219}},\ \bibinfo {pages} {1077--9}
  (\bibinfo {year} {1968})}\BibitemShut {NoStop}%
\bibitem [{\citenamefont {Koch}\ and\ \citenamefont
  {Schaechter}(1962)}]{Koch1962}%
  \BibitemOpen
  \bibfield  {author} {\bibinfo {author} {\bibfnamefont {A~L}\ \bibnamefont
  {Koch}}\ and\ \bibinfo {author} {\bibfnamefont {M}~\bibnamefont
  {Schaechter}},\ }\bibfield  {title} {\enquote {\bibinfo {title} {A model for
  statistics of the cell division process},}\ }\href@noop {} {\bibfield
  {journal} {\bibinfo  {journal} {J. Gen. Microbiol.}\ }\textbf {\bibinfo
  {volume} {29}},\ \bibinfo {pages} {435--54} (\bibinfo {year}
  {1962})}\BibitemShut {NoStop}%
\bibitem [{\citenamefont {Schaechter}\ \emph {et~al.}(1962)\citenamefont
  {Schaechter}, \citenamefont {Williamson}, \citenamefont {Hood},\ and\
  \citenamefont {Koch}}]{Schaechter1962}%
  \BibitemOpen
  \bibfield  {author} {\bibinfo {author} {\bibfnamefont {M.}~\bibnamefont
  {Schaechter}}, \bibinfo {author} {\bibfnamefont {J.~P.}\ \bibnamefont
  {Williamson}}, \bibinfo {author} {\bibfnamefont {Jr}~\bibnamefont {Hood},
  \bibfnamefont {Jr}}, \ and\ \bibinfo {author} {\bibfnamefont {A.~L.}\
  \bibnamefont {Koch}},\ }\bibfield  {title} {\enquote {\bibinfo {title}
  {Growth, cell and nuclear divisions in some bacteria.}}\ }\href@noop {}
  {\bibfield  {journal} {\bibinfo  {journal} {J. Gen. Microbiol.}\ }\textbf
  {\bibinfo {volume} {29}},\ \bibinfo {pages} {421--34} (\bibinfo {year}
  {1962})}\BibitemShut {NoStop}%
\bibitem [{\citenamefont {Cooper}(1993)}]{Cooper1993}%
  \BibitemOpen
  \bibfield  {author} {\bibinfo {author} {\bibfnamefont {Stephen}\ \bibnamefont
  {Cooper}},\ }\bibfield  {title} {\enquote {\bibinfo {title} {The origins and
  meaning of the {Schaechter-Maal{\o}e-Kjeldgaard} experiments},}\ }\href@noop
  {} {\bibfield  {journal} {\bibinfo  {journal} {J. Gen. Microbiol.}\ }\textbf
  {\bibinfo {volume} {139}},\ \bibinfo {pages} {1117} (\bibinfo {year}
  {1993})}\BibitemShut {NoStop}%
\bibitem [{\citenamefont {Campos}\ \emph {et~al.}(2014)\citenamefont {Campos},
  \citenamefont {Surovtsev}, \citenamefont {Kato}, \citenamefont {Paintdakhi},
  \citenamefont {Beltran}, \citenamefont {Ebmeier},\ and\ \citenamefont
  {Jacobs-Wagner}}]{Campos2014}%
  \BibitemOpen
  \bibfield  {author} {\bibinfo {author} {\bibfnamefont {Manuel}\ \bibnamefont
  {Campos}}, \bibinfo {author} {\bibfnamefont {Ivan~V.}\ \bibnamefont
  {Surovtsev}}, \bibinfo {author} {\bibfnamefont {Setsu}\ \bibnamefont {Kato}},
  \bibinfo {author} {\bibfnamefont {Ahmad}\ \bibnamefont {Paintdakhi}},
  \bibinfo {author} {\bibfnamefont {Bruno}\ \bibnamefont {Beltran}}, \bibinfo
  {author} {\bibfnamefont {Sarah~E.}\ \bibnamefont {Ebmeier}}, \ and\ \bibinfo
  {author} {\bibfnamefont {Christine}\ \bibnamefont {Jacobs-Wagner}},\
  }\bibfield  {title} {{\selectlanguage {english}\enquote {\bibinfo {title} {A
  constant size extension drives bacterial cell size homeostasis.}}\ }}\href
  {\doibase 10.1016/j.cell.2014.11.022} {\bibfield  {journal} {\bibinfo
  {journal} {Cell}\ }\textbf {\bibinfo {volume} {159}},\ \bibinfo {pages}
  {1433--46} (\bibinfo {year} {2014})}\BibitemShut {NoStop}%
\bibitem [{\citenamefont {Taheri-Araghi}\ \emph {et~al.}(2014)\citenamefont
  {Taheri-Araghi}, \citenamefont {Bradde}, \citenamefont {Sauls}, \citenamefont
  {Hill}, \citenamefont {Levin}, \citenamefont {Paulsson}, \citenamefont
  {Vergassola},\ and\ \citenamefont {Jun}}]{Taheri-Araghi2014}%
  \BibitemOpen
  \bibfield  {author} {\bibinfo {author} {\bibfnamefont {Sattar}\ \bibnamefont
  {Taheri-Araghi}}, \bibinfo {author} {\bibfnamefont {Serena}\ \bibnamefont
  {Bradde}}, \bibinfo {author} {\bibfnamefont {John~T.}\ \bibnamefont {Sauls}},
  \bibinfo {author} {\bibfnamefont {Norbert~S.}\ \bibnamefont {Hill}}, \bibinfo
  {author} {\bibfnamefont {Petra~Anne}\ \bibnamefont {Levin}}, \bibinfo
  {author} {\bibfnamefont {Johan}\ \bibnamefont {Paulsson}}, \bibinfo {author}
  {\bibfnamefont {Massimo}\ \bibnamefont {Vergassola}}, \ and\ \bibinfo
  {author} {\bibfnamefont {Suckjoon}\ \bibnamefont {Jun}},\ }\bibfield  {title}
  {{\selectlanguage {english}\enquote {\bibinfo {title} {Cell-size control and
  homeostasis in bacteria.}}\ }}\href {\doibase 10.1016/j.cub.2014.12.009}
  {\bibfield  {journal} {\bibinfo  {journal} {Curr Biol}\ }\textbf {\bibinfo
  {volume} {25}},\ \bibinfo {pages} {385--91} (\bibinfo {year}
  {2014})}\BibitemShut {NoStop}%
\bibitem [{\citenamefont {Deforet}\ \emph {et~al.}(2015)\citenamefont
  {Deforet}, \citenamefont {van Ditmarsch},\ and\ \citenamefont
  {Xavier}}]{Deforet2015}%
  \BibitemOpen
  \bibfield  {author} {\bibinfo {author} {\bibfnamefont {Maxime}\ \bibnamefont
  {Deforet}}, \bibinfo {author} {\bibfnamefont {Dave}\ \bibnamefont {van
  Ditmarsch}}, \ and\ \bibinfo {author} {\bibfnamefont {Jo„o~B}\ \bibnamefont
  {Xavier}},\ }\bibfield  {title} {\enquote {\bibinfo {title} {Cell-size
  homeostasis and the incremental rule in a bacterial pathogen.}}\ }\href@noop
  {} {\bibfield  {journal} {\bibinfo  {journal} {Biophys J}\ }\textbf {\bibinfo
  {volume} {109}},\ \bibinfo {pages} {521--528} (\bibinfo {year}
  {2015})}\BibitemShut {NoStop}%
\bibitem [{\citenamefont {Tanouchi}\ \emph {et~al.}(2015)\citenamefont
  {Tanouchi}, \citenamefont {Pai}, \citenamefont {Park}, \citenamefont {Huang},
  \citenamefont {Stamatov}, \citenamefont {Buchler},\ and\ \citenamefont
  {You}}]{Tanouchi2015}%
  \BibitemOpen
  \bibfield  {author} {\bibinfo {author} {\bibfnamefont {Yu}~\bibnamefont
  {Tanouchi}}, \bibinfo {author} {\bibfnamefont {Anand}\ \bibnamefont {Pai}},
  \bibinfo {author} {\bibfnamefont {Heungwon}\ \bibnamefont {Park}}, \bibinfo
  {author} {\bibfnamefont {Shuqiang}\ \bibnamefont {Huang}}, \bibinfo {author}
  {\bibfnamefont {Rumen}\ \bibnamefont {Stamatov}}, \bibinfo {author}
  {\bibfnamefont {Nicolas~E}\ \bibnamefont {Buchler}}, \ and\ \bibinfo {author}
  {\bibfnamefont {Lingchong}\ \bibnamefont {You}},\ }\bibfield  {title}
  {\enquote {\bibinfo {title} {A noisy linear map underlies oscillations in
  cell size and gene expression in bacteria.}}\ }\href@noop {} {\bibfield
  {journal} {\bibinfo  {journal} {Nature}\ }\textbf {\bibinfo {volume} {523}},\
  \bibinfo {pages} {357--360} (\bibinfo {year} {2015})}\BibitemShut {NoStop}%
\bibitem [{\citenamefont {Iyer-Biswas}\ \emph
  {et~al.}(2014{\natexlab{a}})\citenamefont {Iyer-Biswas}, \citenamefont
  {Wright}, \citenamefont {Henry}, \citenamefont {Lo}, \citenamefont {Burov},
  \citenamefont {Lin}, \citenamefont {Crooks}, \citenamefont {Crosson},
  \citenamefont {Dinner},\ and\ \citenamefont {Scherer}}]{Iyer-Biswas2014a}%
  \BibitemOpen
  \bibfield  {author} {\bibinfo {author} {\bibfnamefont {Srividya}\
  \bibnamefont {Iyer-Biswas}}, \bibinfo {author} {\bibfnamefont {Charles~S}\
  \bibnamefont {Wright}}, \bibinfo {author} {\bibfnamefont {Jon~T}\
  \bibnamefont {Henry}}, \bibinfo {author} {\bibfnamefont {Klevin}\
  \bibnamefont {Lo}}, \bibinfo {author} {\bibfnamefont {Stanislav}\
  \bibnamefont {Burov}}, \bibinfo {author} {\bibfnamefont {Yihan}\ \bibnamefont
  {Lin}}, \bibinfo {author} {\bibfnamefont {Gavin~E}\ \bibnamefont {Crooks}},
  \bibinfo {author} {\bibfnamefont {Sean}\ \bibnamefont {Crosson}}, \bibinfo
  {author} {\bibfnamefont {Aaron~R.}\ \bibnamefont {Dinner}}, \ and\ \bibinfo
  {author} {\bibfnamefont {Norbert~F}\ \bibnamefont {Scherer}},\ }\bibfield
  {title} {\enquote {\bibinfo {title} {Scaling laws governing stochastic growth
  and division of single bacterial cells},}\ }\href {\doibase
  10.1073/pnas.1403232111} {\bibfield  {journal} {\bibinfo  {journal} {Proc.
  Natl. Acad. Sci. (U.S.A.)}\ }\textbf {\bibinfo {volume} {111}},\ \bibinfo
  {pages} {15912--7} (\bibinfo {year} {2014}{\natexlab{a}})}\BibitemShut
  {NoStop}%
\bibitem [{\citenamefont {Jun}\ and\ \citenamefont
  {Taheri-Araghi}(2015)}]{Jun2015}%
  \BibitemOpen
  \bibfield  {author} {\bibinfo {author} {\bibfnamefont {Suckjoon}\
  \bibnamefont {Jun}}\ and\ \bibinfo {author} {\bibfnamefont {Sattar}\
  \bibnamefont {Taheri-Araghi}},\ }\bibfield  {title} {{\selectlanguage
  {english}\enquote {\bibinfo {title} {Cell-size maintenance: universal
  strategy revealed.}}\ }}\href {\doibase 10.1016/j.tim.2014.12.001} {\bibfield
   {journal} {\bibinfo  {journal} {Trends Microbiol}\ }\textbf {\bibinfo
  {volume} {23}},\ \bibinfo {pages} {4--6} (\bibinfo {year}
  {2015})}\BibitemShut {NoStop}%
\bibitem [{\citenamefont {Osella}\ \emph {et~al.}(2014)\citenamefont {Osella},
  \citenamefont {Nugent},\ and\ \citenamefont {{Cosentino
  Lagomarsino}}}]{Osella2014}%
  \BibitemOpen
  \bibfield  {author} {\bibinfo {author} {\bibfnamefont {Matteo}\ \bibnamefont
  {Osella}}, \bibinfo {author} {\bibfnamefont {Eileen}\ \bibnamefont {Nugent}},
  \ and\ \bibinfo {author} {\bibfnamefont {Marco}\ \bibnamefont {{Cosentino
  Lagomarsino}}},\ }\bibfield  {title} {\enquote {\bibinfo {title} {Concerted
  control of \emph{Escherichia coli} cell division},}\ }\href@noop {}
  {\bibfield  {journal} {\bibinfo  {journal} {Proc. Natl. Acad. Sci. (U.S.A.)}\
  }\textbf {\bibinfo {volume} {111}},\ \bibinfo {pages} {3431--5} (\bibinfo
  {year} {2014})}\BibitemShut {NoStop}%
\bibitem [{\citenamefont {Wang}\ \emph {et~al.}(2010)\citenamefont {Wang},
  \citenamefont {Robert}, \citenamefont {Pelletier}, \citenamefont {Dang},
  \citenamefont {Taddei}, \citenamefont {Wright},\ and\ \citenamefont
  {Jun}}]{Wang2010a}%
  \BibitemOpen
  \bibfield  {author} {\bibinfo {author} {\bibfnamefont {Ping}\ \bibnamefont
  {Wang}}, \bibinfo {author} {\bibfnamefont {Lydia}\ \bibnamefont {Robert}},
  \bibinfo {author} {\bibfnamefont {James}\ \bibnamefont {Pelletier}}, \bibinfo
  {author} {\bibfnamefont {Wei~Lien}\ \bibnamefont {Dang}}, \bibinfo {author}
  {\bibfnamefont {Francois}\ \bibnamefont {Taddei}}, \bibinfo {author}
  {\bibfnamefont {Andrew}\ \bibnamefont {Wright}}, \ and\ \bibinfo {author}
  {\bibfnamefont {Suckjoon}\ \bibnamefont {Jun}},\ }\bibfield  {title}
  {\enquote {\bibinfo {title} {Robust growth of {\emph{escherichia coli}}},}\
  }\href@noop {} {\bibfield  {journal} {\bibinfo  {journal} {Curr. Biol.}\
  }\textbf {\bibinfo {volume} {20}},\ \bibinfo {pages} {1099--103} (\bibinfo
  {year} {2010})}\BibitemShut {NoStop}%
\bibitem [{\citenamefont {Wakamoto}\ \emph {et~al.}(2013)\citenamefont
  {Wakamoto}, \citenamefont {Dhar}, \citenamefont {Chait}, \citenamefont
  {Schneider}, \citenamefont {{Signorino-Gelo}}, \citenamefont {Leibler},\ and\
  \citenamefont {McKinney}}]{Wakamoto2013}%
  \BibitemOpen
  \bibfield  {author} {\bibinfo {author} {\bibfnamefont {Yuichi}\ \bibnamefont
  {Wakamoto}}, \bibinfo {author} {\bibfnamefont {Neeraj}\ \bibnamefont {Dhar}},
  \bibinfo {author} {\bibfnamefont {Remy}\ \bibnamefont {Chait}}, \bibinfo
  {author} {\bibfnamefont {Katrin}\ \bibnamefont {Schneider}}, \bibinfo
  {author} {\bibfnamefont {Fran\c{c}ois}\ \bibnamefont {{Signorino-Gelo}}},
  \bibinfo {author} {\bibfnamefont {Stanislas}\ \bibnamefont {Leibler}}, \ and\
  \bibinfo {author} {\bibfnamefont {John~D}\ \bibnamefont {McKinney}},\
  }\bibfield  {title} {\enquote {\bibinfo {title} {{Dynamic persistence of
  antibiotic-stressed mycobacteria}},}\ }\href@noop {} {\bibfield  {journal}
  {\bibinfo  {journal} {Science}\ }\textbf {\bibinfo {volume} {339}},\ \bibinfo
  {pages} {91--5} (\bibinfo {year} {2013})}\BibitemShut {NoStop}%
\bibitem [{\citenamefont {Grant}\ \emph {et~al.}(2014)\citenamefont {Grant},
  \citenamefont {Waclaw}, \citenamefont {Allen},\ and\ \citenamefont
  {Cicuta}}]{Grant2014}%
  \BibitemOpen
  \bibfield  {author} {\bibinfo {author} {\bibfnamefont {Matthew A~A}\
  \bibnamefont {Grant}}, \bibinfo {author} {\bibfnamefont {Bartlomiej}\
  \bibnamefont {Waclaw}}, \bibinfo {author} {\bibfnamefont {Rosalind~J}\
  \bibnamefont {Allen}}, \ and\ \bibinfo {author} {\bibfnamefont {Pietro}\
  \bibnamefont {Cicuta}},\ }\bibfield  {title} {\enquote {\bibinfo {title}
  {{The role of mechanical forces in the planar-to-bulk transition in growing
  \emph{Escherichia coli} microcolonies}},}\ }\href@noop {} {\bibfield
  {journal} {\bibinfo  {journal} {J. Roy. Soc.: Interface}\ }\textbf {\bibinfo
  {volume} {11}} (\bibinfo {year} {2014})}\BibitemShut {NoStop}%
\bibitem [{\citenamefont {Giometto}\ \emph {et~al.}(2013)\citenamefont
  {Giometto}, \citenamefont {Altermatt}, \citenamefont {Carrara}, \citenamefont
  {Maritan},\ and\ \citenamefont {Rinaldo}}]{Giometto2013}%
  \BibitemOpen
  \bibfield  {author} {\bibinfo {author} {\bibfnamefont {Andrea}\ \bibnamefont
  {Giometto}}, \bibinfo {author} {\bibfnamefont {Florian}\ \bibnamefont
  {Altermatt}}, \bibinfo {author} {\bibfnamefont {Francesco}\ \bibnamefont
  {Carrara}}, \bibinfo {author} {\bibfnamefont {Amos}\ \bibnamefont {Maritan}},
  \ and\ \bibinfo {author} {\bibfnamefont {Andrea}\ \bibnamefont {Rinaldo}},\
  }\bibfield  {title} {\enquote {\bibinfo {title} {Scaling body size
  fluctuations.}}\ }\href@noop {} {\bibfield  {journal} {\bibinfo  {journal}
  {Proc. Natl. Acad. Sci. (U.S.A.)}\ }\textbf {\bibinfo {volume} {110}},\
  \bibinfo {pages} {4646--50} (\bibinfo {year} {2013})}\BibitemShut {NoStop}%
\bibitem [{\citenamefont {Neidhardt}\ \emph {et~al.}(1974)\citenamefont
  {Neidhardt}, \citenamefont {Bloch},\ and\ \citenamefont
  {Smith}}]{Neidhardt1974}%
  \BibitemOpen
  \bibfield  {author} {\bibinfo {author} {\bibfnamefont {F~C}\ \bibnamefont
  {Neidhardt}}, \bibinfo {author} {\bibfnamefont {P~L}\ \bibnamefont {Bloch}},
  \ and\ \bibinfo {author} {\bibfnamefont {D~F}\ \bibnamefont {Smith}},\
  }\bibfield  {title} {\enquote {\bibinfo {title} {{Culture medium for
  enterobacteria}},}\ }\href@noop {} {\bibfield  {journal} {\bibinfo  {journal}
  {J. Bacteriol.}\ }\textbf {\bibinfo {volume} {119}},\ \bibinfo {pages}
  {736--47} (\bibinfo {year} {1974})}\BibitemShut {NoStop}%
\bibitem [{\citenamefont {Elowitz}\ \emph {et~al.}(2002)\citenamefont
  {Elowitz}, \citenamefont {Levine}, \citenamefont {Siggia},\ and\
  \citenamefont {Swain}}]{Elowitz2002}%
  \BibitemOpen
  \bibfield  {author} {\bibinfo {author} {\bibfnamefont {Michael~B}\
  \bibnamefont {Elowitz}}, \bibinfo {author} {\bibfnamefont {Arnold~J}\
  \bibnamefont {Levine}}, \bibinfo {author} {\bibfnamefont {Eric~D}\
  \bibnamefont {Siggia}}, \ and\ \bibinfo {author} {\bibfnamefont {Peter~S}\
  \bibnamefont {Swain}},\ }\bibfield  {title} {\enquote {\bibinfo {title}
  {{Stochastic gene expression in a single cell}},}\ }\href@noop {} {\bibfield
  {journal} {\bibinfo  {journal} {Science}\ }\textbf {\bibinfo {volume}
  {297}},\ \bibinfo {pages} {1183--6} (\bibinfo {year} {2002})}\BibitemShut
  {NoStop}%
\bibitem [{\citenamefont {Bhattacharjee}\ and\ \citenamefont
  {Seno}(2001)}]{Bhattacharjee2001}%
  \BibitemOpen
  \bibfield  {author} {\bibinfo {author} {\bibfnamefont {Somendra~M}\
  \bibnamefont {Bhattacharjee}}\ and\ \bibinfo {author} {\bibfnamefont
  {Flavio}\ \bibnamefont {Seno}},\ }\bibfield  {title} {\enquote {\bibinfo
  {title} {{A measure of data collapse for scaling}},}\ }\href@noop {}
  {\bibfield  {journal} {\bibinfo  {journal} {J. Phys. A: Math. and Gen.}\
  }\textbf {\bibinfo {volume} {34}},\ \bibinfo {pages} {6375--80} (\bibinfo
  {year} {2001})}\BibitemShut {NoStop}%
\bibitem [{\citenamefont {Henrici}(1928)}]{Henrici1928}%
  \BibitemOpen
  \bibfield  {author} {\bibinfo {author} {\bibfnamefont {Arthur~T}\
  \bibnamefont {Henrici}},\ }\href@noop {} {\emph {\bibinfo {title}
  {{Morphologic Variation and the Rate of Growth of Bacteria}}}},\ edited by\
  \bibinfo {editor} {\bibfnamefont {R~E}\ \bibnamefont {Buchanan}}, \bibinfo
  {editor} {\bibfnamefont {E~B}\ \bibnamefont {Fred}}, \ and\ \bibinfo {editor}
  {\bibfnamefont {S~A}\ \bibnamefont {Waksman}}\ (\bibinfo  {publisher}
  {Bailliere, Tindall, and Cox},\ \bibinfo {address} {London},\ \bibinfo {year}
  {1928})\ p.\ \bibinfo {pages} {194}\BibitemShut {NoStop}%
\bibitem [{\citenamefont {Trueba}\ \emph {et~al.}(1982)\citenamefont {Trueba},
  \citenamefont {Neijssel},\ and\ \citenamefont {Woldringh}}]{Trueba1982}%
  \BibitemOpen
  \bibfield  {author} {\bibinfo {author} {\bibfnamefont {F~J}\ \bibnamefont
  {Trueba}}, \bibinfo {author} {\bibfnamefont {O~M}\ \bibnamefont {Neijssel}},
  \ and\ \bibinfo {author} {\bibfnamefont {C~L}\ \bibnamefont {Woldringh}},\
  }\bibfield  {title} {\enquote {\bibinfo {title} {{Generality of the Growth
  Kinetics of the Average Individual Cell in Different Bacterial
  Populations}},}\ }\href@noop {} {\bibfield  {journal} {\bibinfo  {journal}
  {J. Bacteriol.}\ }\textbf {\bibinfo {volume} {150}},\ \bibinfo {pages} {1048}
  (\bibinfo {year} {1982})}\BibitemShut {NoStop}%
\bibitem [{\citenamefont {Kubitschek}\ and\ \citenamefont
  {Woldringh}(1983)}]{Kubitschek1983}%
  \BibitemOpen
  \bibfield  {author} {\bibinfo {author} {\bibfnamefont {H~E}\ \bibnamefont
  {Kubitschek}}\ and\ \bibinfo {author} {\bibfnamefont {C~L}\ \bibnamefont
  {Woldringh}},\ }\bibfield  {title} {\enquote {\bibinfo {title} {{Cell
  elongation and division probability during the Escherichia coli growth
  cycle}},}\ }\href@noop {} {\bibfield  {journal} {\bibinfo  {journal} {J.
  Bacteriol.}\ }\textbf {\bibinfo {volume} {153}},\ \bibinfo {pages} {1379--87}
  (\bibinfo {year} {1983})}\BibitemShut {NoStop}%
\bibitem [{\citenamefont {{{\AA}kerlund}}\ \emph {et~al.}(1995)\citenamefont
  {{{\AA}kerlund}}, \citenamefont {{Nordstr\"{o}m}},\ and\ \citenamefont
  {Bernander}}]{Akerlund1995}%
  \BibitemOpen
  \bibfield  {author} {\bibinfo {author} {\bibfnamefont {T}~\bibnamefont
  {{{\AA}kerlund}}}, \bibinfo {author} {\bibfnamefont {Kurt}\ \bibnamefont
  {{Nordstr\"{o}m}}}, \ and\ \bibinfo {author} {\bibfnamefont {Rolf}\
  \bibnamefont {Bernander}},\ }\bibfield  {title} {\enquote {\bibinfo {title}
  {{Analysis of cell size and DNA content in exponentially growing and
  stationary-phase batch cultures of \emph{Escherichia coli}}},}\ }\href@noop
  {} {\bibfield  {journal} {\bibinfo  {journal} {J. Bacteriol.}\ }\textbf
  {\bibinfo {volume} {177}},\ \bibinfo {pages} {6791--7} (\bibinfo {year}
  {1995})}\BibitemShut {NoStop}%
\bibitem [{\citenamefont {Wakamoto}\ \emph {et~al.}(2005)\citenamefont
  {Wakamoto}, \citenamefont {Ramsden},\ and\ \citenamefont
  {Yasuda}}]{Wakamoto2005}%
  \BibitemOpen
  \bibfield  {author} {\bibinfo {author} {\bibfnamefont {Yuichi}\ \bibnamefont
  {Wakamoto}}, \bibinfo {author} {\bibfnamefont {Jeremy}\ \bibnamefont
  {Ramsden}}, \ and\ \bibinfo {author} {\bibfnamefont {Kenji}\ \bibnamefont
  {Yasuda}},\ }\bibfield  {title} {\enquote {\bibinfo {title} {Single-cell
  growth and division dynamics showing epigenetic correlations},}\ }\href@noop
  {} {\bibfield  {journal} {\bibinfo  {journal} {Analyst}\ }\textbf {\bibinfo
  {volume} {130}},\ \bibinfo {pages} {311--7} (\bibinfo {year}
  {2005})}\BibitemShut {NoStop}%
\bibitem [{\citenamefont {Stewart}\ \emph {et~al.}(2005)\citenamefont
  {Stewart}, \citenamefont {Madden}, \citenamefont {Paul},\ and\ \citenamefont
  {Taddei}}]{Stewart2005}%
  \BibitemOpen
  \bibfield  {author} {\bibinfo {author} {\bibfnamefont {Eric~J}\ \bibnamefont
  {Stewart}}, \bibinfo {author} {\bibfnamefont {Richard}\ \bibnamefont
  {Madden}}, \bibinfo {author} {\bibfnamefont {Gregory}\ \bibnamefont {Paul}},
  \ and\ \bibinfo {author} {\bibfnamefont {Fran\c{c}ois}\ \bibnamefont
  {Taddei}},\ }\bibfield  {title} {\enquote {\bibinfo {title} {{Aging and death
  in an organism that reproduces by morphologically symmetric division.}}}\
  }\href@noop {} {\bibfield  {journal} {\bibinfo  {journal} {PLoS Biology}\
  }\textbf {\bibinfo {volume} {3}},\ \bibinfo {pages} {e45} (\bibinfo {year}
  {2005})}\BibitemShut {NoStop}%
\bibitem [{\citenamefont {{M\"{a}nnik}}\ \emph {et~al.}(2012)\citenamefont
  {{M\"{a}nnik}}, \citenamefont {Wu}, \citenamefont {Hol}, \citenamefont
  {Bisicchia}, \citenamefont {Sherratt}, \citenamefont {Keymer},\ and\
  \citenamefont {Dekker}}]{Mannik2012}%
  \BibitemOpen
  \bibfield  {author} {\bibinfo {author} {\bibfnamefont {Jaan}\ \bibnamefont
  {{M\"{a}nnik}}}, \bibinfo {author} {\bibfnamefont {Fabai}\ \bibnamefont
  {Wu}}, \bibinfo {author} {\bibfnamefont {Felix J~H}\ \bibnamefont {Hol}},
  \bibinfo {author} {\bibfnamefont {Paola}\ \bibnamefont {Bisicchia}}, \bibinfo
  {author} {\bibfnamefont {David~J}\ \bibnamefont {Sherratt}}, \bibinfo
  {author} {\bibfnamefont {Juan~E}\ \bibnamefont {Keymer}}, \ and\ \bibinfo
  {author} {\bibfnamefont {Cees}\ \bibnamefont {Dekker}},\ }\bibfield  {title}
  {\enquote {\bibinfo {title} {{Robustness and accuracy of cell division in
  \emph{Escherichia coli} in diverse cell shapes.}}}\ }\href@noop {} {\bibfield
   {journal} {\bibinfo  {journal} {Proc. Natl. Acad. Sci. (U.S.A.)}\ }\textbf
  {\bibinfo {volume} {109}},\ \bibinfo {pages} {6957--62} (\bibinfo {year}
  {2012})}\BibitemShut {NoStop}%
\bibitem [{\citenamefont {Soifer}\ \emph {et~al.}(2014)\citenamefont {Soifer},
  \citenamefont {Robert}, \citenamefont {Barkai},\ and\ \citenamefont
  {Amir}}]{Soifer2014}%
  \BibitemOpen
  \bibfield  {author} {\bibinfo {author} {\bibfnamefont {I.}~\bibnamefont
  {Soifer}}, \bibinfo {author} {\bibfnamefont {L.}~\bibnamefont {Robert}},
  \bibinfo {author} {\bibfnamefont {N.}~\bibnamefont {Barkai}}, \ and\ \bibinfo
  {author} {\bibfnamefont {A.}~\bibnamefont {Amir}},\ }\bibfield  {title}
  {\enquote {\bibinfo {title} {Single-cell analysis of growth in budding yeast
  and bacteria reveals a common size regulation strategy},}\ }\href@noop {}
  {\bibfield  {journal} {\bibinfo  {journal} {arXiv:1410.4771}\ } (\bibinfo
  {year} {2014})}\BibitemShut {NoStop}%
\bibitem [{\citenamefont {Fisher}\ and\ \citenamefont
  {Barber}(1972)}]{Fisher1972}%
  \BibitemOpen
  \bibfield  {author} {\bibinfo {author} {\bibfnamefont {Michael~E}\
  \bibnamefont {Fisher}}\ and\ \bibinfo {author} {\bibfnamefont {Michael~N}\
  \bibnamefont {Barber}},\ }\bibfield  {title} {\enquote {\bibinfo {title}
  {{Scaling Theory for Finite-Size Effects in the Critical Region}},}\
  }\href@noop {} {\bibfield  {journal} {\bibinfo  {journal} {Phys. Rev. Lett.}\
  }\textbf {\bibinfo {volume} {28}},\ \bibinfo {pages} {1516--9} (\bibinfo
  {year} {1972})}\BibitemShut {NoStop}%
\bibitem [{\citenamefont {Cardy}(1988)}]{cardy1988}%
  \BibitemOpen
  \bibfield  {author} {\bibinfo {author} {\bibfnamefont {J.L.}\ \bibnamefont
  {Cardy}},\ }\href@noop {} {\emph {\bibinfo {title} {Finite-Size Scaling}}},\
  Current physics\ (\bibinfo  {publisher} {North-Holland},\ \bibinfo {year}
  {1988})\BibitemShut {NoStop}%
\bibitem [{\citenamefont {Rinaldo}\ \emph {et~al.}(2002)\citenamefont
  {Rinaldo}, \citenamefont {Maritan}, \citenamefont {Cavender-Bares},\ and\
  \citenamefont {Chisholm}}]{Rinaldo2002}%
  \BibitemOpen
  \bibfield  {author} {\bibinfo {author} {\bibfnamefont {Andrea}\ \bibnamefont
  {Rinaldo}}, \bibinfo {author} {\bibfnamefont {Amos}\ \bibnamefont {Maritan}},
  \bibinfo {author} {\bibfnamefont {Kent~K}\ \bibnamefont {Cavender-Bares}}, \
  and\ \bibinfo {author} {\bibfnamefont {Sallie~W}\ \bibnamefont {Chisholm}},\
  }\bibfield  {title} {\enquote {\bibinfo {title} {{Cross-scale ecological
  dynamics and microbial size spectra in marine ecosystems}},}\ }\href@noop {}
  {\bibfield  {journal} {\bibinfo  {journal} {Proc. R. Soc. Lond. B: Biol.
  Sci.}\ }\textbf {\bibinfo {volume} {269}},\ \bibinfo {pages} {2051--9}
  (\bibinfo {year} {2002})}\BibitemShut {NoStop}%
\bibitem [{\citenamefont {Banavar}\ \emph {et~al.}(2007)\citenamefont
  {Banavar}, \citenamefont {Damuth}, \citenamefont {Maritan},\ and\
  \citenamefont {Rinaldo}}]{Banavar2007}%
  \BibitemOpen
  \bibfield  {author} {\bibinfo {author} {\bibfnamefont {Jayanth~R}\
  \bibnamefont {Banavar}}, \bibinfo {author} {\bibfnamefont {John}\
  \bibnamefont {Damuth}}, \bibinfo {author} {\bibfnamefont {Amos}\ \bibnamefont
  {Maritan}}, \ and\ \bibinfo {author} {\bibfnamefont {Andrea}\ \bibnamefont
  {Rinaldo}},\ }\bibfield  {title} {{\selectlanguage {english}\enquote
  {\bibinfo {title} {Scaling in ecosystems and the linkage of macroecological
  laws.}}\ }}\href@noop {} {\bibfield  {journal} {\bibinfo  {journal} {Phys Rev
  Lett}\ }\textbf {\bibinfo {volume} {98}},\ \bibinfo {pages} {068104}
  (\bibinfo {year} {2007})}\BibitemShut {NoStop}%
\bibitem [{\citenamefont {Banavar}\ \emph {et~al.}(1999)\citenamefont
  {Banavar}, \citenamefont {Green}, \citenamefont {Harte},\ and\ \citenamefont
  {Maritan}}]{Banavar1999}%
  \BibitemOpen
  \bibfield  {author} {\bibinfo {author} {\bibfnamefont {Jayanth~R}\
  \bibnamefont {Banavar}}, \bibinfo {author} {\bibfnamefont {Jessica~L}\
  \bibnamefont {Green}}, \bibinfo {author} {\bibfnamefont {John}\ \bibnamefont
  {Harte}}, \ and\ \bibinfo {author} {\bibfnamefont {Amos}\ \bibnamefont
  {Maritan}},\ }\bibfield  {title} {\enquote {\bibinfo {title} {{Finite Size
  Scaling in Ecology}},}\ }\href
  {http://link.aps.org/doi/10.1103/PhysRevLett.83.4212} {\bibfield  {journal}
  {\bibinfo  {journal} {Physical Review Letters}\ }\textbf {\bibinfo {volume}
  {83}},\ \bibinfo {pages} {4212--4} (\bibinfo {year} {1999})}\BibitemShut
  {NoStop}%
\bibitem [{\citenamefont {Kiviet}\ \emph {et~al.}(2014)\citenamefont {Kiviet},
  \citenamefont {Nghe}, \citenamefont {Walker}, \citenamefont {Boulineau},
  \citenamefont {Sunderlikova},\ and\ \citenamefont {Tans}}]{Kiviet2014}%
  \BibitemOpen
  \bibfield  {author} {\bibinfo {author} {\bibfnamefont {Daniel~J}\
  \bibnamefont {Kiviet}}, \bibinfo {author} {\bibfnamefont {Philippe}\
  \bibnamefont {Nghe}}, \bibinfo {author} {\bibfnamefont {Noreen}\ \bibnamefont
  {Walker}}, \bibinfo {author} {\bibfnamefont {Sarah}\ \bibnamefont
  {Boulineau}}, \bibinfo {author} {\bibfnamefont {Vanda}\ \bibnamefont
  {Sunderlikova}}, \ and\ \bibinfo {author} {\bibfnamefont {Sander~J.}\
  \bibnamefont {Tans}},\ }\bibfield  {title} {\enquote {\bibinfo {title}
  {Stochasticity of metabolism and growth at the single-cell level},}\
  }\href@noop {} {\bibfield  {journal} {\bibinfo  {journal} {Nature}\ }\textbf
  {\bibinfo {volume} {514}},\ \bibinfo {pages} {376--9} (\bibinfo {year}
  {2014})}\BibitemShut {NoStop}%
\bibitem [{\citenamefont {{Di Talia}}\ \emph {et~al.}(2007)\citenamefont {{Di
  Talia}}, \citenamefont {Skotheim}, \citenamefont {Bean}, \citenamefont
  {Siggia},\ and\ \citenamefont {Cross}}]{DiTalia2007}%
  \BibitemOpen
  \bibfield  {author} {\bibinfo {author} {\bibfnamefont {Stefano}\ \bibnamefont
  {{Di Talia}}}, \bibinfo {author} {\bibfnamefont {Jan~M}\ \bibnamefont
  {Skotheim}}, \bibinfo {author} {\bibfnamefont {James~M}\ \bibnamefont
  {Bean}}, \bibinfo {author} {\bibfnamefont {Eric~D}\ \bibnamefont {Siggia}}, \
  and\ \bibinfo {author} {\bibfnamefont {Frederick~R}\ \bibnamefont {Cross}},\
  }\bibfield  {title} {\enquote {\bibinfo {title} {{The effects of molecular
  noise and size control on variability in the budding yeast cell cycle}},}\
  }\href@noop {} {\bibfield  {journal} {\bibinfo  {journal} {Nature}\ }\textbf
  {\bibinfo {volume} {448}},\ \bibinfo {pages} {947--51} (\bibinfo {year}
  {2007})}\BibitemShut {NoStop}%
\bibitem [{\citenamefont {Skotheim}(2013)}]{Skotheim2013a}%
  \BibitemOpen
  \bibfield  {author} {\bibinfo {author} {\bibfnamefont {Jan~M.}\ \bibnamefont
  {Skotheim}},\ }\bibfield  {title} {\enquote {\bibinfo {title} {Cell growth
  and cell cycle control},}\ }\href@noop {} {\bibfield  {journal} {\bibinfo
  {journal} {Mol. Biol. Cell}\ }\textbf {\bibinfo {volume} {24}},\ \bibinfo
  {pages} {678} (\bibinfo {year} {2013})}\BibitemShut {NoStop}%
\bibitem [{\citenamefont {Hosoda}\ \emph {et~al.}(2011)\citenamefont {Hosoda},
  \citenamefont {Matsuura}, \citenamefont {Suzuki},\ and\ \citenamefont
  {Yomo}}]{Hosoda2011a}%
  \BibitemOpen
  \bibfield  {author} {\bibinfo {author} {\bibfnamefont {Kazufumi}\
  \bibnamefont {Hosoda}}, \bibinfo {author} {\bibfnamefont {Tomoaki}\
  \bibnamefont {Matsuura}}, \bibinfo {author} {\bibfnamefont {Hiroaki}\
  \bibnamefont {Suzuki}}, \ and\ \bibinfo {author} {\bibfnamefont {Tetsuya}\
  \bibnamefont {Yomo}},\ }\bibfield  {title} {\enquote {\bibinfo {title}
  {Origin of lognormal-like distributions with a common width in a growth and
  division process},}\ }\href@noop {} {\bibfield  {journal} {\bibinfo
  {journal} {Phys. Rev. E Stat. Nonlin. Soft Matter Phys.}\ }\textbf {\bibinfo
  {volume} {83}},\ \bibinfo {pages} {031118} (\bibinfo {year}
  {2011})}\BibitemShut {NoStop}%
\bibitem [{\citenamefont {Amir}(2014)}]{Amir2014}%
  \BibitemOpen
  \bibfield  {author} {\bibinfo {author} {\bibfnamefont {Ariel}\ \bibnamefont
  {Amir}},\ }\bibfield  {title} {\enquote {\bibinfo {title} {Cell size
  regulation in bacteria},}\ }\href@noop {} {\bibfield  {journal} {\bibinfo
  {journal} {Phys. Rev. Lett.}\ }\textbf {\bibinfo {volume} {112}},\ \bibinfo
  {pages} {208102} (\bibinfo {year} {2014})}\BibitemShut {NoStop}%
\bibitem [{\citenamefont {Robert}\ \emph {et~al.}(2014)\citenamefont {Robert},
  \citenamefont {Hoffmann}, \citenamefont {Krell}, \citenamefont {Aymerich},
  \citenamefont {Robert},\ and\ \citenamefont {Doumic}}]{Robert2014}%
  \BibitemOpen
  \bibfield  {author} {\bibinfo {author} {\bibfnamefont {Lydia}\ \bibnamefont
  {Robert}}, \bibinfo {author} {\bibfnamefont {Marc}\ \bibnamefont {Hoffmann}},
  \bibinfo {author} {\bibfnamefont {Nathalie}\ \bibnamefont {Krell}}, \bibinfo
  {author} {\bibfnamefont {St\'{e}phane}\ \bibnamefont {Aymerich}}, \bibinfo
  {author} {\bibfnamefont {J\'{e}r\^{o}me}\ \bibnamefont {Robert}}, \ and\
  \bibinfo {author} {\bibfnamefont {Marie}\ \bibnamefont {Doumic}},\ }\bibfield
   {title} {\enquote {\bibinfo {title} {{Division in \emph{Escherichia coli} is
  triggered by a size-sensing rather than a timing mechanism}},}\ }\href@noop
  {} {\bibfield  {journal} {\bibinfo  {journal} {BMC Biology}\ }\textbf
  {\bibinfo {volume} {12}},\ \bibinfo {pages} {17} (\bibinfo {year}
  {2014})}\BibitemShut {NoStop}%
\bibitem [{\citenamefont {Powell}(1955)}]{Powell1955}%
  \BibitemOpen
  \bibfield  {author} {\bibinfo {author} {\bibfnamefont {E~O}\ \bibnamefont
  {Powell}},\ }\bibfield  {title} {\enquote {\bibinfo {title} {{Some Features
  of the Generation Times of Individual Bacteria}},}\ }\href@noop {} {\bibfield
   {journal} {\bibinfo  {journal} {Biometrika}\ }\textbf {\bibinfo {volume}
  {42}},\ \bibinfo {pages} {16--44} (\bibinfo {year} {1955})}\BibitemShut
  {NoStop}%
\bibitem [{\citenamefont {Voorn}\ and\ \citenamefont
  {Koppes}(1998)}]{Voorn1998}%
  \BibitemOpen
  \bibfield  {author} {\bibinfo {author} {\bibfnamefont {W~J}\ \bibnamefont
  {Voorn}}\ and\ \bibinfo {author} {\bibfnamefont {L~J}\ \bibnamefont
  {Koppes}},\ }\bibfield  {title} {\enquote {\bibinfo {title} {{Skew or third
  moment of bacterial generation times}},}\ }\href@noop {} {\bibfield
  {journal} {\bibinfo  {journal} {Archiv. Microbiol.}\ }\textbf {\bibinfo
  {volume} {169}},\ \bibinfo {pages} {43--51} (\bibinfo {year}
  {1998})}\BibitemShut {NoStop}%
\bibitem [{\citenamefont {Ullman}\ \emph {et~al.}(2013)\citenamefont {Ullman},
  \citenamefont {Wallden}, \citenamefont {Marklund}, \citenamefont
  {Mahmutovic}, \citenamefont {Razinkov},\ and\ \citenamefont
  {Elf}}]{Ullman2013}%
  \BibitemOpen
  \bibfield  {author} {\bibinfo {author} {\bibfnamefont {G}~\bibnamefont
  {Ullman}}, \bibinfo {author} {\bibfnamefont {M}~\bibnamefont {Wallden}},
  \bibinfo {author} {\bibfnamefont {E~G}\ \bibnamefont {Marklund}}, \bibinfo
  {author} {\bibfnamefont {A}~\bibnamefont {Mahmutovic}}, \bibinfo {author}
  {\bibfnamefont {Ivan}\ \bibnamefont {Razinkov}}, \ and\ \bibinfo {author}
  {\bibfnamefont {J}~\bibnamefont {Elf}},\ }\bibfield  {title} {\enquote
  {\bibinfo {title} {{High-throughput gene expression analysis at the level of
  single proteins using a microfluidic turbidostat and automated cell
  tracking}},}\ }\href@noop {} {\bibfield  {journal} {\bibinfo  {journal}
  {Philos. Trans. R. Soc. Lond. B Biol. Sci.}\ }\textbf {\bibinfo {volume}
  {368}},\ \bibinfo {pages} {20120025} (\bibinfo {year} {2013})}\BibitemShut
  {NoStop}%
\bibitem [{\citenamefont {Pugatch}(2015)}]{Pugatch2015}%
  \BibitemOpen
  \bibfield  {author} {\bibinfo {author} {\bibfnamefont {Rami}\ \bibnamefont
  {Pugatch}},\ }\bibfield  {title} {\enquote {\bibinfo {title} {Greedy
  scheduling of cellular self-replication leads to optimal doubling times with
  a log-fr\'echet distribution.}}\ }\href@noop {} {\bibfield  {journal}
  {\bibinfo  {journal} {Proc Natl Acad Sci U S A}\ }\textbf {\bibinfo {volume}
  {112}},\ \bibinfo {pages} {2611--2616} (\bibinfo {year} {2015})}\BibitemShut
  {NoStop}%
\bibitem [{\citenamefont {Iyer-Biswas}\ \emph
  {et~al.}(2014{\natexlab{b}})\citenamefont {Iyer-Biswas}, \citenamefont
  {Crooks}, \citenamefont {Scherer},\ and\ \citenamefont
  {Dinner}}]{Iyer-Biswas2014}%
  \BibitemOpen
  \bibfield  {author} {\bibinfo {author} {\bibfnamefont {Srividya}\
  \bibnamefont {Iyer-Biswas}}, \bibinfo {author} {\bibfnamefont {Gavin~E}\
  \bibnamefont {Crooks}}, \bibinfo {author} {\bibfnamefont {Norbert~F}\
  \bibnamefont {Scherer}}, \ and\ \bibinfo {author} {\bibfnamefont {Aaron~R}\
  \bibnamefont {Dinner}},\ }\bibfield  {title} {\enquote {\bibinfo {title}
  {Universality in stochastic exponential growth},}\ }\href@noop {} {\bibfield
  {journal} {\bibinfo  {journal} {Phys. Rev. Lett.}\ }\textbf {\bibinfo
  {volume} {113}},\ \bibinfo {pages} {028101} (\bibinfo {year}
  {2014}{\natexlab{b}})}\BibitemShut {NoStop}%
\bibitem [{\citenamefont {Long}\ \emph {et~al.}(2013)\citenamefont {Long},
  \citenamefont {Nugent}, \citenamefont {Javer}, \citenamefont {Cicuta},
  \citenamefont {Sclavi}, \citenamefont {{Cosentino Lagomarsino}},\ and\
  \citenamefont {Dorfman}}]{Long2013}%
  \BibitemOpen
  \bibfield  {author} {\bibinfo {author} {\bibfnamefont {Zhicheng}\
  \bibnamefont {Long}}, \bibinfo {author} {\bibfnamefont {Eileen}\ \bibnamefont
  {Nugent}}, \bibinfo {author} {\bibfnamefont {Avelino}\ \bibnamefont {Javer}},
  \bibinfo {author} {\bibfnamefont {Pietro}\ \bibnamefont {Cicuta}}, \bibinfo
  {author} {\bibfnamefont {Bianca}\ \bibnamefont {Sclavi}}, \bibinfo {author}
  {\bibfnamefont {Marco}\ \bibnamefont {{Cosentino Lagomarsino}}}, \ and\
  \bibinfo {author} {\bibfnamefont {Kevin~D}\ \bibnamefont {Dorfman}},\
  }\bibfield  {title} {\enquote {\bibinfo {title} {{Microfluidic chemostat for
  measuring single cell dynamics in bacteria}},}\ }\href@noop {} {\bibfield
  {journal} {\bibinfo  {journal} {Lab Chip}\ }\textbf {\bibinfo {volume}
  {13}},\ \bibinfo {pages} {947--54} (\bibinfo {year} {2013})}\BibitemShut
  {NoStop}%
\bibitem [{\citenamefont {Efron}(1987)}]{Efron1987}%
  \BibitemOpen
  \bibfield  {author} {\bibinfo {author} {\bibfnamefont {Bradley}\ \bibnamefont
  {Efron}},\ }\bibfield  {title} {\enquote {\bibinfo {title} {Better bootstrap
  confidence intervals},}\ }\href
  {http://www.tandfonline.com/doi/abs/10.1080/01621459.1987.10478410}
  {\bibfield  {journal} {\bibinfo  {journal} {Journal of the American
  statistical Association}\ }\textbf {\bibinfo {volume} {82}},\ \bibinfo
  {pages} {171--85} (\bibinfo {year} {1987})}\BibitemShut {NoStop}%
\bibitem [{\citenamefont {Wheals}(1982)}]{Wheals1982}%
  \BibitemOpen
  \bibfield  {author} {\bibinfo {author} {\bibfnamefont {A~E}\ \bibnamefont
  {Wheals}},\ }\bibfield  {title} {\enquote {\bibinfo {title} {Size control
  models of saccharomyces cerevisiae cell proliferation},}\ }\href@noop {}
  {\bibfield  {journal} {\bibinfo  {journal} {Mol. Cell. Biol.}\ }\textbf
  {\bibinfo {volume} {2}},\ \bibinfo {pages} {361--8} (\bibinfo {year}
  {1982})}\BibitemShut {NoStop}%
\end{thebibliography}%

\clearpage

\newpage

\onecolumngrid

\appendix

\section{Theoretical arguments on finite-size scaling and division
  control}

This Appendix presents a general formulation of the process of growth
and division as a stochastic process, and discusses the constraints
that the empirical finite-size scaling of doubling time and size
distributions impose on possible models of division control.  

In particular, using a simple analytical calculation, \rev{we will
  show that the linear scaling of size and doubling time distributions
  with their mean values is equivalent to the scaling of the division
  rate hazard function and the collapse of the size-growth plots.}
Limiting the class of models compatible with the experimental data
gives indications on the microscopic scheme at the basis of the
observed phenomenology.

\subsection{Theoretical description of the  growth and division process}
\label{sec:TheorySetUp}
As presented in detail in~\cite{Osella2014}, the growth and division
of single cells can be represented as a stochastic process defined by
the two functions, representing the rates of growth ($h_g$) and the
division hazard rate ($h^*$), i.e. the rate per unit time of cell
division as a function of the measurable variables.  A linear
dependence on cell size $V$ of the growth rate, $h_g=\alpha V$
implements the observed exponential growth of single cells.
Empirically $\alpha$ follows an approximately Gaussian distribution
with a mean value dependent on the strain and nutrient conditions
(Fig.~\ref{fig:AlphaDistribution}). 
\rev{The division hazard rate $h^*$ may be a function of all the
  growth parameters, and its form can be inferred from the
  data~\cite{Osella2014}.  In general, it can be described as a
  function of all the state variables, e.g., initial cell size and
  time elapsed in the cell cycle $h^*(t,V_0,\alpha)$, or of current
  size and initial size $h^*(V,V_0,\alpha)$.  Under the constraint of
  exponential growth $V_f=V_0 e^{\alpha\tau}$, different choices of
  parameters, such as the ones just given, are equivalent.  The
  probablity of division at time $t$ for a cell with initial size
  $V_0$ and growth rate $\alpha$ can be expressed as:
\begin{equation}
  p(t|V_0,\alpha)  = h^*(t,V_0,\alpha) e^{-\int_{0}^{t} ds h^*(s,V_0,\alpha)} =
  -\frac{d}{dt} P_0(t|V_0,\alpha),  
  \label{ptx0}
\end{equation}
where $P_0(t|V_0,\alpha)$ is the cumulative probability that a cell
born at $t=0$ is not divided at time $t$, given that its initial size
is $V_0$ and its growth rate $\alpha$.  Alternatively, the size $V$
can be used as a coordinate
\begin{equation}
  p(V|V_0,\alpha) = 
  h(V,V_0,\alpha)e^{-\int_{V_0}^{V} dv h(v,V_0,\alpha)} = 
  -\frac{d}{dV} P_0(V|V_0,\alpha). 
       \label{pxx0}
\end{equation}
Here, $h(V,V_0,\alpha) dV$ is the probability of cell division in the size
interval $[V,V+dV]$.  The two rates $h$ and $h^*$ are simply related
by $h(V,V_0,\alpha) dV = h^*(t,V_0,\alpha)dt $, where $dV/dt =h_g(V) =\alpha V$,
and therefore
\begin{equation}
h^*(t,V_0,\alpha) =
h(V(t),V_0,\alpha) \alpha V(t) =
h(V_0 e^{\alpha t},V_0,\alpha) \alpha V_0 e^{\alpha t} 
\ .
       \label{hstarh}
\end{equation}
The difference between the hazard functions $h^\ast$ and $h$ is that
the former is a probability per unit of time (i.e. a proper rate)
while the latter is a probability per unit of volume. Note that both
of them can be expressed as a function of size or time. In particular,
in the main text we considered $h^\ast(V,V_0)$, i.e. the probability
per unit of time of cell division at size $V$ given an initial size
$V_0$.  }

\rev{For simplicity, in the following we will neglect fluctuations of
$\alpha$ in a given condition, assuming $\alpha = \langle \alpha \rangle$.
We will indicate the rates obtained under this assumption as
$h^*_{\langle\alpha\rangle}(t,V_0)$ and $h^*_{\langle\alpha\rangle}(V,V_0)$.}
\rev{In this formulation of the process, the stationary distribution of
initial cell sizes $\rho_{\langle\alpha\rangle}(V_0)$ (if it exists) must satisfy
\begin{equation}
  \rho_{\langle\alpha\rangle}(V_0) = 
  2 \int_{0}^{\infty} \theta(2V_0-V'_0)  \rho_{\langle\alpha\rangle}(V'_0)
  P_{\langle\alpha\rangle}(2V_0|V'_0) dV'_0 , 
\label{pv0_equation}
\end{equation}
as described previously \cite{Osella2014}, where the Heaviside
function $\theta(2V_0-V'_0)$ is written explicitly to show the bounds.
The equation above is fully defined given a functional form of the
division rate $h$ (which defines
$\rho_{\langle\alpha\rangle}(V=2V_0|V'_0)$ in Eq.~\ref{pxx0}).}
\rev{Once $\rho_{\langle\alpha\rangle}(V_0)$ is known, the
  interdivision time distribution at steady state can in principle be
  calculated from the condition
\begin{equation}
\rho_{\langle\alpha\rangle}(\tau) = \int_{0}^{\infty}
p_{\langle\alpha\rangle}(t=\tau|V_0) 
\rho_{\langle\alpha\rangle}(V_0) dV_0 \ .
\label{tau_distribution}
\end{equation}}
Since the nutrient conditions define the average growth rate and the
average cell size (Fig.~\ref{fig:SchaechterFigure}), division control
is expected to change with nutrient conditions.  Moreover, in this
modeling framework, the functional form of the division rate sets the
mean values and the level of fluctuations of the observables, and must
induce the observed finite-size scaling of both doubling time and cell
size distributions.




%
%

\subsection{\revtwo{General scaling form of the division hazard rate
    function.}}

\rev{ This section addresses the constraints imposed by the observed
  collapse of interdivision time and initial size distributions on the
  division hazard rate function $h$ (or equivalently $h^*$).  The
  initial size distribution $\rho_{\langle \alpha \rangle}(V_0)$ in a
  given condition characterized by mean growth rate $\langle \alpha
  \rangle$ is given by
\begin{equation}
  \label{eq:cond}
  \rho_{\langle \alpha \rangle}(V_0) = 2\int_0^{\infty} 
  \theta(2V_0-V_0') \rho_{\langle \alpha \rangle}(V_0') 
   p_{\langle \alpha \rangle}(2V_0| V_0') \mathrm{d}V_0'  \ ,
 \end{equation}
 where $\theta$ is the Heaviside function, and $p_{\langle \alpha
   \rangle}(V_f| V_0)$ is the conditioned distribution of final sizes
 given initial ones. 
}

\rev{
 The collapse of initial sizes implies that $\rho_{\langle \alpha
   \rangle}(y) = \rho(y)$, with $y=V_0/\langle V_0 \rangle_{\langle
   \alpha \rangle}$. Imposing this condition in Eq~\eqref{eq:cond}
 implies that
 \begin{equation}
   \label{eq:collapse}
   \rho(y) = 2\int_0^{\infty} 
  \theta(2y-y') \rho(y') 
   p_{\langle \alpha \rangle}(2y| y') \mathrm{d}y'  \ .
 \end{equation}
 This equation immediately shows that a necessary and sufficient
 condition for the collapse is that the conditioned distribution 
 \begin{equation}
   p_{\langle \alpha \rangle}(y_f| y_0) =  f(y_f| y_0) \ ,  
 \label{eq:conditcond}
\end{equation}
i.e., it does not depend on $\langle \alpha \rangle$.
}

\rev{ 
  This condition immediately translates into a constraint for the
  division rate $h_d(V,V_0)$, which is related to the above
  conditional distribution by the following equation
\begin{multline}
  h_{\langle \alpha \rangle}(V,V_0) = - \frac{d}{d V} \log
  \int_{V}^{V_0}
  p_{\langle \alpha \rangle}(V| V_0) \mathrm{d}V_0 \\
   = - \frac{1}{\langle V_0\rangle_{\langle \alpha \rangle}} \frac{d}{d
    (V/\langle V_0\rangle_{\langle \alpha \rangle})} \log
  \int_{V_0/\langle V_0 \rangle_{\langle \alpha \rangle}}^{V/\langle
    V_0 \rangle_{\langle \alpha \rangle}} p(y|V_0/\langle V_0
  \rangle_{\langle \alpha \rangle} ) \mathrm{d}y \ .
\label{eq:hcollapse}
\end{multline}
This shows that the collapse of initial size distributions is
equivalent to the fact that the division hazard rate is universal when
rescaled by mean initial sizes, i.e. that
\begin{equation}
  h_{\langle \alpha \rangle}(V,V_0) = \frac{1}{\langle V_0\rangle_{\langle \alpha \rangle}}
  f\left( 
    \frac{V}{  \langle V_0\rangle_{\langle \alpha \rangle}}, 
    \frac{V_0}{\langle V_0\rangle_{\langle \alpha \rangle}} 
   \right) 
\label{Eq:hdcond}  
\end{equation}
}
%
%
\rev{ The equivalent condition for $h^\ast$, follows directly from the
  fact that $ h^\ast_{\langle \alpha \rangle}(V,V_0) = \langle \alpha
  \rangle V h_{\langle \alpha \rangle}(V,V_0) $. 
 \begin{equation}
   h^\ast_{\langle \alpha \rangle}(V,V_0) = \langle \alpha \rangle
   \frac{V}{\langle V_0\rangle_{\langle \alpha \rangle}}  
   f\left( 
     \frac{V}{  \langle V_0\rangle_{\langle \alpha \rangle}}, 
     \frac{V_0}{\langle V_0\rangle_{\langle \alpha \rangle}} 
   \right) \ ,
 \label{Eq:hdastcond}  
 \end{equation}
 implying that $h^\ast_{\langle \alpha \rangle}(V,V_0)/\langle \alpha
 \rangle$ is a function only of the rescaled variable.  }

\rev{
  We now consider the collapse of interdivision-time distributions and
  the size-growth plot.  Introducing a change of variables in
  eq.~\ref{eq:conditcond}, the conditional distribution for final
  sizes can be written as
\begin{equation}
  p_{\langle \alpha \rangle}(V_f| V_0) = \frac{1}{\langle
    V_0\rangle_{\langle \alpha \rangle}} 
  g_1 \left( 
    \frac{V_f}{\langle V_0 \rangle_{\langle\alpha \rangle}}, 
    \frac{V_0}{\langle V_0 \rangle_{\langle\alpha \rangle}} 
  \right) \ .
\label{eq:conditionalfinal}
\end{equation}
Since $\log(V_f/V_0) =\langle \alpha \rangle \tau $, the above
expression, combined with Eq~\eqref{eq:conditcond}, immediately gives
the following condition for the collapse of the distribution of
interdivision times
\begin{equation}
  \label{eq:condtimes}
    p_{\langle \alpha \rangle}(\tau| V_0) = \langle \alpha \rangle
    \frac{V_f}{\langle V_0\rangle_{\langle \alpha \rangle}} 
   g_1\left(
     \frac{V_f}{\langle V_0\rangle_{\langle \alpha \rangle}}, 
     \frac{V_0}{\langle V_0 \rangle_{\langle \alpha \rangle}} 
   \right) 
                                      = \langle \alpha \rangle 
   g_2\left(
     \langle \alpha\rangle \tau, 
     \frac{V_0}{\langle V_0 \rangle_{\langle \alpha \rangle}} 
   \right)
 \ . 
\end{equation}
The above condition implies the joint collapse of the distribution of
interdivision times and initial cell sizes.  }

\rev{
  Additionally, the same condition also implies a collapse of the
  size-growth plot - essentially given by an average of the
  conditional distribution $ p_{\langle \alpha \rangle}(\tau| V_0)$.
  Neglecting the variability of $\alpha$ within a single condition we
  have that
\begin{equation}
  \langle\alpha \tau \rangle = \langle \alpha \rangle \int_0^\infty d
  \tau \ \tau \  p_{\langle \alpha \rangle}(\tau|V_0) \ .
\end{equation}
If Eq.~\eqref{eq:condtimes} holds, then
\begin{equation}
  \langle\alpha \tau \rangle = \langle \alpha \rangle \int_0^\infty d
  \tau \ \tau \ f(\alpha 
  \tau|V_0/\langle V_0 \rangle_{\alpha}) \ ,
\end{equation}
and the change of variable $u=<\alpha >\tau$ gives
\begin{equation}
  \langle\alpha \tau \rangle =  
  \int_0^\infty \mathrm{d} u \ u \ g(u|V_0/\langle V_0\rangle_{\alpha})  \ , 
\end{equation}
i.e. the mean net volume change is a function of the sole ratio
$V_0/\langle V_0\rangle_{\alpha}$, therefore implying that size-growth
plots obtained with different conditions collapse when the sizes are
rescaled relatively to the average initial size. 
}

\rev{ Importantly, Eq.~\eqref{eq:condtimes} and~\eqref{Eq:hdastcond}
  are necessary and sufficient conditions for the collapse of
  interdivision time and initial size distributions. Therefore the
  collapse of the size-growth plots (which is a direct consequence of
  Eq.~\eqref{eq:condtimes}), is a necessary condition for the
  universality of interdivision time and size distribution.
These conditions are obtained neglecting the fluctuations of $\alpha$,
and are approximately valid if these are sufficiently
small. Growth-rate fluctuations introduce a new time scale
(\revtwo{proxied for example by the inverse standard deviation of
  individual growth rate} fluctuations), making
Eq.~\eqref{eq:condtimes} not strictly applicable. Hence, these
fluctuations are not compatible with a perfect collapse of the
size-growth plot and the size and doubling time distribution. This
fact could explain the small deviations across conditions that are
observed when the size-growth plots are rescaled.
}

\revthree{The recent study by Taheri-Araghi and
  coworkers~\cite{Taheri-Araghi2014} has shown that, within the adder
  model, the collapse of the added size, the initial size and the
  final size are equivalent (i.e.  the scaling of one of this quantity
  implies the scaling of the other ones). The more general calculation
  performed here shows that this result is model independent. Indeed,
  our calculation allows to recover their result in our more general
  setting.
  For binary divisions, it is trivial to show that the scaling of the
  initial and final size are equivalent: since the division ratio is
  independent of the size, the probability of the initial sizes is
  just given by the one of the final sizes under the change of
  variables $V_f = 2 V_0$.
  In order to show that the scaling property of the added size and the
  ones of final and initial size are equivalent, we can use the result
  of the previous section. There we showed that the scaling of
  final/initial sizes is equivalent to the scaling of the probability
  $p_{\langle \alpha \rangle}(V|V_0)$. The probability of the added
  size can be obtained from this one from a simple change of variable
  $V = V_0 + \Delta$:
  \begin{equation}
    p^{add}_{\langle \alpha \rangle}(\Delta|V_0) = p_{\langle \alpha \rangle}(V_0+\Delta|V_0) \ ,
    \label{eq:padd}
  \end{equation}
  where $p^{add}$ is the probability of the added size given an
  initial size $V_0$. The case of an adder would correspond to
  $p^{add}_{\langle \alpha \rangle}(\Delta|V_0)$ being independent of
  $V_0$.  From equation~\ref{eq:padd} and \eqref{eq:conditionalfinal},
  we have that the scaling of initial, final and added size are
  equivalent even if the division control is not an adder. }
\revthree{This result can be understood by a generic dimensional
  argument using the simple observation that the scaling of the sizes
  are a consequence of the fact that the division control depends on a
  single size scale. If a single size scale exists, it follows
  immediately, just from dimensional analysis, that all the size
  distributions collapse when rescaled by that scale or any quantity
  with the same dimension.}

\subsection{\revtwo{Relationship between fluctuations around the
    Shaechter growth law and universal distributions of interdivision
    times and initial sizes.}}

\revtwo{ This section derives the slope of fluctuations of individual
  cells logarithmic size around the Schaechter-Maal\o{}e-Kjeldgaard law
  (Fig.~\ref{fig:SchaechterFigure}a) directly from the collapse
  condition on the division hazard rate (Eq.~\eqref{eq:collapseh} and
  \eqref{Eq:hdcond}).  The fluctuations around Schaecter law have the
  form
\begin{equation}
\displaystyle
\log(\langle V_0\rangle_{\tau,\alpha}))  =  \log(A) + \frac{B}{ \tau  } \ , 
\label{eq:fig5aformula}
\end{equation}
where $\langle V_0\rangle_{\tau,\alpha})$ stands for the average
initial size $V_0$ for a given growth condition $\langle\alpha \rangle
$ and single-cell interdivision time $\tau$.  The quantities $A$ and
$B$ have respectively the dimensions of a size and a time. The
collapse implies that the only size and time scales of the system are
$ \langle V_0\rangle_{\alpha}$ and $1/\langle\alpha \rangle$, and
therefore the only dependence compatible with the collapse is $A = a
\langle V_0\rangle_{\alpha}$ and $B = b/\langle\alpha \rangle$, where
$a$ and $b$ are two dimensionless constants, independent of the
condition. We have therefore
\begin{equation}
  \displaystyle
  log(\langle V_0\rangle_{\tau,\alpha}))  =  \log(\langle V_0\rangle_{\alpha})
  + \log(a) + \frac{b}{ \langle\alpha \rangle \tau  } \ .  
\end{equation}
Stationarity implies that when $V_0 = \langle V_0\rangle_{\alpha}$,
$\langle\alpha \rangle \tau = \log(2)$. Under this condition
\begin{equation}
  \displaystyle
  log(\langle V_0\rangle_{\tau,\alpha}))  =  \log(\langle V_0\rangle_{\alpha})
  - \frac{b}{ \log(2)  } + \frac{b}{ \langle\alpha \rangle \tau  } \ .  
\label{eq:fig5aformulafinal}
\end{equation}
The parameter $b$ can take different values depending on the mechanism
of size control.  We observe, in agreement with
Fig.~\ref{fig:SchaechterFigure}a, that the slope of the fluctuations
decreases for fast-growth conditions as $1/\langle\alpha \rangle$.  }

\revtwo{ The same result can be obtained without dimensional
  considerations, from the conditional probability of initial size and
  interdivision times, applying Bayes' formula, as follows,
  $p_{\langle\alpha\rangle}(V_0|\tau) =
  p_{\langle\alpha\rangle}(\tau|V_0)
  p_{\langle\alpha\rangle}(V_0)/p_{\langle\alpha\rangle}(\tau)$, which
  gives
$$
p_{\langle\alpha\rangle}(V_0|\tau) = \frac{1}{\langle V_0
  \rangle_{\langle\alpha\rangle}} g_3\left( \frac{V_0}{\langle V_0
  \rangle_{\langle\alpha\rangle}}, \langle\alpha\rangle \tau \right) \ .
$$ 
Since $\langle V_0\rangle_{\tau,\alpha}$ is defined as the mean of
this distribution, if we impose it to have a linear dependence on
$1/\tau$ as in equation~\ref{eq:fig5aformula}, we immediately recover
the dependence of $A$ and $B$ on $\langle V_0
\rangle_{\langle\alpha\rangle}$ and $\langle\alpha\rangle$ obtained
above.  }

\subsection{Inference of division hazard rate from data}

Recently, we have introduced a simple method to estimate directly the
dependency of hazard-rate function from measurable variables such as
size, cell-cycle time and initial size~\cite{Osella2014}. 
%
%
Under the simplifying assumption of a division rate only dependent of
current size $V$, the division hazard $h(V)$ can be directly estimated
from the cumulative fraction $P_0(V)$ of surviving cells at size $V$
using Eq.~\eqref{pxx0}.
Considering our data, in every growth condition the estimated division
rates shows a functional dependence on size characterized by a steep
increase at small sizes, followed by a relaxation of control for
larger sizes (Supplementary Fig.~\ref{figShd}),
in good agreement with previous results~\cite{Osella2014}.

However, a cell's decision to divide may not depend solely on its
current size~\cite{Osella2014,Robert2014}. To test whether variables
other than cell size are used to determine cell division, we applied
the inference method considering the division rate dependence of both
current size and an additional variable.
As a coarse test of this additional dependence, we defined two bins of
initial sizes and estimated division rates $h_>(V,\Xi)$ and a
$h_<(V,\Xi)$ respectively from the cumulative fractions $P_{0>}(V |
V_0 > \Xi) = P_{0>}(\Xi)$ and $P_{0<}(V | V_0 < \Xi) = P_{0<}(\Xi)$ of
surviving cells at size $V$, and with initial size $V_0$ larger or
smaller than $\Xi$ respectively. Specifically, we chose for each
condition $\Xi=\langle V_0 \rangle$ and defined $h_> = h_>(V,\langle
V_0\rangle)$ and $h_< = h_<(V,\langle V_0\rangle)$.

These functions, as estimated from data, are plotted in
Fig.~\ref{fig:SizeControl}.
Under the assumption that $h$ depends only on size $V$, these two
curves would be equal for data from the same experimental
condition. The fact that the two curves deviate indicates that
additional variables, summarized by $V_0$, control division, a
condition that can be defined ``concerted control''~\cite{Osella2014}.
In other words, cell division is not determined solely by the
instantaneous size, but may contain a memory of a landmark size, or
elapsed time from a given cell cycle
event. Fig.~\ref{fig:SizeControl3} in the main text reports the same
estimate for $h^*$.
%
%
We also performed two-sample Kolmogorov-Smirnov tests comparing the
cumulative histograms $P_{0>}(\langle V_0 \rangle)$ and $P_{0<}(\langle V_0
\rangle)$, obtaining P-values lower than $10^{-4}$ for all growth
conditions for the null hypothesis that the underlying distributions
were equal. Since these small P-values may be affected by the large
sample sizes, we also performed the test on survival histograms
obtained from two random sub-samples of the same data set, composed of
a list of 1000 or 1500 dividing cells chosen randomly. In all cases
the P-values were higher, between 0.18 and 0.75, meaning that the null
hypothesis that the underlying distribution is the same could not be
rejected in this case.
%
%
This analysis indicates that size-based control is similar at
different growth rates (and is consistent with concerted control).
Conversely pure sizer or timer of division control are not consistent
with the \emph{E.~coli} data, and support a control, where at least
one extra variable, in addition to size, determines division. This
variable could be recapitulated equivalently by age in the cell cycle
or initial size~\cite{Osella2014}, in line with the results of recent
studies~\cite{Osella2014,Tzur2009}, and as argued in less recent
ones~\cite{Voorn1998}.   

In addition, the shapes of the functions $h_<$ and $h_>$ are also
similar at different growth rates. Furthermore, upon rescaling by
average initial size $\langle V_0 \rangle$ the $h_<$ and $h_>$ curves
appear to collapse (Fig.~\ref{fig:SizeControl}b and
Fig.~\ref{fig:SizeControl3}), suggesting that the mechanism of
division control is universal across conditions, as expected from
Eq.~\eqref{Eq:hdcond}. Finally, the distance between $h_<$ and $h_>$
is constant across conditions (Fig.~\ref{fig:SizeControl}c).

\subsection{Connection between scaling and division control in
  specific models}

In the minimal assumption of a division rate only dependent on size
$V$, the functional form of the divison rate $h^*(V)$ (or equivalently
$h(V)$) can be estimated from empirical data starting from
Eq.~\ref{ptx0} (or~\ref{pxx0})~\cite{Osella2014}.
More specifically, Supplementary Fig.~\ref{figShd} 
shows $h^*(V)$ for each environmental condition and \textit{E.~coli}
strain used in experiments.  The functional form is compatible with
the result of the analysis of \textit{E.~coli} cells growing in a
microfluidic device~\cite{Osella2014}.  In particular, in every
condition the division rate is characterized by a steep increase with
cell size for small sizes with respect to the average one, and a
subsequent plateau in division rate, indicating relaxation of control.
Therefore, the empirical division rate $h^*(V)$ as a function of size
$V$ can be well represented by a nonlinear saturating function such as
a Hill function in which the parameters are all in principle dependent
on the average growth rate $\alpha$:
\begin{equation}
h^*(V)= k( \alpha) \frac{1}{1 + (\frac{g(\alpha)} {V})^{n(\alpha)}}. 
\label{hill}
\end{equation}
In the above expression, the Hill coefficient $n$ sets the strength of
division control, i.e. a sharper increase of the division rate with
cell size. In the limit of $n\to\infty$ the Hill function tends to a
step function, and the model becomes equivalent to a ``perfect''
sizer, defined as a fixed size threshold at which division occurs.
The parameter $g$ is the half-maximum position of the division rate,
setting an intrinsic size scale. In the $n\to\infty$ perfect sizer
limit this parameter becomes the size threshold for division.
Finally, $k$ is the maximum value of the division rate, defining the
plateau level of the Hill function, and dimensionally defining an
intrisic time scale.
%
%
With this functional form for the division rate, the stationary
distribution of initial cell sizes (Eq.~\ref{pv0_equation}) can be
calculated analytically~\cite{Osella2014}
\begin{equation}
 p(V_0) =  
  \frac{k}{\alpha} \frac{1}{V_0} \frac{1 } { (\frac{g}{2 V_0})^n +1 } 
\left[ 1 + \left(\frac{2V_0}{g}\right)^n \right]^{ -\frac{k}{\alpha n}},
\label{p0_hill}
\end{equation}
 and consequently the coefficient of variation $CV_{V_0}=\sigma_{V_0}/
\langle V_0\rangle$ of initial cell size is
\begin{equation}
  CV_{V_0}^2 = 2 n \frac{\Gamma(\frac{2}{n}) \Gamma(\frac{k}{\alpha n}) \Gamma(\frac{k-2\alpha}{\alpha n})} {\Gamma(\frac{1}{n}^2) \Gamma(\frac{k-\alpha}{\alpha n})^2} -1. 
\label{cv}
\end{equation}
(Here the dependence of $g$, $n$, and $k$ has been omitted for
clarity). The empirical linear scaling of cell size shown in
Fig.~\ref{fig:Rescaling} implies a constant level of relative
fluctuations $CV_{V_0}$.  In the model, this noise level depends on
the Hill coefficient $n$, and on the ratio $k/\alpha$, but does not
depend on the intrinsic size scale in the division rate defined by its
half-maximum position $g$.  Therefore, a sizer mechanism with a
constant strength of control $n$ (i.e., independent of $\alpha$)
naturally leads to a constant $CV_{V_0}$ if the only intrinisc time
scale is simply set by $\alpha$ (i.e., $k/\alpha$ is a constant).  In
fact, the parameter $k$ in the division rate is the only one with the
dimensions of time, and has to be linear in $\alpha$ to keep the
relative fluctuations constant in every growth condition.  This is a
constraint on the possible mechanisms of size control.

Supplementary Fig.~\ref{figShd}a and \ref{fig:SizeControl}
strongly suggest an independence of $n$ on growth conditions,
supporting the picture of a constant strength of size
control. Similarly, Supplementary Fig.~\ref{figShd}b shows that the
maximum division rate is simply proportional to the growth rate, i.e.,
$k=A~\alpha$ where $A$ is a constant.  Note that, due to the relation
$h^* =h \alpha V$, this is equivalent to an independence from $\alpha$
of the plateau value of the rate $h$ shown in
Fig.~\ref{fig:SizeControl}.
Therefore, the empirical division rates increase with cell size with
the same steepness across growth conditions, and hence are compatible
with a constant parameter $n$.  Additionally, the only time scale in
the model, set by the plateau level $k$ of the division rate, is
simply proportional to the growth rate $\alpha$.  These two
observations imply a level of relative size fluctuations completely
independent from the average growth rate induced by the nutrient
conditions.  Moreover, Eq.~\ref{cv} shows that this level of
fluctuations is completely independent from the intrinsic size scale
in the model, defined by the half-maximum position $g$.  In turn, the
size scale $g$ defines the average initial cell size, which is
described by the expression
%
%
%
%
%
\begin{equation}
  \langle V_0\rangle = g \frac{k}{2 \alpha n^2}
  \frac{\Gamma(\frac{2}{n})\Gamma(\frac{k-\alpha}{\alpha n}) }{\Gamma(1+
    \frac{k}{\alpha n})}.  
\label{meanv0}
\end{equation}
Fig.~\ref{figShd}c confirms the linear proportionality $g=B~\langle
V_0\rangle$, where $B$ is a constant, in the data analysed.  Note that
this implies an exponential dependence of $g$ on growth rate, in
agreement with the Schaechter \emph{et al.} law.
The different division rates can be collapsed on a universal division
control function if size is rescaled by the average initial size and
the rate is rescaled by the average growth rate (Supplementary
Fig.~\ref{figShd}d).
This opens the possibility of accumulating statistics using data
collected for different strains and in different nutrients conditions
to infer more precisely this universal function.
With the two established relations $k(\alpha) =A \alpha$ and
$g(\langle V_0 \rangle)= B \langle V_0 \rangle $, the size
distribution in Eq.~\ref{p0_hill} can be rewritten as
\begin{equation}
 p(V_0) V_0 =   
  A  \frac{1 } { \left(\frac{B}{2}\right)^n \left(\frac{\langle V_0\rangle } {V_0}\right)^n  +1 }  
\left[ 1 + \left(\frac{2}{B}\right)^n  \left(\frac{V_0}{\langle V_0\rangle}\right)^n \right]^{ -\frac{A}{n}},
\label{p0_hill_resc}
\end{equation}
which represents the model prediction for the rescaled size
distributions in Fig.~\ref{fig:Rescaling}a.  Supplementary
Fig.~\ref{figScollapse}a shows that Eq.~\ref{p0_hill_resc} with the
estimated values of the constants $A$ and $B$ is indeed in good
agreement with the empirical distributions.

Even for this simplified model in which the division rate is a
function of size only, the stationary doubling time distribution is
hard to calculate analytically.  However, simulations of the process
show that the model predicts a finite-size scaling also for the
doubling time distribution (Supplementary Fig.~\ref{figScollapse}b),
as it is observed in empirical data (Fig.~\ref{fig:Rescaling}).  In
this case, the empirical and the simulated distributions cannot be
compared quantitatively.  Indeed, the model is neglecting the presence
of concerted control, i.e. the dependence of the division rate on an
additional control variable ($V_0$ or $t$), which is supported by the
data (Fig.~\ref{fig:SizeControl3} and Supplementary
Fig.~\ref{fig:SizeControl}).  As shown in~\cite{Osella2014}, this
concerted control has the effect of reducing the fluctuations in the
doubling time distributions (as well as altering some correlations
between variables) but does not influence substantially the size
distributions.  For this reason, a simple sizer model can predict well
the empirical size distributions (Supplementary
Fig.~\ref{figScollapse}a) but fails to capture, even qualitatively,
the interdivision time distributions.

\clearpage


\newpage

\setcounter{page}{1}

\setcounter{figure}{0}
\renewcommand{\figurename}{Supplementary Figure}
\renewcommand{\thefigure}{S\arabic{figure}}

\setcounter{table}{0}
\renewcommand{\tablename}{Supplementary Table}
\renewcommand{\thetable}{S\arabic{table}}

\renewcommand{\theequation}{S\arabic{equation}}


\vspace{3cm}

\begin{center}
  {\Large \bf Supplementary Figures for Kennard \emph{et.~al.}}
\end{center}

\vspace{3cm}



\begin{figure*}[h!]
  \includegraphics[width=0.7\textwidth]{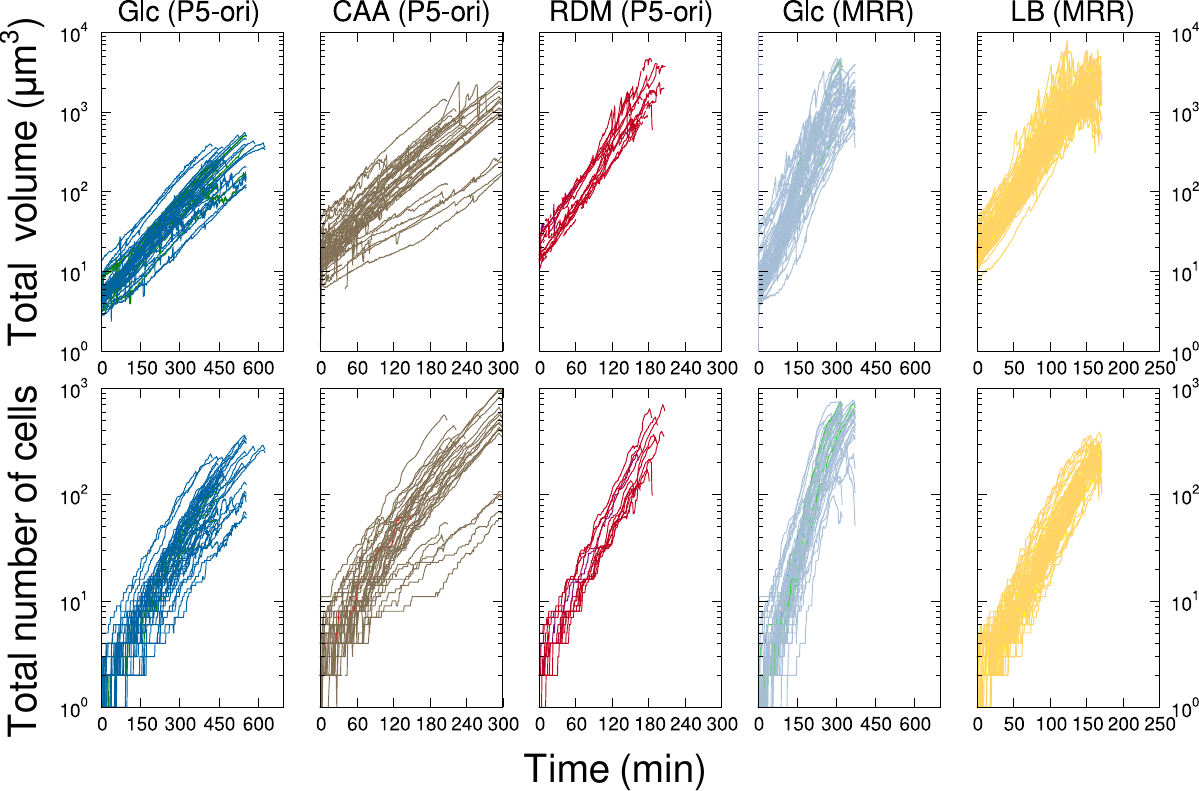}
  \caption{Microcolony growth is exponential with respect to both
    total cell volume and cell number. \textit{Top panels}: Plot of
    total cell volume over time for each microcolony in each growth
    condition.  Each line represents a single field of view (from left
    to right $n=48,39,15,63,88$, and 109 fields of view).
    \textit{Bottom panels}: total number of cells over time for the
    same experiments. Because filters tend to exclude many cells at
    the end of an experiment (because these cells might not finish
    dividing before the end of the experiment), the tracks shown here
    are for unfiltered data.}
\label{fig:S-balanced_microcol_growth}
\end{figure*}
\clearpage


\begin{figure*}
  \includegraphics[width=0.6\textwidth]{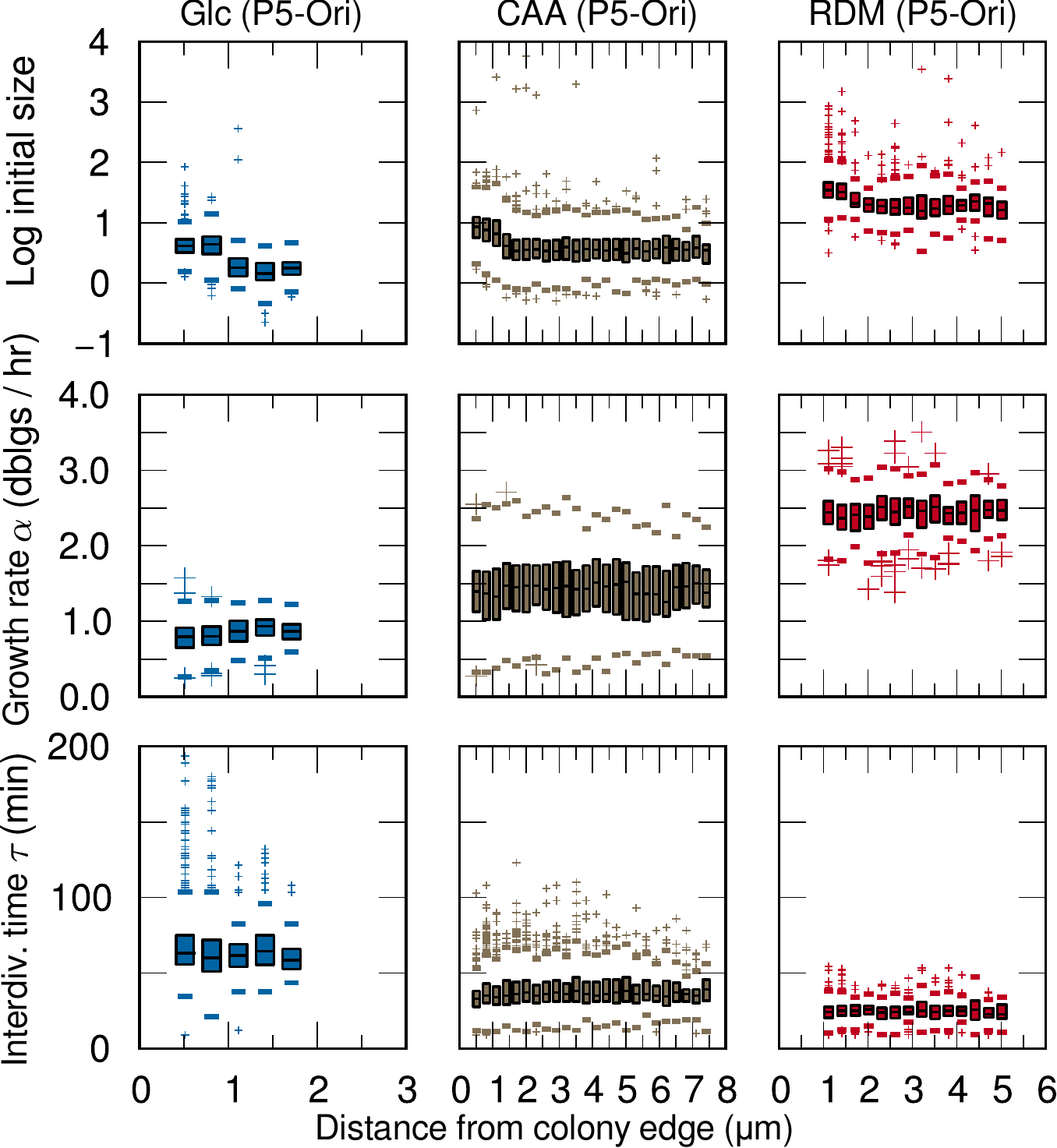}
  \caption{Stability of measured parameters with respect to position
    of the cell in the microcolony. The plots are boxplots
    representing distributions, binned by distance of a cell from the
    microcolony edge (bin width is 0.3~$\mu$m).  \textit{Top panel}: logarithm
    of the initial volume $\log V_0$; the scored sizes are biased
    towards the colony edge, possibly because of the asymmetry of the
    image. \textit{Mid/bottom panel}: the same bias is absent from measured
    interdivision times and individual-cell growth rates.  The three
    colums refer to three different growth conditions.  }
\label{fig:S-size-bias-colny-edge}
\end{figure*}

\clearpage

\begin{figure}[t]
  \centering
  \includegraphics[width=0.7\textwidth]{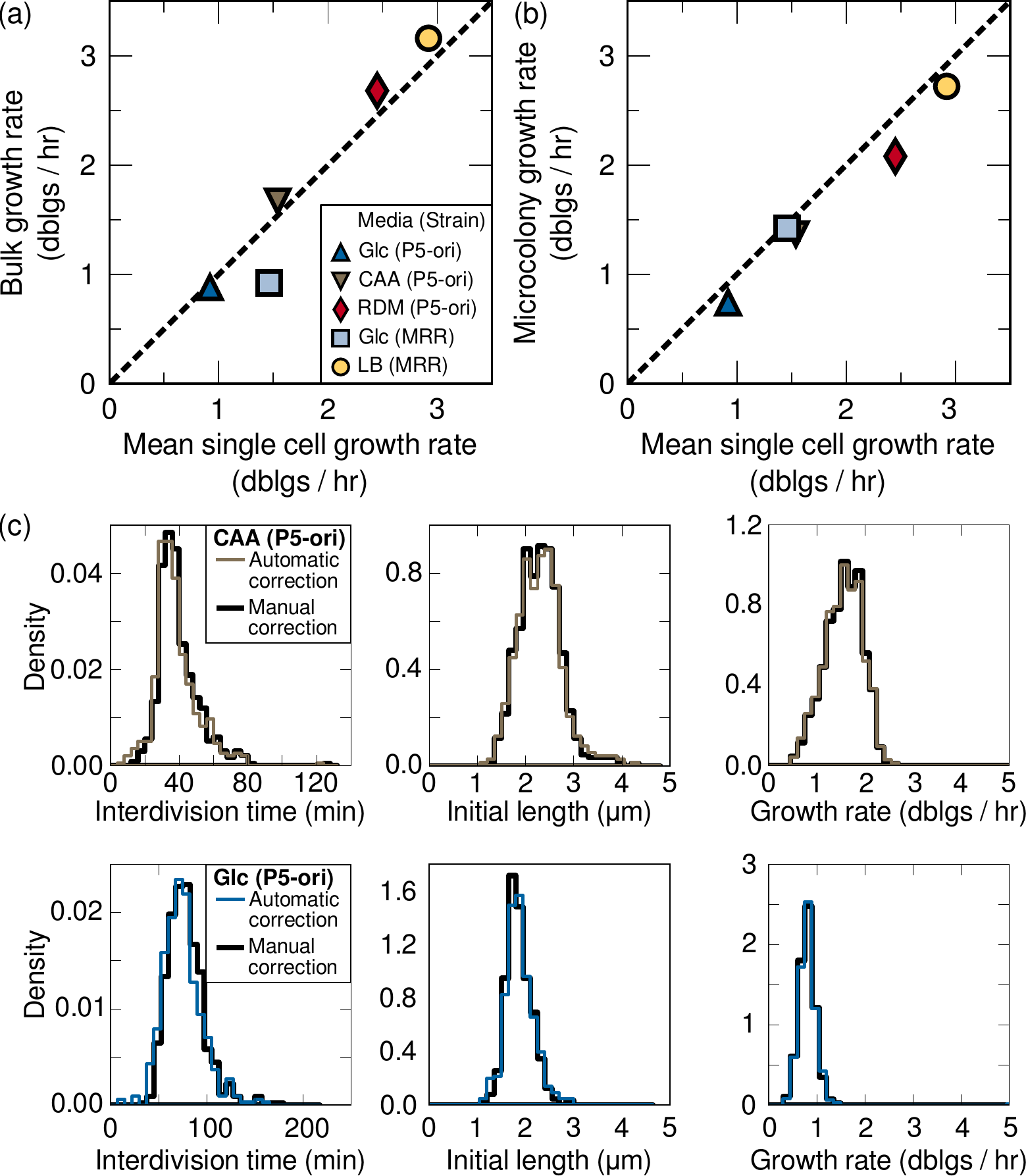}
  \caption{\revtwo{Absence of biases from the segmentation/tracking
      analysis.  (a and b) Growth rates of analyzed cells on agar pads
      are consistent with measured bulk and colony growth rates.}  (a)
    Correlation between bulk growth rate and the average growth rate
    in each condition. Symbols show the average across all biological
    replicate
    experiments.  
    \revtwo{All conditions show close agreement between bulk and agar
      growth, except for the MRR strain in M9 + Glc; these differences
      are reproducible, and suggest the possibility of unexplored
      physiological differences between growth on agar and in bulk
      culture. Nevertheless, these differences do not significantly
      affect the mechanism of cell size control.}
 \revtwo{(b) Comparison of the single-cell growth
      rate with the growth rate of individual microcolonies.  The
      average single-cell growth rate (after all filters from the
      segmentation algorithm are applied) is compared to the average
      growth rate of microcolony area across each data set. Note that
      this growth rate is calculated \emph{without} any filtering of
      cells due to the segmentation and tracking algorithm.  The lines
      show $y=x$ as a guide to the eye. (c) Technical filters to
      correct for tracking errors do not bias the distributions of the
      main observables.  The analysis compares the algorithm to to
      manually corrected data with the Schnitzcells software package
      (Young JW \textit{et al.}  \textit{Nature Protocols} 7, 80–88,
      2012). The resulting distributions of length, growth rate, and
      interdivision times are indistinguishable in the two cases. Top
      row shows the distributions for the P5 CAA condition, while the
      bottom row shows P5 Glc.  }}
  \label{fig:S-BulkGrowthCompare}
\end{figure}

\clearpage


\begin{figure*}
  \includegraphics[width=0.8\textwidth]{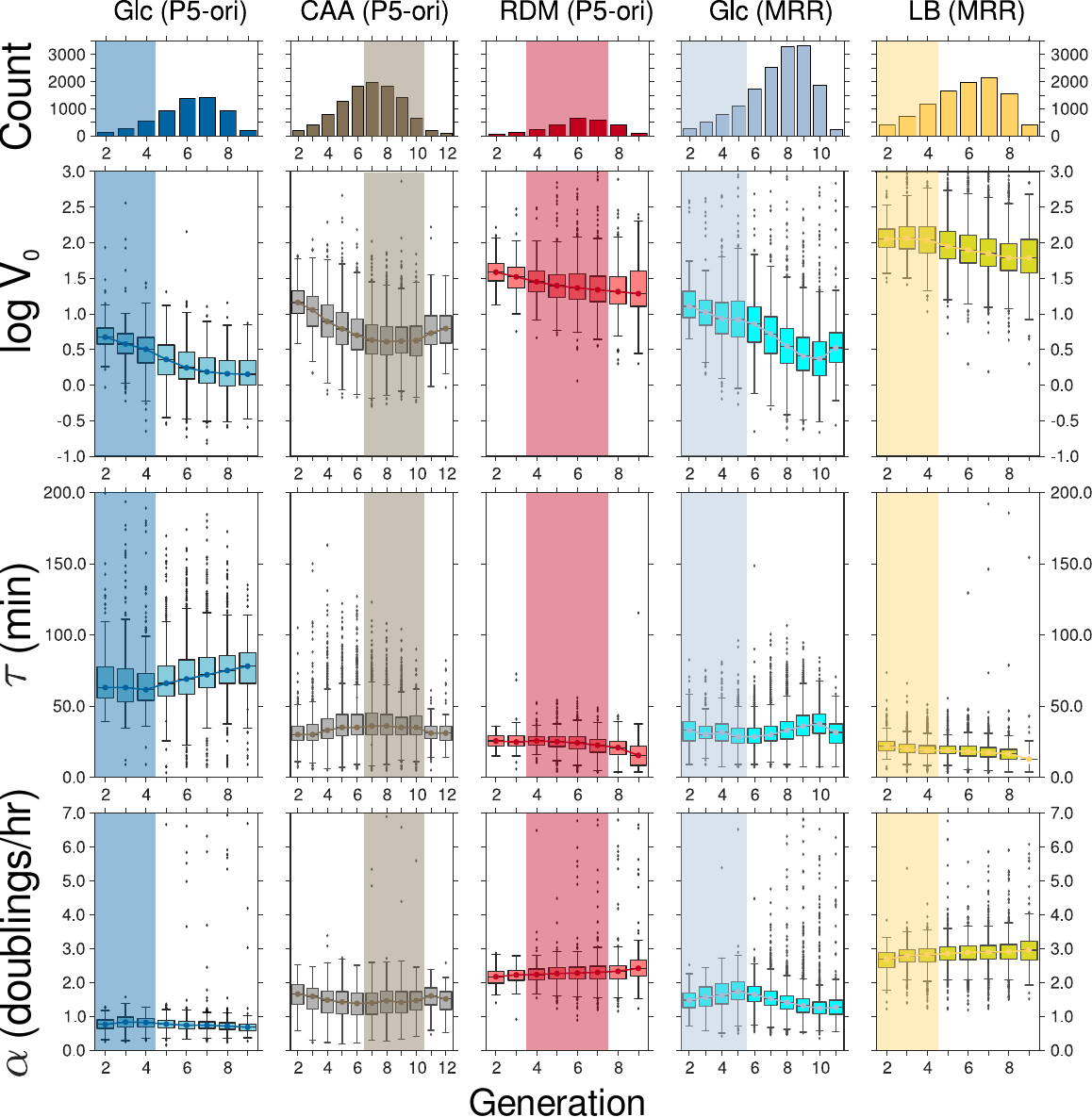}
  \caption{ Distributions of main observables for each growth
    condition binned by generation.  \textit{Top row}: total number of cells in
    each generation.  \textit{Second row}: distribution of log cell size for
    each generation.  \textit{Third row}: distribution of interdivision time
    $\tau$ for each generation.  \textit{Bottom row}: distribution of
    growth rate $\alpha$ for each generation. Trend lines
    represent the median.  Box limits mark the inner quartile range
    ($IQR$). Whiskers extend to lowest and highest data point within
    $1.5\times IQR$ of the box boundaries. Vertical axes are common to
    all plots in a row. Box plot conventions are the same in all
    rows. The highlighted regions mark the filter on the range of
    generations with steadier growth used in further analyses.}
\label{fig:S-steadyness_by_gen}
\end{figure*}
\clearpage

\begin{figure*}
  \includegraphics[width=0.8\textwidth]{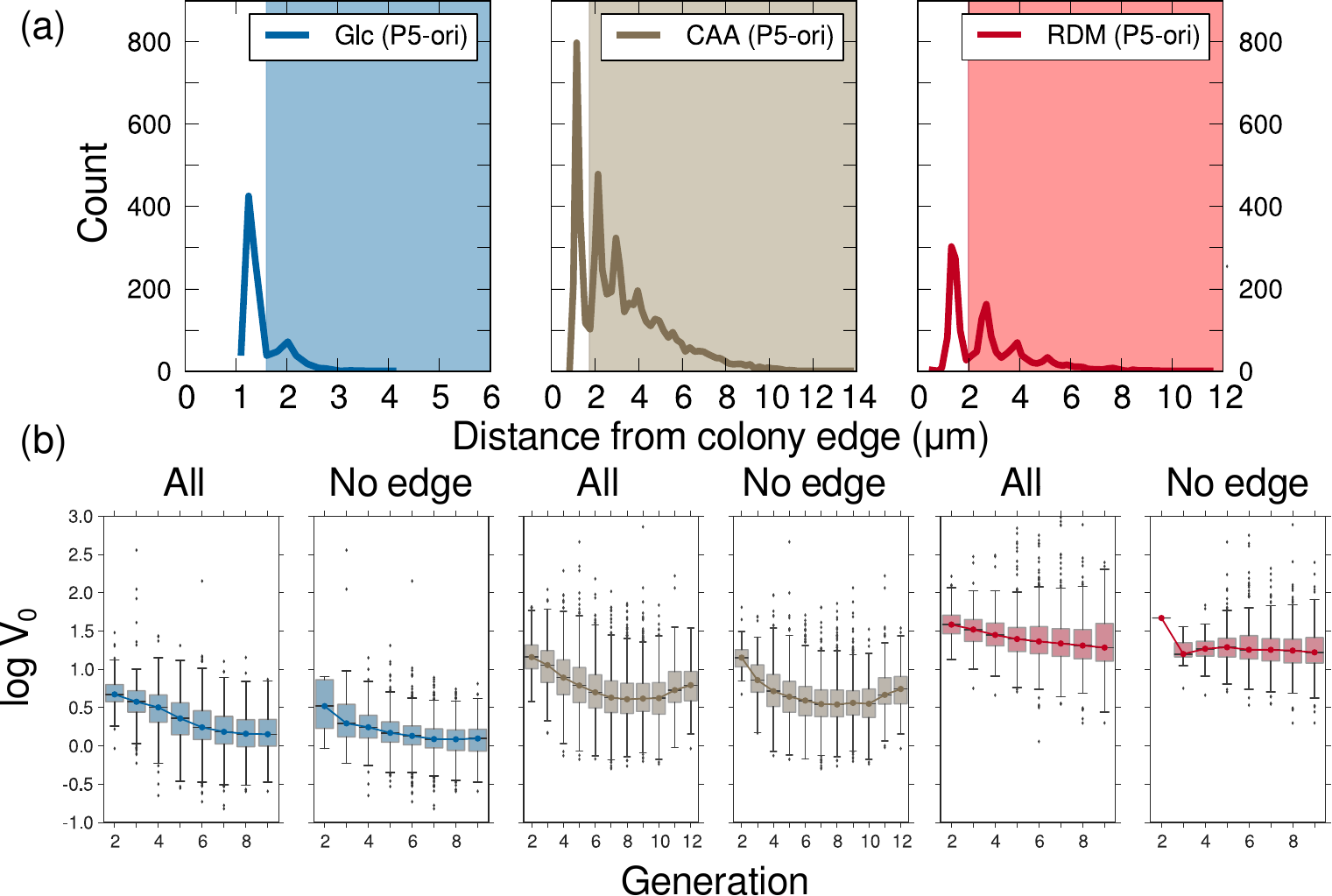}
  \caption{Effects of colony edge segmentation bias on steadiness of
    initial cell size by generation. \textit{Top panel}: overall distribution
    of minimal distances from colony edge in three different growth
    conditions. Shaded areas indicate filtered regions. \textit{Bottom panel}:
    comparison of distributions of log initial size $\log V_0$ binned
    by generation (shown as boxplots as in figure
    \ref{fig:S-steadyness_by_gen}), filtered to exclude cells on the
    colony edge (``no edge'') or unfiltered (``all''). These plots
    show that removing the cells close to colony edges improves the
    steadiness of initial size by generation, but does not fully
    account for the observed increasing trends in later generations.  }
\label{fig:S-bias_and_steadiness}
\end{figure*}
\clearpage


\begin{figure}
  \includegraphics[width=0.8\textwidth]{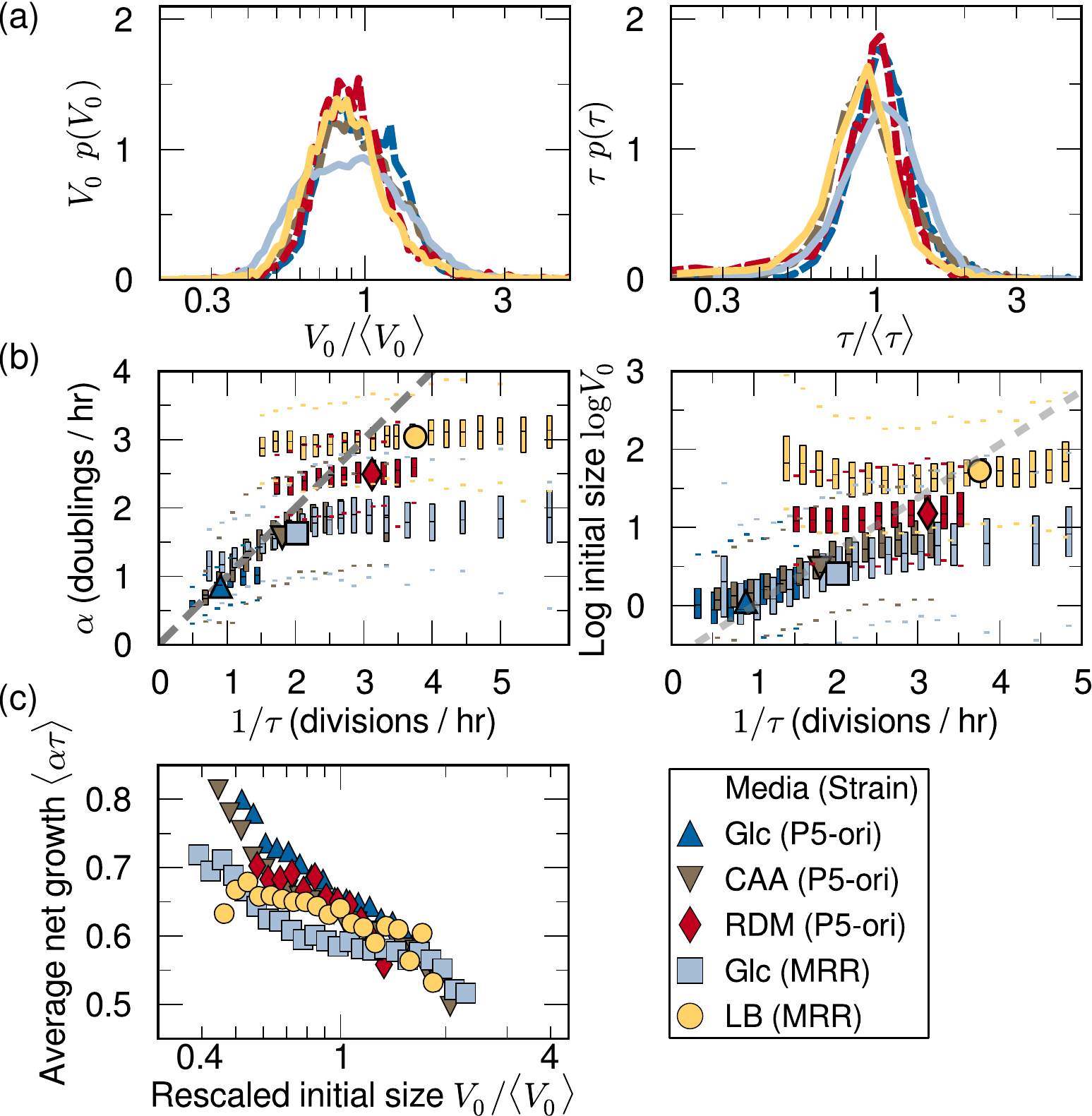}
  \caption{All the results of this work are robust with respect to
    releasing the filter on generation range used in the main
    analysis (Fig.~\ref{fig:S-steadyness_by_gen}). The plots
    illustrate the main results without this filter applied.  
    (a) Scaling of the initial size and
    doubling time distributions (Fig.~\ref{fig:Rescaling}). (b)
    Crossover in the fluctuations around the mean behavior and
    fluctuations around the Schaechter-Maal\o{}e-Kjeldgaard plot
    (Fig.~\ref{fig:AlphaInvTau} and \ref{fig:SchaechterFigure}). (c)
    Scaling properties of size-growth plot  (Fig.~\ref{fig:SizeControl}) }
\label{fig:S-unfiltered-results}
\end{figure}
\clearpage




\begin{figure*}
  \includegraphics[width=0.8\textwidth]{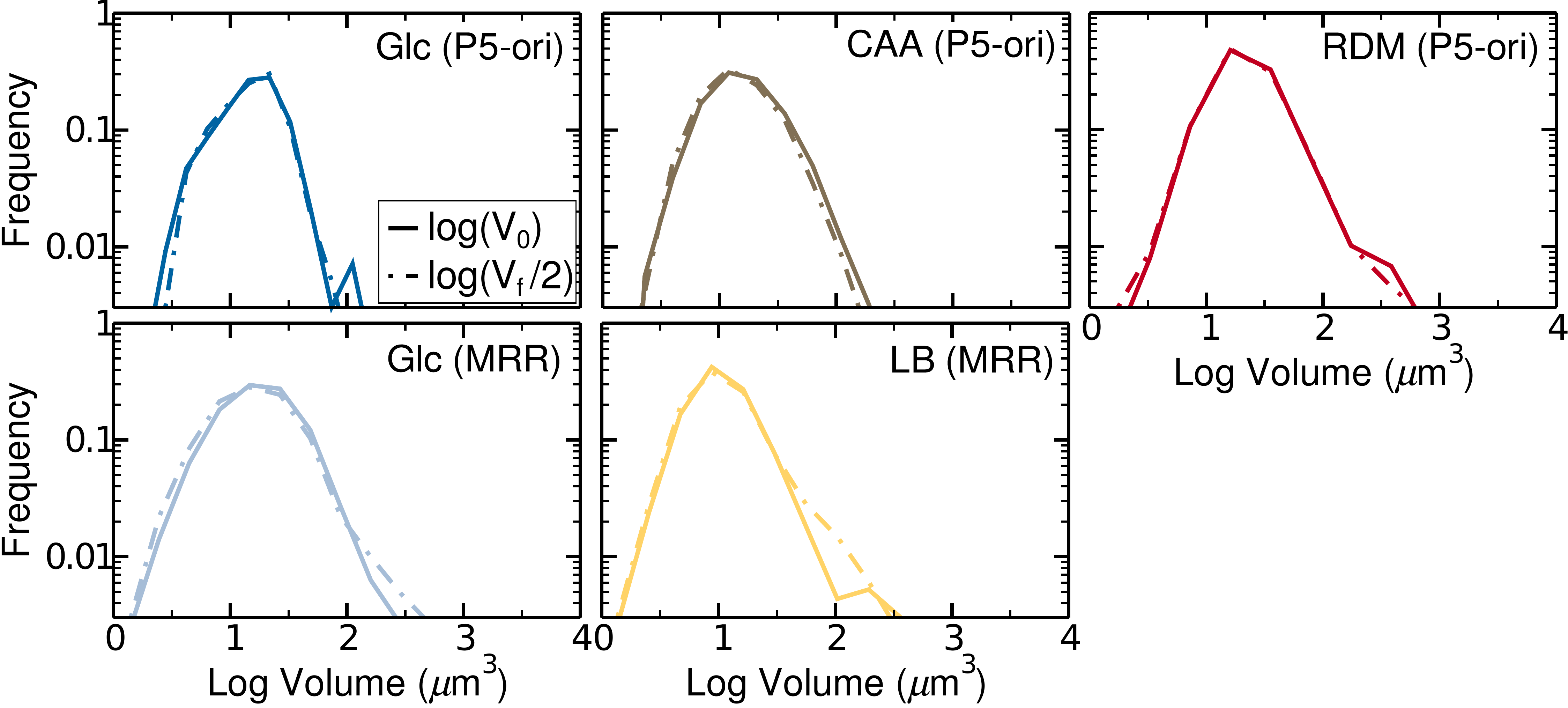}
  \caption{The distribution of final sizes divided by two matches that of initial
    sizes. Distributions of initial size (dashed lines) or half the
    final size (solid lines) plotted for each growth condition (and suitably normalized). The
    good overlap in all conditions suggests that each population is in
    a nearly steady state of growth.
  }
\label{fig:S-steadyV0distr}
\end{figure*}
\clearpage

\begin{figure}
  \includegraphics[width=0.35\textwidth]{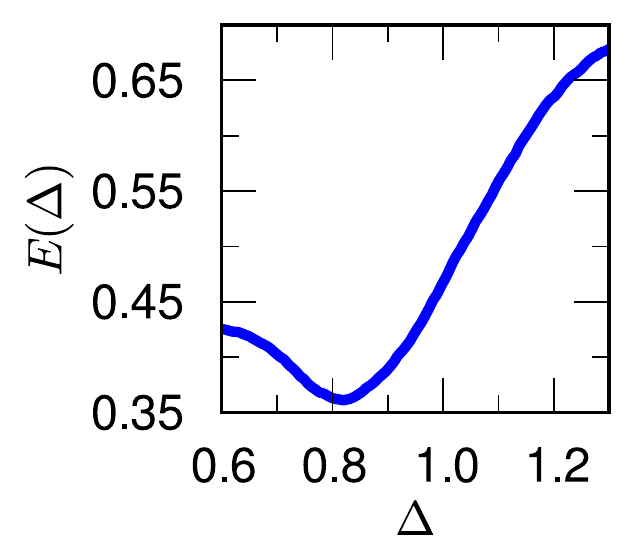}
  \caption{\rev{Most parsimonious scaling collapse of the single-cell
      growth rate distributions is not at $\Delta=1$. To measure the
      goodness-of-collapse, a scaling exponent $\Delta$ is chosen and
      the histograms $p(\alpha)$ from each condition are rescaled
      according to
      $\alpha^{\Delta}p(\alpha)=F(\alpha/\langle\alpha\rangle^{1/(2-\Delta)})$
      in order to obtain the curves $F$ as a function of the rescaled
      single-cell growth rate
      $\alpha/\langle\alpha\rangle^{1/(2-\Delta)}$. The functional
      $E(\Delta)$ is then defined as the total area of overlap between
      each pair of rescaled curves $F$ in the dataset, evaluated on
      their common support and normalized by the total number of
      overlapping pairs~\cite{Giometto2013,Bhattacharjee2001}. The
      value of $\Delta$ for which $E(\Delta)$ is minimized is the most
      parsimonious scaling exponent; the uncertainty can be inferred
      from the width around the minimum.  Unlike the most parsimonious
      scaling collapses for the interdivision time and initial size,
      the most parsimonious scaling collapse for the \revtwo{growth}
      rate is $\Delta=0.82\pm0.004$ (1\% error).}}
    \label{fig:Alpha_GoodnessofScaling}
  \end{figure}

\clearpage



\begin{figure*}
  \includegraphics[width=0.5\textwidth]{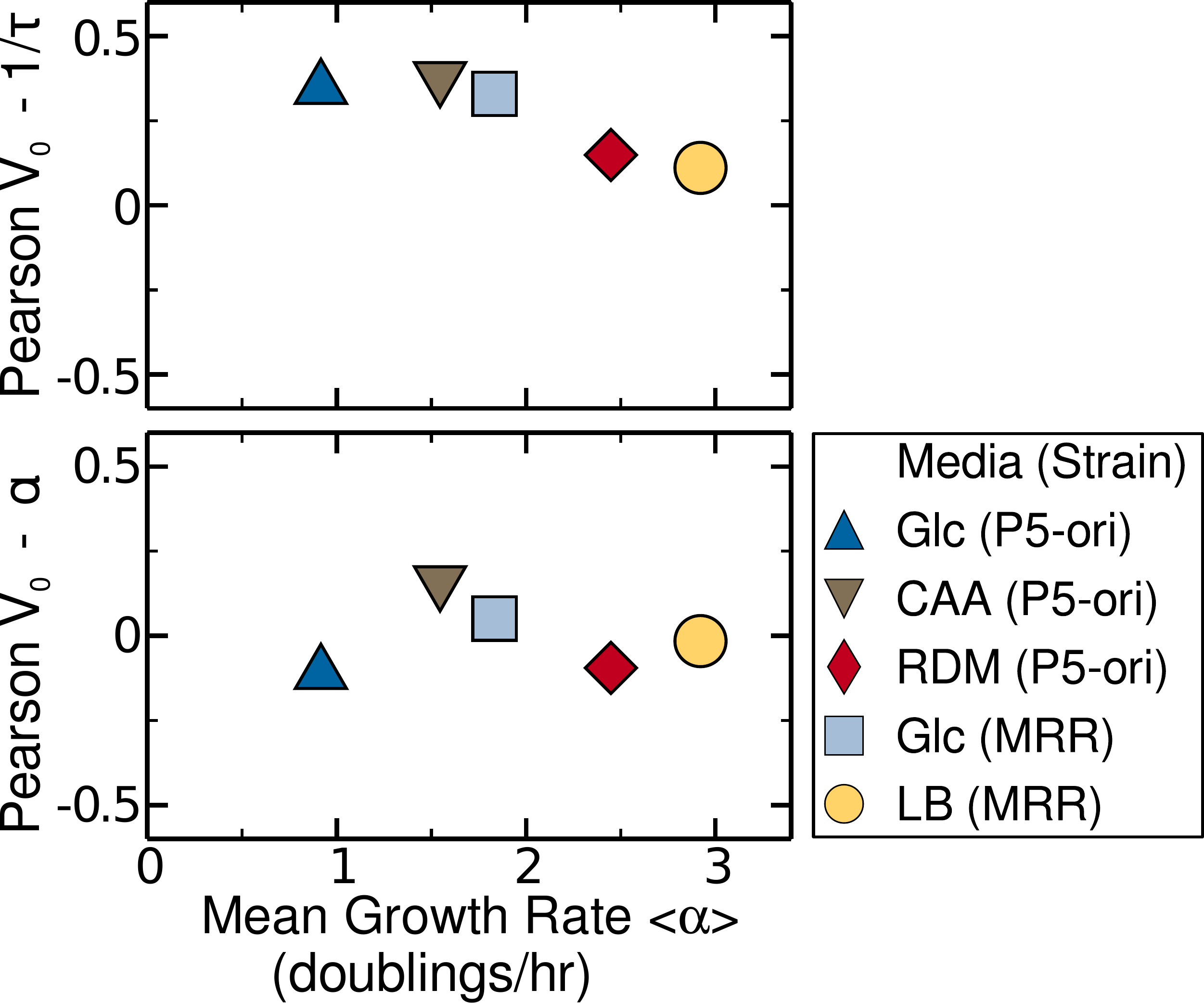}
  \caption{Correlation between initial size and inverse interdivision
    time (top panel) or growth rate (bottom panel). Pearson
    correlation between the two quantities as a function of mean
    growth rate $\langle\alpha\rangle$.}
\label{fig:S-Pearsonate}
\end{figure*}
\clearpage


\begin{figure*}
  \includegraphics[width=0.45\textwidth]{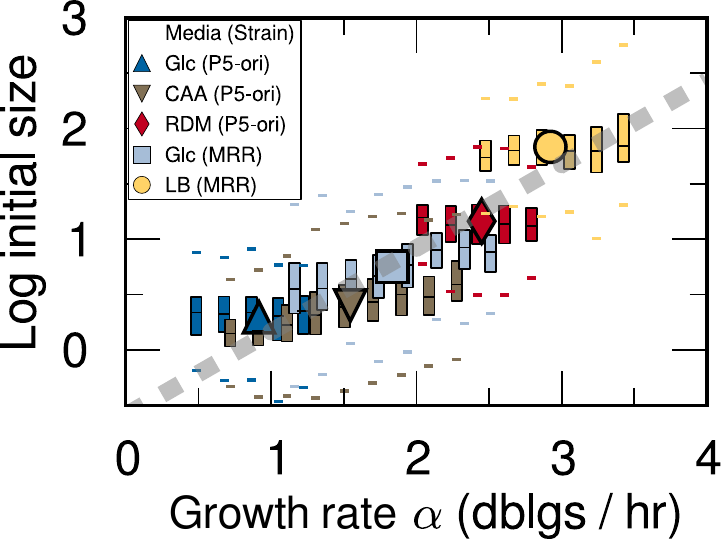}
  \caption{Schaechter-Maal\o{}e-Kjeldgaard plot of initial size as a
    function of growth rate, rather than inverse interdivision time
    (Fig.~\ref{fig:SchaechterFigure}). Bin width is 0.2 doublings /
    hr. Large symbols represent population medians. Gray line is the
    fit of the population medians, with a slope of 66.3 minutes.}
\label{fig:S-Schaechter_alpha}
\end{figure*}
\clearpage

\begin{figure}
  \includegraphics[width=0.45\textwidth]{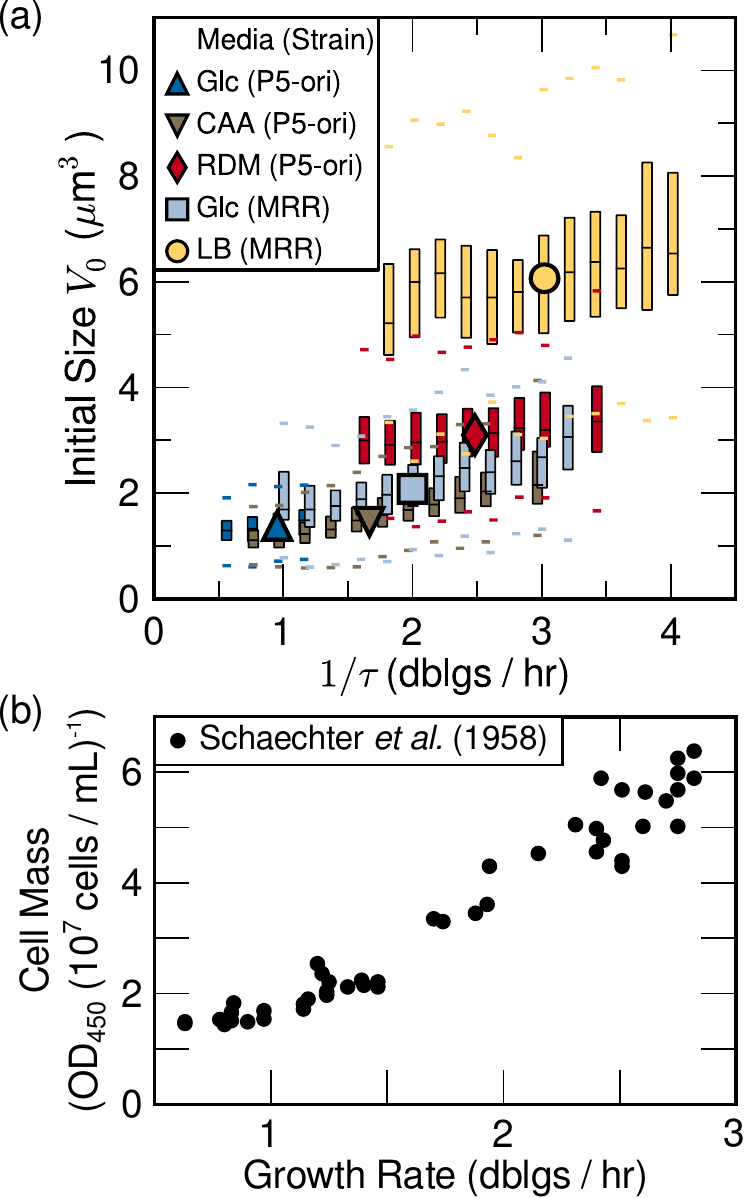}
  \caption{\rev{Data is consistent with an exponential SKM
      law. \textbf{(a)} Plot of initial size (on a linear axis) as a
      function of inverse interdivision time
      (cf. Fig.~\ref{fig:SchaechterFigure}(a), which plots size on a
      logarithmic axis). The trend in the mean is consistent with a
      super-linear dependence (e.g. exponential) dependence of size on
      $1/\tau$. Furthermore, the spread in size for a given value of
      $1/\tau$ increases with increasing values of $1/\tau$
      (demonstrated by the increasing length of the boxes, which
      represent the interquartile range). However, perturbations of
      growth rate are limited in dynamic range, and thus these trends
      could also be consistent with other functional dependencies of
      size on $1/\tau$, such as a linear or polynomial
      dependence. \textbf{(b)} For comparison, the original data from
      the 1958 paper of Schaechter and coworkers~\cite{SCHAECHTER1958}
      plotting average cell mass versus bulk growth rate on linear
      axes. Each point represents the average of a culture growing in
      different nutrient conditions.  Establishing the exact
      functional dependence of the SKM law is an open question, and
      alternate fits are possible also with the original
      Schaechter~\emph{et al.} data. }}
\label{fig:SchaechterLinLin}
\end{figure}
\clearpage

\begin{figure}
  \includegraphics[width=0.45\textwidth]{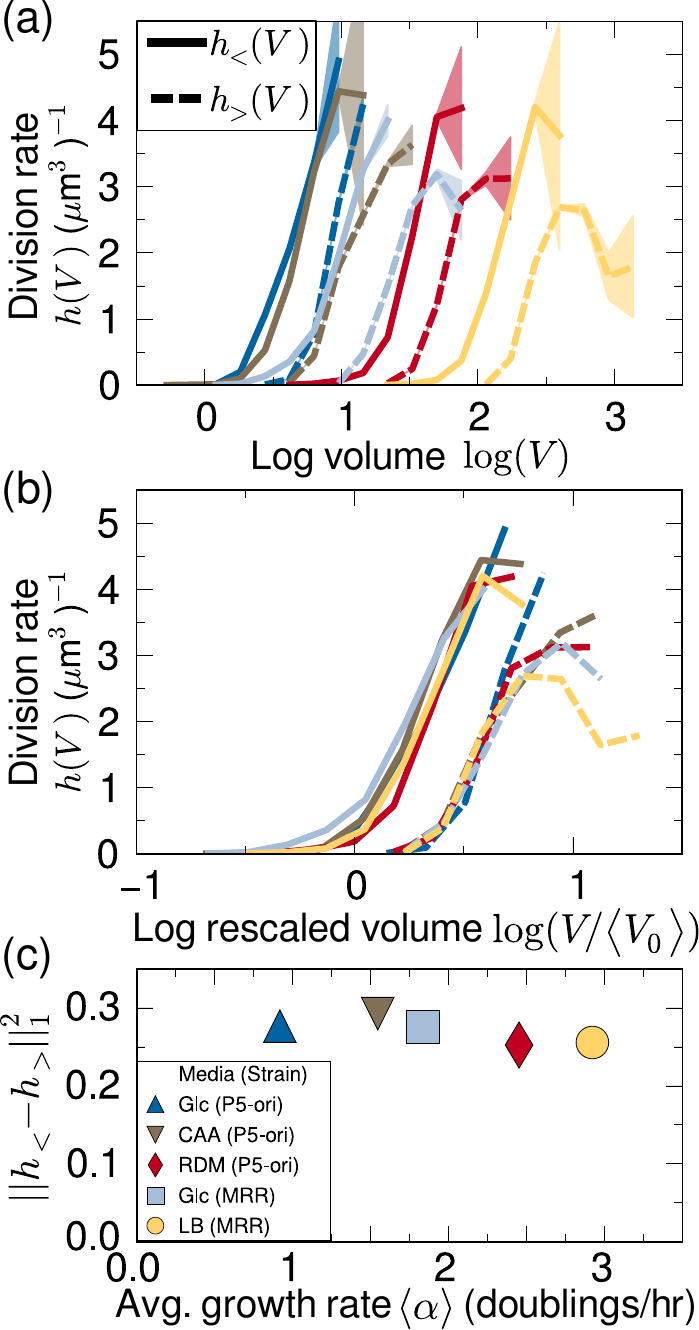}
  \caption{\rev{Division control variations across different growth
      conditions are intimately linked to the universal size and
      doubling-time distributions, and consistent with a ``concerted''
      control, where current size is not the only variable determining
      division. \textbf{(a)} Division (hazard) rates per unit volume
      conditional on initial size, plotted as a function of size
      alone. $h_<(V)$ (solid line) is the rate of cell division for
      cells whose initial size was smaller than the average initial
      size; $h_>(V)$ (dashed line) is the rate of cell division for
      cells whose initial size was larger than the average initial
      size.  If size control depended only on current size then these
      curves should be the same. Shaded regions represent the standard
      error as in~\cite{Wheals1982}. Colors represent different
      conditions as listed in the legend of (c). \textbf{(b)} As in
      (a) with size rescaled by average initial size. Error omitted
      for clarity.  \textbf{(c)} $L_1$ distance between $h_<$ and
      $h_>$ on their common support, normalized by the length of the
      common support, and plotted as a function of growth rate. }}
\label{fig:SizeControl} 
\end{figure}

\clearpage

\begin{figure}
  \centering
  \includegraphics[width=0.45\textwidth]{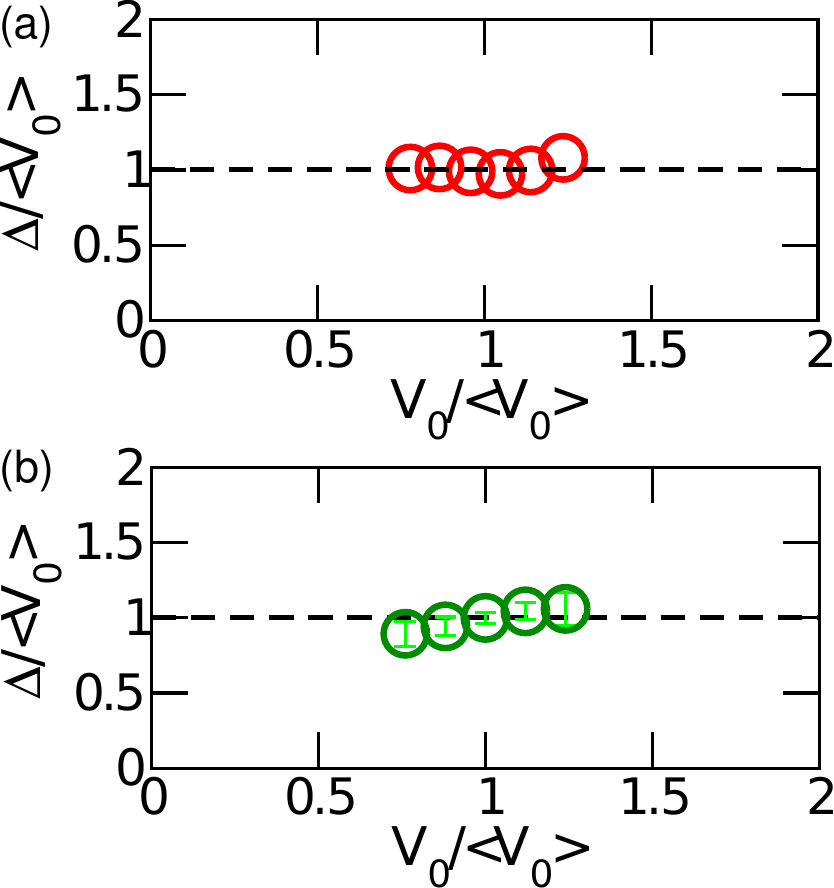}
  \caption{\revtwo{Cell size change is consistent with an adder
      mechanism~\cite{Campos2014,Taheri-Araghi2014}. \textbf{(a)} The
      relative change in cell size (relative to initial size) plotted
      as a function of rescaled initial size for the MRR LB
      dataset. \textbf{(b)} Relative change in cell size plotted as a
      function of rescaled initial size for all data sets pooled
      together.}}
  \label{fig:Sadder}
\end{figure}

\clearpage

\begin{figure*}
\centering
\includegraphics[width=0.6\textwidth]{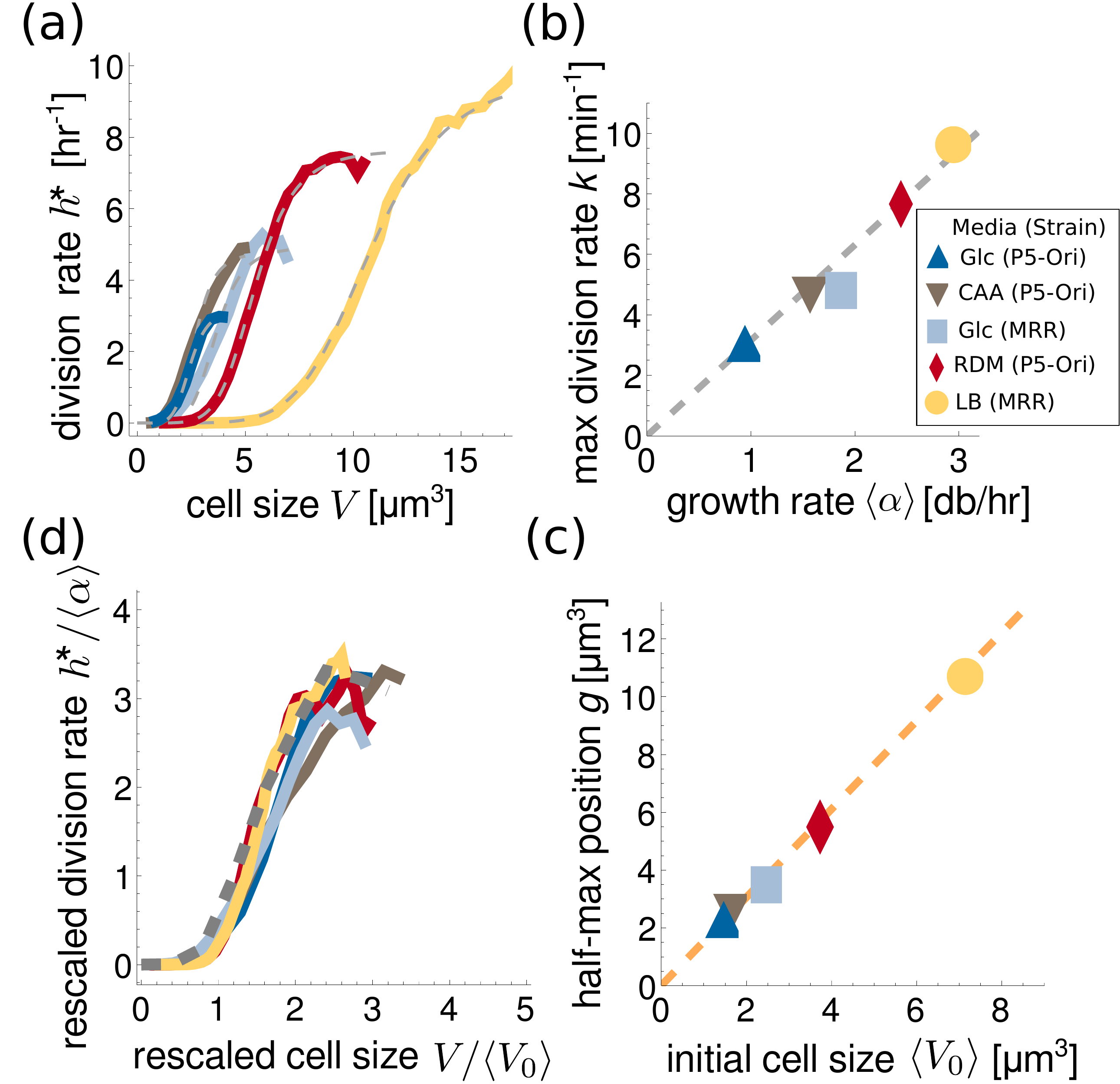}
  \caption{ \textbf{Inference of division rate $h^*(V)$.}  (a) The
    dependence on 
cell size $V$ of the division rate $h^*(V)$ for cells  growing in
different conditions (see legend), thus with different average growth
rate.  The functional 
dependence is  compatible with the results of the analysis of
fast growing cells in a microfluidic device~\cite{Osella2014}. In
particular, the division rate in every condition can be represented
by a nonlinear saturating function $h^*=\frac{k}{1+(g/V)^n}$ (dashed
lines) with a constant Hill coefficient $n$, while the other two
parameters $g$,$k$ show a dependence on conditions.
(b) Linear dependence of the maximum division rate on average growth
rate $k=A\langle\alpha\rangle$. The values of the parameter $k$ are
obtained by fitting the empirical division rates in (a) with a Hill
function with $n=6$.
(c) Direct proportionality between the half-maximum position of the
division rate and the average cell size $g=B \langle V_0\rangle$.  The
$g$ values are obtained by fitting as in (b).
(d) The division rates corresponding to different conditions collapse
in a universal curve if the size is rescaled with its average value,
and the division rate is rescaled with the corresponding average
growth rate.  Therefore, data from different strains and nutrient
conditions can be in principle merged, if appropriately rescaled, and
used to infer the universal division rate function (dashed line in the
plot) with larger statistics.  }
\label{figShd}
\end{figure*}

\clearpage).

\begin{figure*}
\includegraphics[width=0.6\textwidth]{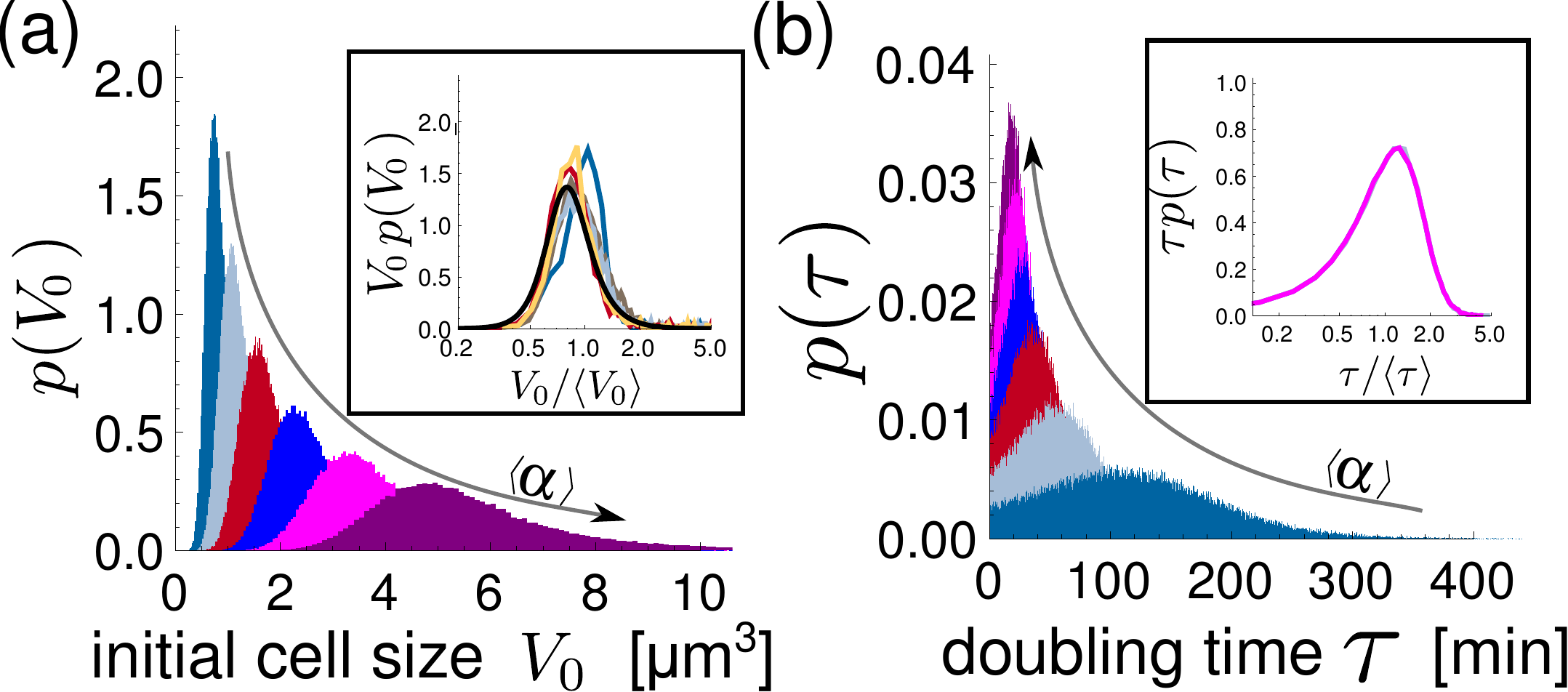}
\caption{ \textbf{Finite-size scaling in a sizer model, with
    Hill-function division hazard.}
  (a) Histograms of initial size distributions obtained with direct
  simulations of the growth-division process for different values of
  $\langle\alpha\rangle$ (from 0.5 to 2 doublings per hour), using the
  two linear relations described in Fig.~\ref{figShd}b,c to estimate
  the parameters of the division rate function.  The inset shows how
  the model prediction corresponding to Eq.~\ref{p0_hill_resc} (black
  line) well captures the empirical rescaled size distributions (same
  as Fig.~\ref{fig:Rescaling}a) (b) Histograms of doubling time
  distributions obtained with direct simulations as in (a).  The inset
  shows that the finite-size scaling is predicted by the model also
  for the doubling time distribution. In fact, the distributions
  $p(\tau) \tau$ as a function of $\tau/\langle\tau\rangle$ perfectly
  collapse on a universal distribution. This distribution can not be
  quantitatively compared to the empirical ones in
  Fig.~\ref{fig:Rescaling}a, since concerted control is neglected in
  the model, resulting in broader and wrongly skewed predicted
  distributions of doubling times.
 \label{figScollapse}
}
\end{figure*}

\clearpage

\begin{figure}
  \includegraphics[width=0.45\textwidth]{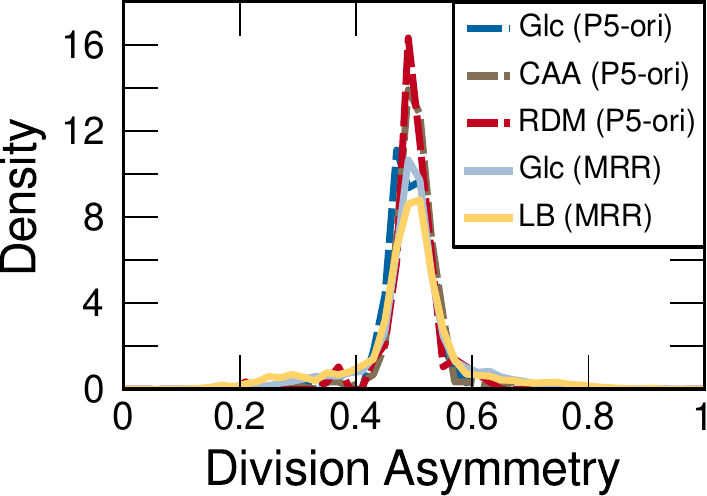}
  \caption{\rev{Cell division is close to symmetric. Histograms of
      ``division asymmetry'' scores for cells. Division asymmetry is
      calculated from the initial lengths of a cell's daughters
      according to $L^{D1}_0/(L^{D1}_0 + L^{D2}_0)$, where $L^{D1}_0$
      is the initial length of daughter 1 and $L^{D2}_0$ is the
      initial length of daughter 2; a division asymmetry of 0.5
      indicates that division was perfectly symmetric.
  In the P5-ori strain over 93\% of cells in each condition have a
  division asymmetry between 0.4 and 0.6, suggesting close to
  symmetric division. This plots compares well to similar plots
  reported in other
  studies~\cite{Campos2014,Taheri-Araghi2014,Soifer2014}. The MRR
  strain had more division asymmetry (16-20\% of cells had a division
  asymmetry score outside of the interval [0.4, 0.6]), perhaps due to
  a higher rate of filamentation in this strain.}}
\label{fig:DivAsymmetry}
\end{figure}

\clearpage

\begin{figure}
  \includegraphics[width=0.75\textwidth]{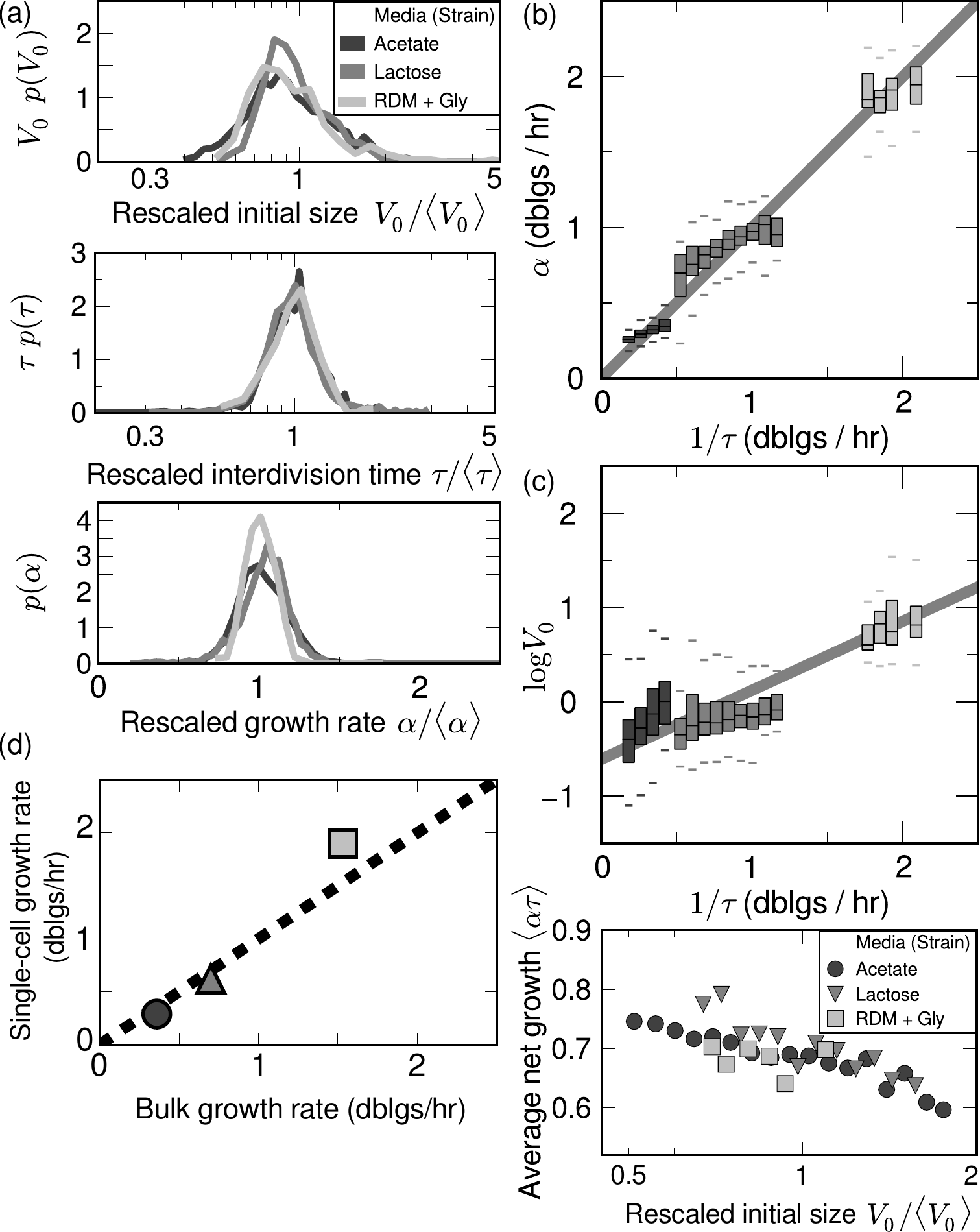}
  \caption{\rev{The main results of this work are consistent with
      previously acquired data from different nutrient
      conditions. \revtwo{Data was taken and analyzed as described
        previously \citet{Kiviet2014}} Nutrient conditions: M9 +
      Acetate, M9 + Lactose, and Neidhardt's Rich Defined Media (RDM)
      + Glycerol, spanning growth rates from between $\approx0.25$ to
      $\approx 1.8$ doubling per hour~\cite{Kiviet2014}.
      Each data set has between 500-1000 cells. \textbf{(a)} Rescaled
      histograms of initial size $V_0$ (\textit{top}), interdivision
      time $\tau$ (\textit{middle}) and \revtwo{growth} rate $\alpha$
      (\textit{bottom}), as in Figs~\ref{fig:Rescaling}
      and~\ref{fig:AlphaDistribution}.  \textbf{(b)} Correlation
      between \revtwo{growth} rate $\alpha$ and inverse interdivision
      time, binned by $1/\tau$ (cf. Fig.~\ref{fig:AlphaInvTau}a).
      \textbf{(c)} Shaechter-Maal\o{}e-Kjeldgaard plot (top) showing
      correlation between log initial size and inverse interdivision
      time (cf. Fig.~\ref{fig:SchaechterFigure}a), and Average net
      growth $\alpha\tau$ binned by log initial size (bottom,
      cf. Fig.~\ref{fig:SchaechterFigure}d)}. \revtwo{ \textbf{(d)}
      comparison of bulk and agar growth rates
      (cf. Fig.~\ref{fig:S-BulkGrowthCompare})}. }
    \label{fig:TansData}
  \end{figure}

\vspace{3cm}

\clearpage
 \newpage

\begin{center}
  {\Large \textbf{Supplementary Tables}}
\end{center}

\vspace{3cm}


\begin{table}[h]
  \centering
{\scriptsize
  \begin{tabular}{|p{6cm}||c|c|c||c|c|}
    \hline
    \textbf{Data set} & \multicolumn{3}{c||}{\textbf{P5-ori}} &
    \multicolumn{2}{c|}{\textbf{MRR}} \\ 
    \cline{2-6}
    & Glc     & CAA    & RDM     & Glc       & LB \\
    \hhline{|=||=|=|=||=|=|}
    \textit{Segmentation and tracking algorithm} &
    \multicolumn{3}{c||}{ } & \multicolumn{2}{c|}{ }  \\ 
    \hhline{|=||=|=|=||=|=|}
    \textbf{Objects touch border or beginning/end of movie$^a$} & 456
    &7,725  &812  &14,578  & 7.136  \\
    \hline
    \textbf{Unsuccessful tracking$^b$} & 10,203  & 7,882
    & 7,167  & 35,608  & 35,000 
    \\ 
    \textbf{Unsuccessful tracking / \emph{estimated cells} $^+$} &6,428
    & 5,178 & 3,705 & 28,842 & 15,225 \\ 
    \hline
    \textbf{No mother$^c$} & 127  & 422  & 2
    & 6  & 28  \\ 
    \hline
    \textbf{Negative growth rate} &22 & 14  & 9
    & 236  & 85 \\ 
    \hline
    \textbf{Low $r^2$ value$^d$} & 398  & 517 & 147
    & 1,644 & 580 \\ 
    \hline
    \textbf{Division time $<8.6$ min$^f$} & 0  & 28 
    & 78  & 6  & 23 \\ 
    \hhline{|=||=|=|=||=|=|}
    \textbf{Final} & \textbf{5,765} & \textbf{10,802}  & \textbf{2,579}
    & \textbf{19,564}  & \textbf{9,489} \\ 
    \hhline{|=||=|=|=||=|=|}
    \textit{Steadiness filters} & \multicolumn{3}{c||}{ } & \multicolumn{2}{c|}{ }\\
    \hhline{|=||=|=|=||=|=|}
    \textbf{Restriction of the interval of generations considered$^e$
    } &4,803  & 4,897     & 744  & 16,861  & 7,839  \\ 
    \hhline{|=||=|=|=||=|=|}
    \textbf{Final} & \textbf{962} & \textbf{5,905}  & \textbf{1,835}
    & \textbf{2,703}  & \textbf{1,650} \\ 
    \hline
  \end{tabular}
}
\caption{\revtwo{Summary of the effects of  data-analysis
    filters.  Each row of the table counts the objects 
    {}$^a$ Excluded objects were touching the border of the image for at
    least one frame, or  were present at either the start or the end
    of the move, precluding assignment of initial or final 
    size. {}$^b$ Excluded objects were lost for at least a frame during
    tracking due to image segmentation errors. This was often due to
    cells being lost for a frame which disrupted
    tracking. Importantly, most of these excluded tracks artificially
    inflate the number of objects excluded with respect to the number
    of excluded \emph{cells}; for example, if the track of a single
    cell is interrupted   3 times during its cell cycle, then 4
    objects representing the  same cell is discarded by the 
    filter. {}$^+$ Estimate of the number of cells excluded by this filter 
    as the number of excluded objects times the  
    ratio of the average track length for excluded objects relative
    to  passing objects. {}$^c$ Not assigned a mother by the tracking algorithm,  
    due to errors in segmentation. {}$^d$ Growth of excluded cells
    with a goodness-of-fit ($r^2$) 
    to an exponential of less than 0.8. Excluded objects were typically
    incorrect segmentations lasting several frames, affecting estimated
    size and growth rate.} 
  \revtwo{{}$^e$ Cells were restricted by  generation based on examination of 
    the steadiness of the data set (see  Fig.~\ref{fig:S-steadyness_by_gen} and
    \ref{fig:S-unfiltered-results}). {}$^f$ Exclusion of  
    track lengths shorter than 8.6 minutes (see  Methods for an explanation). This filter
    impacts very few objects.}}    
  \label{tab:DataSummary}
\end{table}

\end{document}